\documentclass{aa}
\usepackage{amsmath}
\usepackage{wasysym}

\usepackage{marvosym}
\usepackage{txfonts}
\pdfoutput=1
\usepackage[pdftex]{color,xcolor}
\usepackage[pdftex]{graphicx}
\usepackage[]{hyperref}               
\hypersetup{pdfauthor=Christoph Mordasini}
\hypersetup{breaklinks=true,colorlinks=true,urlcolor=blue, linkcolor=blue,citecolor=blue}
\usepackage{mathtools}
\usepackage{natbib}
\usepackage{breqn}
\usepackage{setspace}

\bibpunct{(}{)}{;}{a}{}{,}

\def\mearth{M_\oplus}
\def\rearth{R_\oplus}

\def\fpg{f_{\rm D/G}} 

\def\f1{f_{\rm I}}

\newcommand{\rj}{R_{\textrm{\tiny \jupiter}}}

\def\beq{\begin{equation}}
\def\eeq{\end{equation}}

\def\fopa{f_{\rm opa}}
\def\t2{\tau_{\rm II}}

\def\sigmas0{\Sigma_{\rm s,0}}

\def\apjl{ApJ}                
\def\apjs{ApJS}               
\def\ao{Appl.Optics}          
\def\aap{A\&A}                

\def\({\left(}
\def\){\right)}
\def\<{\left<}
\def\>{\right>}

\defcitealias{movshovitzpodolak2008}{MP08}
\defcitealias{movshovitzbodenheimer2010}{MBPL10}
\defcitealias{mordasiniklahr2014}{Paper I}

\begin{document}

\title{Grain opacity and the bulk composition of extrasolar planets. II.}
\subtitle{An analytical model for the grain opacity in protoplanetary atmospheres}

\author{C. Mordasini\thanks{Reimar-L\"ust Fellow of the MPG}}

\institute{Max-Planck-Institut f\"ur Astronomie, K\"onigstuhl 17, D-69117 Heidelberg, Germany }

\offprints{C. Mordasini, \email{mordasini@mpia.de}}

\date{Received  24.02.2014  / Accepted 25.6.2014}

\abstract
{We investigate the grain opacity $\kappa_{\rm gr}$ in the  {atmosphere (outer radiative zone) of forming planets}. This is important  {for the observed} planetary mass-radius relationship since  $\kappa_{\rm gr}$ affects the primordial H/He envelope mass of low-mass planets and the critical core mass of giant planets.}  
{The goal of this study is to derive a simple analytical model for  $\kappa_{\rm gr}$  {and to explore its implications for the atmospheric structure and resulting gas accretion rate.}  }
{Our model is based on the comparison of the timescales of the most important microphysical processes. We consider grain settling in the Stokes and Epstein drag regime, growth by Brownian motion coagulation and differential settling, grain evaporation in hot layers, and grain advection due to the contraction of the envelope. With these timescales and the assumption of a radially constant grain flux, we derive the typical grain size, abundance, and opacity.}
{We find that the dominating growth process is differential settling. In this regime, $\kappa_{\rm gr}$ has a simple functional form and is given as $27Q/8H\rho$ in the Epstein regime in the outer atmosphere and as $2Q/H\rho$ for Stokes drag in the deeper layers. Grain growth leads to a typical radial structure of $\kappa_{\rm gr}$ with high ISM-like values in the outer layers but a strong decrease towards the deeper parts where $\kappa_{\rm gr}$ becomes so low that the grain-free molecular opacities take over.}
{In agreement with earlier results, we find that $\kappa_{\rm gr}$ is typically much lower than in the ISM. In retrospect, this suggests that classical giant planet formation models should have considered the grain-free case as  {equally} meaningful as the full ISM opacity case. The equations also show that a higher dust input  {in the top layers} does not  {strongly} increase $\kappa_{\rm gr}$. This has two important  implications. First, for the formation of giant planet cores via pebbles, there could be the  {adverse effect} that pebbles tend to increase the grain input  {high} in the atmosphere due to ablation. This could potentially increase the opacity, making giant planet formation difficult. Our study shows that this  {potentially adverse effect should not be important}. Second, it means that a higher stellar [Fe/H]  which presumably leads to a higher surface density of planetesimals only favors giant planet formation without being detrimental to it due to an increased $\kappa_{\rm gr}$. This corroborates the result that core accretion can explain the observed increase of the giant planet frequency with stellar [Fe/H].}
 \keywords{opacity --  planets and satellites: formation  --  planets and satellites: atmospheres -- planets and satellites: interiors -- planets and satellites: individual: Jupiter -- methods: analytical} 

\titlerunning{Analytical model for grain opacity}
\authorrunning{C. Mordasini}

\maketitle

\section{Introduction}\label{sect:grainopa}
Thanks to the precise determination of the mass and radius of many extrasolar planets, it has recently become possible to study the bulk composition of a rapidly growing number of exoplanets. These planets have masses ranging from super-Earth to Jovian masses. This allows to investigate statistically how the bulk composition depends on fundamental planetary parameters like the mass or semi-major axis, and how this compares to the Solar System with its  basic types of planets (terrestrial, ice giants, and gas giants). A specific question that has  recently attracted a lot of attention is the transition from solid to gas dominated planets, or in other words, how the hydrogen/helium mass fraction depends on a planet's properties \citep[e.g.,][]{millerfortney2011,weissmarcy2013,lopezfortney2013b,marcyisaacson2014}.

Planet formation theory based on the  core accretion paradigm \citep{perricameron1974,mizuno1980,stevenson1982,bodenheimerpollack1986} predicts that the H/He mass fraction is an increasing function of the planet's mass  (see \citealt{rogersbodenheimer2011} and \citealt{mordasinialibert2012c}). This is because the Kelvin-Helmholtz timescale for the contraction of the envelope which controls the gas accretion rate in the sub-critical regime is a decreasing function of the core mass \citep{ikomanakazawa2000}. Population synthesis simulations based on global core accretion models have therefore found a synthetic mass-radius relationship (or in other words, a bulk composition) that resembles the observed one \citep{mordasinialibert2012c}. These calculations also predict that the population of  planets with radii larger than $\sim2\rearth$ can be characterized by planets possessing a H/He envelope, whereas other planetary types dominate at smaller radii. Similar conclusions were reached from observational data \citep{howardmarcy2011,gaidosfischer2012,wolfganglaughlin2012,marcyisaacson2014}.   

The Kelvin-Helmholtz timescale is, however, not only a function of the core mass, but also of the opacity $\kappa$ in the protoplanetary envelope \citep[e.g.,][]{ikomanakazawa2000}\footnote{In this work, we mean with opacity always the Rosseland mean except when otherwise stated.}. The latter is thought to be mainly due to tiny dust grains suspended in the protoplanet's outer radiative zone.  The lower the opacity, the shorter the cooling timescale, so that for a fixed core mass a more massive envelope can be accreted during the lifetime of the protoplanetary gas nebula. This eventually translates into different planetary mass-radius relationships. This is because at a given total mass, the H/He mass fraction will be higher if $\kappa$ was low during formation which causes a larger planetary radius during the evolutionary phase after the dispersion of the nebula \citep[e.g.,][]{fortneymarley2007a,valenciaguillot2013}.  Interestingly, it means that it is at least in principle possible to relate an observable quantity like the mass-radius relationship to the microphysics of grain growth during   formation (for other effects, see \citealt{vazankovetz2013}). This link was studied in \citet{mordasiniklahr2014}, hereafter \citetalias{mordasiniklahr2014}, with the conclusion that in order to reproduce the observed planetary bulk compositions, the grain opacity $\kappa_{\rm gr}$ in protoplanetary atmospheres must be much smaller than in the interstellar medium (ISM). 

The magnitude of the grain opacity is not only central for the H/He content of low-mass planets. It is also a key factor for giant planet formation. This is because the critical core mass beyond which rapid runaway gas accretion occurs, depends on the opacity\footnote{This holds for strictly static calculations like in \citet{mizuno1980} or the corresponding analytical solution \citet{stevenson1982}. In time dependent calculations as in \citet{pollackhubickyj1996}, the crossover mass (which is the equivalent of the critical mass) is independent of $\kappa$, but the time until the crossover mass is reached depends on it \citepalias{mordasiniklahr2014}. This means that qualitatively, the implications are the same.}. For example, as already found in the early work of \citet{mizuno1980}, the critical core mass can vary from 1.5 $\mearth$ at grain-free opacity to 12 $\mearth$ at full ISM opacity. This is  a very significant difference.

The potentially very important consequences of different grain opacities outlined above motivated \citet{podolak2003} to develop a detailed but numerically expensive model for the microphysics  of the grains in protoplanetary atmospheres that yields a physically motivated $\kappa_{\rm gr}$. It numerically solves the Smoluchowski equation in each atmospheric layer, taking into account the effects of grain growth, settling, and vaporization.  This model was further elaborated in \citet{movshovitzpodolak2008}, hereafter \citetalias{movshovitzpodolak2008}, and applied to an atmospheric structure of a protoplanet calculated by \citet{hubickyjbodenheimer2005}. In \citet{movshovitzbodenheimer2010}, hereafter \citetalias{movshovitzbodenheimer2010}, and later \citet{rogersbodenheimer2011} this grain evolution model was finally self-consistently coupled to the giant planet formation model of Bodenheimer and collaborators. \citetalias{movshovitzbodenheimer2010} then simulated the in situ formation of Jupiter as in \citet{pollackhubickyj1996} but using now a physically motivated grain opacity, instead of an arbitrarily scaled ISM opacity.

These calculations showed that first, the conditions in protoplanetary atmospheres are such that the aforementioned processes may substantially alter the properties of the grains. Second, it was found that the grain opacity and resulting optical depth of the atmosphere can be substantially reduced relative to the ISM case. This results in a short formation timescale of giant planets, and allows low-mass cores of only about 4 $\mearth$  to trigger gas runaway accretion during the typical lifetime of a protoplanetary disk. For comparison, at full ISM opacity, this would take several 100 Myr \citepalias{mordasiniklahr2014}.  Finally, the radial structure of the opacity in the protoplanet's atmosphere is characterized by a high, ISM-like opacity in the outer layers ($\kappa_{\rm gr}\sim1$ cm$^{2}$/g) that falls to very low values  ($\kappa_{\rm gr}\sim10^{-3}$ cm$^{2}$/g) in the deep layers close to the radiative-convective boundary. This substantially differs from the radial opacity structure that is obtained when simply scaling the ISM opacity (see Sect. \ref{sect:compism}).

Despite this, with the exception of the aforementioned works, it has been customary in the literature to model the effect of a lower grain opacity by simply reducing the ISM opacity by some arbitrary uniform reduction factor $\fopa$. This is the case for very different kinds of studies ranging from analytical work to 3D hydrodynamic simulations (e.g., \citealt{mizunonakazawa1978,stevenson1982,pollackhubickyj1996,papaloizounelson2005,hubickyjbodenheimer2005,tanigawaohtsuki2010,levisonthommes2010,horiikoma2011,ayliffebate2012,dangelobodenheimer2013}). Due to the lack of knowledge of even just the magnitude of the grain opacity $\kappa_{\rm gr}$, very large ranges for $\fopa$ were often considered, typically from $10^{-3}$ to 1.  The same basic approach was also taken in \citetalias{mordasiniklahr2014}, with the improvement that $\fopa$ was  calibrated by determining which reduction factor leads to the same formation timescales for Jupiter as in \citetalias{movshovitzbodenheimer2010}. A best-fitting value of $\fopa\approx0.003$ was found in this way, and used as the nominal  value in population synthesis calculations.  It is however clear that one global reduction factor that was calibrated for certain conditions only like a semi-major axis, core mass, or pressure and temperature in the protoplanetary disk is in principle not applicable for an entire population of planets which cover a very large range in planetary properties. 

This  shows that it is desirable to develop a better analytical understanding of the mechanisms governing the grain evolution in protoplanetary atmospheres, and to be able to estimate the expected magnitude of the grain opacity. The goal of this work is therefore to derive a first (and  simple) analytical model for the grain opacity based on microphysical processes. Despite the simplicity, this model should be able to predict $\kappa_{\rm gr}$ in a way that also quantitatively, it agrees with the results of the numerical model of \citetalias{movshovitzpodolak2008} and \citetalias{movshovitzbodenheimer2010}. On one hand, this fosters the physical understanding  how important parameters like the planet's mass or the dust-to-gas ratio in the disk (which could be related to the stellar [Fe/H]) influence the  grain opacity and therefore the growth history of a protoplanet. On the other hand, from a practical point of view, the model  makes it possible to efficiently calculate a physically motivated $\kappa_{\rm gr}$. This is because the expressions for $\kappa_{\rm gr}$ derived below can be easily integrated in the calculation of the standard internal structure equations of a protoplanetary envelope without requiring special, time consuming measures like sub-stepping or iterations. This in turn means that this model can be used in numerical simulations of the accretion of gas by protoplanets in general, and in population synthesis calculations in particular.  {In a contemporaneous manuscript, \citet{ormel2014} has presented a model of intermediate complexity. Similar to the analytical model presented here, it assumes that per layer there is only one (or two) typical grain sizes. Similar to the numerical model, the grain size is found numerically by integrating an additional fifth equation besides the normal planetary structure equations. This in particular allows to take into account the effect of a bimodal grain size distribution and of a radially varying grain input from planetesimal ablation. }

In future work, we will therefore run population syntheses using the analytical grain opacity model  {and the work of \citet{ormel2014}}, and compare with the results of \citetalias{mordasiniklahr2014}. We will also take into account other effects like the concurrent formation of several embryos \citep{alibertcarron2013}, the effect of envelope enrichment by heavy elements \citep{fortneymordasini2013}, or the evaporation of the primordial H/He envelope of close-in planets  \citep{jinmordasini2014}. This will help to better understand the role of the grain opacity in shaping the observed mass-radius relationship and thus bulk composition of the extrasolar planets.  

The contents of this paper are as follows: in Sect. \ref{sect:analyticalmodel} we first write down the general equations describing the dynamics of the grains due to dust settling, growth (by coagulation due to Brownian motion or differential settling), evaporation, and due to the contraction of the gaseous envelope itself.  In Sect. \ref{sect:grainaccrrateplanetesimal} we discuss the two mechanisms that bring new grains into the protoplanetary atmosphere, which are the accretion together with newly accreted nebular gas, and second the breakup of planetesimals flying through the atmosphere. We then derive analytical expressions for $\kappa_{\rm gr}$  in Sect. \ref{sect:calctypsizeopa}, treating the five relevant regimes (Epstein/Stokes drag, Brownian coagulation/differential settling, plus grain advection) separately.  In Sect. \ref{sect:compMPforgivenrhoT}, the analytical model is  applied to calculate the grain opacity in the same atmospheric structure of \citet{hubickyjbodenheimer2005} that was already considered in \citetalias{movshovitzpodolak2008}. A detailed analysis of the processes regulating the growth of the grains is made, as well as different comparisons of the results of the analytical and numerical model.
In Sect. \ref{sect:combsims}, we show the results of coupling the analytical grain opacity model with our core accretion model. This section is in parallel with the work of \citetalias{movshovitzbodenheimer2010}, and extensive comparisons are made between the analytical and numerical result.  Finally, in Sect. \ref{sect:summaryconclusions} we summarize our results and present our conclusions.

\section{Analytical model}\label{sect:analyticalmodel}
The basic approach of the analytical model is to calculate for each layer the typical size of the grains by comparing the  timescales of the governing microphysical processes. Based on the typical size of the grains, and the (supposed) conservation of the radial flux of the dust grains across the atmosphere  (due to settling), it is then possible to calculate the grain opacity in each layer. This fundamental approach is very similar to the one of \citet{rossow1978} who calculated the microphysics of cloud formation in the atmospheres of different planets in the Solar System. This approach was later used by \citet{coopersudarsky2003} to study cloud formation and the resulting opacity in atmospheres of brown dwarfs. In the context of grain opacity in protoplanetary atmospheres, estimations based on timescale arguments were also made in \citet{podolak2004a,movshovitzbodenheimer2010,nayakshin2010},  and \citet{helledbodenheimer2011}.  {In the context of grain growth in protoplanetary disks, timescale arguments were for example used in \citet{birnstielklahr2012}.} Here we develop a simple, but complete model to dynamically calculate the grain opacity in protoplanetary atmospheres as a function of local atmospheric properties. It can be used in numerical simulations of (giant) planet formation (Sect. \ref{sect:combsims}).

\subsection{Relevant timescales}\label{sect:relevanttimescales}
For the calculation of the grain size, the most important timescales are the settling and growth timescale. Additional timescales that must be considered are the advection timescale of the grains (because the gas envelope is itself not static) and the evaporation timescale. While the  specific form of the settling and  growth timescale depends on the drag regime (Epstein and Stokes regime) and the growth mechanism (Brownian motion coagulation or differential settling), they share the property that the growth timescale increases with increasing grain size, while the settling timescale decreases with increasing grain size. This means that in a given layer, very small grains will only be found in small quantities, because they quickly grow to larger sizes. Very large grains will also be present only in small quantities because they quickly fall  out of the layer.  The typical size is therefore found by equating the two timescales.

In this section, we first define the general equations necessary for the analytical model, including the different aerodynamic regimes, the expression for the grain accretion rate, and the general expressions for the opacity.  We then calculate the typical grain size for five regimes: Brownian coagulation in the Epstein regime, Brownian coagulation in the Stokes regime, differential settling in the Epstein regime, differential settling in the Stokes regime. And finally, we study  the advection regime where the radial motion due to the contraction of the gas envelope is dominant over the settling of the grains. Once the grain size (and abundance) is known, the opacity can be calculated. We then address the effect of grain evaporation, give an approximation for the extinction coefficient, and show how the different regimes are coupled to obtain the final expression for $\kappa$. 

\subsection{General equations}
The settling timescale is found as the timescale it takes the grains to cross a typical length scale in the atmosphere if they are settling at a velocity $v_{\rm set}$ relative to the gas. The natural length scale in an atmosphere is the scale height $H$, so that the settling timescale $\tau_{\rm set}$ can be estimated in the limit of a static gas envelope as $\tau_{\rm set}=H/v_{\rm set}$.  The scale height in the atmosphere is 
\beq
H=\frac{k_{\rm B} T}{\mu m_{\rm H} g}
\eeq
where $k_{\rm B}$ is the Boltzmann constant, $T$ the temperature of the gas, $\mu$ its mean molecular weight ($\approx 2.4$ for solar composition and molecular hydrogen), $m_{\rm H}$ the mass of a hydrogen atom, and $g$ is the gravitational acceleration that is given as $g=G M(R)/R^{2}$. In this equation, $G$ is the gravitational constant, $R$ the radial distance measured from the center of the planet (called below the ``height''), and $M(R)$ the mass inside of $R$ (core plus envelope gas).  The structure of a protoplanetary envelope usually consists of a deep convective zone and an upper radiative zone, even though more complicated structures with several convective zones occur \citep[see, e.g.,][ or Fig. \ref{fig:crossM10}]{bodenheimerpollack1986}. Our model for the grain opacity applies to the upper radiative zone that we call (as \citetalias{movshovitzbodenheimer2010}) the atmosphere. 

Typically, the mass in the radiative zone is much smaller than the mass of the core and the envelope mass in the convective zone, so that $M(R)$ is nearly constant. However, a constant $M(R)$ is not required for the analytical model for $\kappa_{\rm gr}$.
 
\subsubsection{Aerodynamic regimes}\label{sect:aeroregimes}
The settling and growth of the grains depends on the aerodynamic regime they are in. Two dimensionless numbers characterize the aerodynamic regimes. First, the Knudsen number Kn describes whether the grains are in the free molecular flow regime where the grains interact with single molecules (large Kn), or if the gas acts as a continuous, hydrodynamic fluid  (small Kn). The Knudsen number is given as  $\mathrm{Kn}=\ell/a$ where $a$ is the radius of a grain and $\ell$ is the mean free  path of a gas molecule that is calculated as 
\beq
\ell=\frac{\mu m_{\rm H} }{ \sqrt{2} \rho \sigma_{\rm m}}.
\eeq
In this expression, $\rho$ is the gas density and $\sigma_{\rm m}$ the collisional cross section of a molecule, $\sigma_{\rm m}=\pi d_{\rm m}^{2}$. For simplicity, we set the molecular diameter $d_{\rm m}$ equal to the  one of molecular hydrogen H$_{2}$ at standard conditions $\approx2.7\times 10^{-8}$ cm \citep{waldmann1958}. It is clear that,  in reality, $d_{\rm m}$ would depend  on both the composition of the gas as well as its pressure and temperature. We set the critical Knudsen number for the transition from free molecular  to continuum flow to Kn=4/9 (e.g., \citealt{stepinskivalageas1996}). 

The second dimensionless number is the Reynolds number. It describes whether the flow is laminar (small Re) or turbulent (large Re) and is given as $\mathrm{Re}=2 a \rho v/\eta$ where $v$ is the settling velocity of the grains while $\eta$ is the dynamic viscosity of the gas.  It is given in the approximation of hard-sphere molecules with a Maxwellian velocity distribution as $\rho v_{\rm th} \ell/3$  where $v_{\rm th}$ is the mean thermal velocity of the gas,
\beq
v_{\rm th}=\sqrt{\frac{8 k_{\rm B} T}{\pi \mu m_{\rm H}}}.
\eeq
More accurate expression for $\eta$ based on, e.g., the Lennard-Jones potential instead of hard spheres exist (see, e.g., \citealt{podolak2003}), but given the other simplifications in the model, we stick to this basic expression.

We will find below that the grains are typically in the free molecular flow regime in the outer parts of the atmosphere, and in the continuum flow in the inner parts. The Reynolds number on the  other hand is always small (typically much less than unity) throughout the atmosphere, so that the grains are always in the laminar flow regime (see Fig. \ref{fig:mp2008fpgknre}). This justifies the choice of drag laws used below. 

In the derivation below, we will further assume that the grains are always settling at the terminal velocity where the drag and gravitational force balance each other. This holds for particles with a short stopping timescale and  can be quantified by a third dimensionless number that is the ratio of the stopping time $\tau_{\rm stop}$ of the particles to the characteristic flow time\footnote{A similar dimensionless number is the Stokes number. It is usually defined in the context of turbulent flows \citep[e.g.,][]{birnstielklahr2012}, while we are dealing with a situation where in principle there is no background flow of  gas (at least if we neglect the effect of envelope contraction). Therefore, we do not call this quantity directly Stokes number despite its similar nature.}. In the current situation, the background motion is the grain settling, therefore we define $\mathrm{RSS}=\tau_{\rm stop}/\tau_{\rm set}$ where small RSS mean that the terminal velocity is rapidly reached. In this expression, the stopping time is $\tau_{\rm stop}=m_{\rm gr} v/F_{\rm D}$ where $F_{\rm D}$ is the drag force and $m_{\rm gr}$ is the mass of a dust grain. As \citetalias{movshovitzpodolak2008} we assume for simplicity spherical dust grains (which might not be justified, see Sect. \ref{sect:limitations}) so that $m_{\rm gr}=4/3 \pi \rho_{\rm gr} a^{3}$ where $\rho_{\rm gr}$ is the material density of the grains. Like \citetalias{movshovitzpodolak2008} we only  consider  silicate grains in this work,  and follow them in setting the material density of the grains to $\rho_{\rm gr}=2.8$ g/cm$^{3}$. 
  
We will find below (Sect. \ref{timescales}) that the grains in protoplanetary atmospheres  have very small RSS$\sim$$10^{-7}$,  and very short stopping times of a few seconds to minutes. The assumption of settling at the terminal velocity in the settling regime or of a motion at the velocity of the gas in the advection regime (Sect. \ref{sect:advect}) is therefore  justified. 

\subsubsection{Planetesimal mass deposition, grain accretion rate, and dust-to-gas ratio}\label{sect:grainaccrrateplanetesimal}
In this section we specify the rate at which grains are accreted into the protoplanetary atmosphere, $\dot{M}_{\rm gr}$. Two different sources exist: first, grains that are brought into the atmosphere together with the newly accreted nebular gas and second grains that are due to the accretion of planetesimals. During their flight through the envelope, planetesimals undergo mass loss via thermal ablation. Large impactors are additionally aerodynamically disrupted by large dynamic pressures (see \citealt{podolakpollack1988} and \citealt{mordasinialibert2006}). The aerodynamic disruption fragments big impactors into small pieces that are then easily digested by thermal ablation since fragmentation greatly increases the ablating surface. 

Grains that are accreted by the gas are injected into the planet's envelope at its outer radius. Grains originating from planetesimal ablation can in contrast be injected into the envelope at different heights depending on the aforementioned mechanism that determine the radial mass deposition profile of the planetesimals. This mass deposition profile can be complex, and depends on the planetesimal size, the impact velocity, the tensile strength of the body, the impact geometry and so on. In this work, we consider two idealized, limiting cases. 

First, in the ``deep deposition'' case, it is assumed that the planetesimals fly through the atmosphere (outer radiative zone) without losing any mass. The planetesimals deposit their mass only in the deeper parts of the envelope. If the mass deposition occurs at a radius where the grain-free gas opacity dominates over the grain opacity in any case, and/or where the temperature in the background envelope is higher than the evaporation temperature of the grains, then the grains brought in by the planetesimals do not influence the opacity and therefore the atmospheric structure. The later depends of course on the opacity, therefore the mass deposition profile and atmospheric profile are interdependent in both directions. The only source of grains in the ``deep deposition'' case is then the accretion of gas that occurs at a rate $\dot{M}_{\rm XY}$. Grains are mixed into this gas at a dust-to-gas mass ratio in the protoplanetary disk, $f_{\rm D/G, disk}$. Typically  $f_{\rm D/G, disk}\approx0.01$, but this ratio could also be much lower if  most solids have already been incorporated into large bodies. The  dust-to-gas  ratio  in the disk could also scale with the stellar [Fe/H], establishing an interesting link between opacity (and in the end the planetary H/He content, i.e., bulk composition) and stellar properties. We discuss the impact of [Fe/H] on $\kappa_{\rm gr}$ in Sect. \ref{sect:implicationsfehpebble}.

Second, in the ``shallow deposition'' case, it is in contrast assumed that the planetesimals break up into grains at the very top of the atmosphere, so that the full accretion rate of planetesimals $\dot{M}_{Z}$ contributes to the grain accretion rate. In equations, the two cases are
\beq\label{eq:grainaccrationrate}
 \dot{M}_{\rm gr} =
  \begin{cases}
   f_{\rm D/G, disk} \dot{M}_{\rm XY}& \textrm{for deep  deposition}   \\
   f_{\rm D/G, disk} \dot{M}_{\rm XY}  +\dot{M}_{Z}     & \text{for shallow  deposition.} 
  \end{cases}
\eeq

\begin{figure}
\begin{center}
\includegraphics[width=\columnwidth]{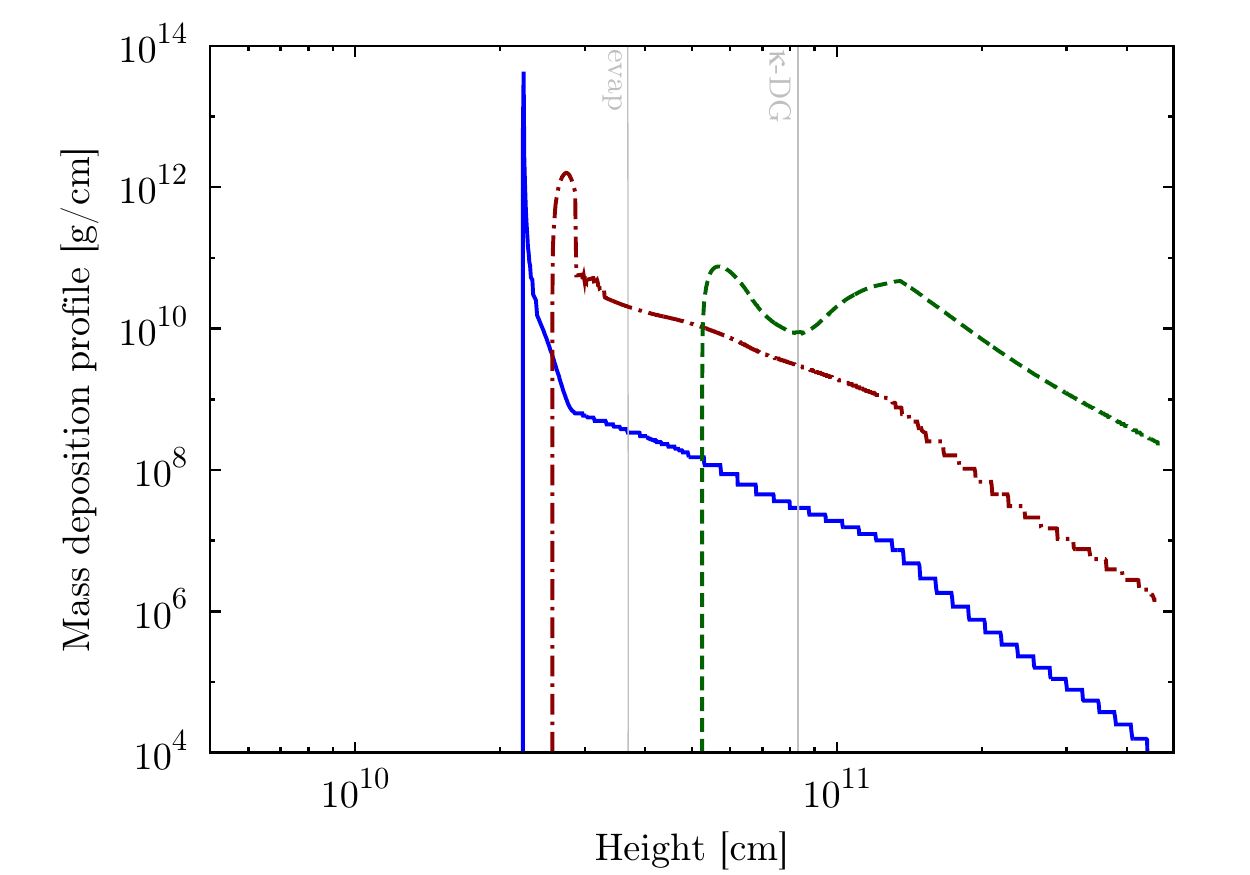}
\caption{Mass deposition profile (ablated mass per unit length) as a function of height for 100 km planetesimals (solid blue line). The brown dashed-dotted line is the mass deposition for 1 km planetesimals, multiplied by a factor $10^{6}$ while the green dashed line shows the result for  10 m planetesimals, multiplied by a factor $10^{12}$ so that the total mass deposited is identical in the three cases. The gray vertical lines show the radius where the envelope temperature becomes so high that silicate grains evaporate (``evap'') and where the molecular opacities alone become larger than the opacity due to dust in the deep deposition case (``$\kappa$-DG''). The steps in the right part of the lines are a numerical artifact without physical meaning. }\label{fig:massdepo}
\end{center}
\end{figure}

The larger the planetesimals, the deeper they  deposit in general their mass, even though the actual behavior is complex due to different mechanism leading to mass loss (pure thermal ablation versus aerodynamic disruption; see \citealt{mordasinialibert2006}). Still, to first order, the ``deep (shallow)'' deposition case can be associated with the accretion of big (small) planetesimals. This is illustrated in Fig. \ref{fig:massdepo}. The figure shows the radial mass deposition profile (ablated mass per unit length) of planetesimals of different sizes as a function of height in the protoplanetary atmosphere in the \citetalias{movshovitzbodenheimer2010} comparison calculation (see Sect. \ref{sect:tempevoSigma10} below for a description). The profiles were obtained by simulating the impact of a planetesimal with an impact model similar to \citet{podolakpollack1988}, but using an updated model for the heat transfer coefficient, and a multi-staged aerodynamic fragmentation model (see \citealt{mordasinialibert2006} and \citealt{alibertmordasini2005} for a short description;  a full model description will be given in a future work). 

Figure \ref{fig:massdepo} shows the mass deposition by a planetesimal with an initial size of 100 km, 1 km, and 10 m  {(see also the Appendices \ref{sect:impactgeometry} and \ref{sect:impactcompo})}. The curves for 1 km and 10 m were scaled so that the total deposited mass be the same in the three cases (of order $10^{21}$ g). The planetesimals hit head-on with an initial velocity approximately equal to the planet's escape velocity. Material parameters appropriate for water ice were used which is strictly speaking inconsistent with the assumption of silicate grains made above.  {The consequences of the planetesimal composition on the mass deposition profile is studied in Appendix \ref{sect:impactcompo}. It is found that rocky planetesimals deposit their mass in deeper layers, as expected. The extent of the difference between icy and rocky impactors depends however significantly on the impactor size, with a larger difference for small planetesimals.}
 
One notes that the bigger the planetesimal, the deeper the peak mass deposition, as expected. For the 100 and 1 km sized planetesimals, the mass deposition is very low in the outer parts since pure thermal ablation is not efficient for large bodies \citep{svetsovnemtchinov1995}, but then shoots up by many orders of magnitudes as soon as the impactor is aerodynamically fragmented. This mass (and energy) deposition concentrated over a comparatively small radial domain ($\approx0.05\rj$ in the 100 km case) is reminiscent of the terminal explosion that was seen in the simulations of the collision of comet SL-9 with Jupiter \citep[e.g.,][]{lowzahnle1994} even if the absolute scales are much longer here. The mass deposition profile of the 10 m planetesimal is in contrast much smoother because no violent fragmentation occurs. Much more mass is deposited in the outer layers. Figure \ref{fig:massdepo} also contains two vertical lines. The right line shows where the grain-free molecular opacities for a solar composition gas becomes dominant over the grain opacity under the assumption of  ``deep deposition'' (Sects. \ref{sect:mp08structopa} and \ref{sect:Sigma10structcrossover}).  The left line corresponds to the radius where the temperature in the undisturbed envelope is sufficiently high to evaporate silicate grains on a short timescale (see Sect. \ref{sect:evap}). The radiative-convective boundary is found at nearly the same radius, meaning that most of the mass deposited by 100 and 1 km planetesimals goes in vaporized form into the deep convective zone without contributing much to the grain opacity in the outer atmosphere. The 10 m planetesimals in contrast deposits a significant fraction of their mass in the atmosphere, contribution in an important way to the grain influx into the atmosphere.  {It is clear that in reality, most impacts will not be head-on as assumed here. The consequences of different impact geometries are studied in Appendix \ref{sect:impactgeometry}. As expected, off-center collisions typically increase the mass deposition in the upper layers. However, for 100 km planetesimals, the large majority of the mass is still deposited in the terminal explosion the height of which varies by  $\sim$1 $\rj$ depending on the geometry.}

The plot also indicates that the two extreme cases for $\dot{M}_{\rm gr}$ given by Eq. \ref{eq:grainaccrationrate} represent useful limiting cases bracketing the actual behavior, even if they are clearly strong simplifications of the actual mass deposition profile. Since we assume  that the planetesimals have a size of 100 km in the nominal model, we will consider  ``deep deposition'' as the nominal case, but we will also investigate the ``shallow'' case, which should be more appropriate for the accretion of small bodies (pebbles, cf. \citealt{ormelklahr2010,lambrechtsjohansen2012,morbidellinesvorny2012}).

 {We} will see below  (Sects. \ref{sect:diffsetepstein} and \ref{sect:diffsetstokes}) that somewhat surprisingly, in the most important grain growth regime, the assumption of a ``deep'' or ``shallow'' deposition is not important. We will in fact even find that the grain opacity is generally independent of  $\dot{M}_{\rm gr}$ for the dominant range of conditions. This is clearly the result of this study that has the most important implications for general planet formation theory (Sect. \ref{sect:implicationsfehpebble}). 

Since the grain opacity is important in the outer radiative zone that typically contains only a very small fraction of the total envelope mass \citep[e.g., ][]{bodenheimerpollack1986}, we assume that grains do not accumulate locally in the atmosphere, but that all grains that enter the atmosphere at its outer edge will also settle out of it at the radiative-convective boundary.  As demonstrated by \citetalias{movshovitzpodolak2008}, we can further  assume that the processes governing grain evolution (growth, settling, evaporation) are so fast that the grains quasi instantaneously assume a steady state. This is justified by the fact that these processes  occur on timescales of just 1-100 years (see Fig. \ref{fig:mp2008taucompcoagcoal}), which is typically much smaller than  the timescales on which the properties of the protoplanetary envelope itself change.
  
 We can then assume that the (instantaneous) grain accretion rate is radially constant in the atmosphere,  i.e., $\mathrm{d} \dot{M}_{\rm gr}/{\mathrm{d}  R}=0$. Mass conservation then directly yields the dust-to-gas ratio $\fpg$ (by mass) in an atmospheric layer as 
\beq\label{eq:fpggeneral}
\fpg=\frac{\dot{M}_{\rm gr}}{4 \pi R^{2} \rho v_{\rm set}}
\eeq
where $v_{\rm set}$ is the vertical settling velocity of the grains. Because both $\rho$ and $v_{\rm set}$ vary with $R$, $\fpg$ will in general also be a function of the radius inside the atmosphere and differ from the value at the top $f_{\rm D/G, disk}$, except for the advection regime (Sect. \ref{sect:advect}).

\subsubsection{Basic expression for the opacity}
Equating the growth and settling time yields the typical grain size in each layer. Once this grain size $a$ is found, the settling velocity can also be calculated, so that, together with Eq. \ref{eq:fpggeneral}, and the assumption of spherical grains, one can calculate the number density of the grains $n_{\rm gr} =\fpg \rho/m_{\rm gr}$.  The contribution of grains of size $a$ to the  (wavelength dependent) grain opacity is then \citepalias{movshovitzpodolak2008}
\beq\label{eq:kappax}
\kappa_{x}=\frac{Q(x) \pi a^{2} n_{\rm gr}}{\rho}
\eeq
where $Q(x)$ is the scattering efficiency. Note that in this work, with grain opacity we always mean the opacity per unit mass of envelope material (gas and grains), and not per unit mass of  grains only. The scattering efficiency depends on grain material properties and on the size parameter $x=2 \pi a/\lambda$ where $\lambda$ is the wavelength of the impinging photons. The calculation of $Q$ is described below in Sect. \ref{sect:extinctioncoeff}. In the limit of large grains relative to the typical wavelength $(x\gg 1)$, $Q(x)$  is simply equal to 2. This regime is found below to be the most important one. 

In our simple analytical model, there is only one grain size per layer. If additionally the wavelength dependency of $Q$ over the relevant wavelength domain is weak, then $\kappa$ given by Eq. \ref{eq:kappax} can directly be identified with the Rosseland mean opacity which is the quantity needed to solve the internal structure equations.  With 
\beq\label{eq:ngr}
n_{\rm gr}=\frac{3 \fpg \rho}{4 \pi \rho_{\rm gr} a^{3}}
\eeq
we can also write
\beq\label{eq:kappaxfpga}
\kappa_{x}=\frac{3 Q(x) \fpg}{4 \rho_{\rm gr} a}.
\eeq
This shows that for a nearly wavelength independent $Q$ only the ratio of the dust-to-gas ratio to the typical particle size enters the opacity. 

\subsection{Drag regimes}
We now give the equations for the dust-to-gas ratio and the settling velocity in the Epstein and Stokes drag regimes (see, e.g., \citealt{weidenschilling1977,stepinskivalageas1996}), assuming that the Reynolds number is always small (Sect. \ref{sect:aeroregimes}). We treat the two regimes separately, but note that it is in principle possible to write the drag law in a combined form that approaches the two separate expressions in the limit of  large or small Knudsen numbers  \citep{nayakshin2010}. The reason for treating the regimes separately is that it is then possible to solve analytically for $a$.  {The combined case is treated in Appendix \ref{sect:numsolve}.}

We find that the impact of treating the two regimes separately has no negative consequences, because the transition behaves in a benign way even if, a priori, it is unknown which regime applies to a given atmospheric level. Operationally, one therefore derives the grain size for both drag regimes, and then checks a posteriori which regime is self-consistent (i.e., if the a priori assumed Kn regime is the one that is actually found a posteriori for the resulting grain size). One finds that not only the resulting grain radius is continuous across the transition between the two regimes at Kn=4/9 (as it must be), but also that the overlapping (radial) domain where both the Epstein and Stokes regime lead to self-consistent results (so that it is unclear which regime actually applies) is  small. 

\subsubsection{Epstein: Kn$\geq$4/9 (small grains)}
In the Epstein regime which applies to small grains in the upper part of the atmosphere, the drag force is given as
\beq\label{eq:dragepstein}
F_{\rm D,E}=\frac{4 \pi}{3} \rho a^{2} v v_{\rm th}.
\eeq
This leads to a stopping time  of 
\beq
\tau_{\rm stop,E}=\frac{\rho_{\rm gr} a}{\rho v_{\rm th}}.
\eeq
The terminal settling velocity found by equating the drag and gravitational force in the Epstein regime  and the corresponding settling timescale through an atmospheric scale height $H$ is
\beq\label{eq:vsetEtausetE}
v_{\rm set,E}=\frac{ a g \rho_{\rm gr}}{\rho v_{\rm th}} \, \, \, \, \, \, \, \ \ \ \tau_{\rm set,E}=\frac{H v_{\rm th} \rho}{a g \rho_{\rm gr}}.
 \eeq
Together with Eq. \ref{eq:fpggeneral}, the dust-to-gas ratio is 
\beq\label{eq:fpgepstein}
f_{\rm D/G,E}=\frac{\dot{M}_{\rm gr} v_{th}}{4 \pi a g  R^2 \rho_{\rm gr}}.
\eeq

\subsubsection{Stokes: Kn$<$4/9 (large grains)}
The Stokes regime applies to grains that are large in comparison to the free mean path of the gas molecules. The drag force in this regime is 
\beq\label{eq:dragstokes}
F_{\rm D,S}=6 \pi \eta a v
\eeq
leading to a stopping time of 
\beq
\tau_{\rm stop,S}=\frac{2 \rho_{\rm gr} a^{2}}{9 \eta}. 
\eeq
The terminal settling velocity and the timescale to cross one scale height is 
\beq\label{eq:vsettausetstokes}
v_{\rm set,S}=\frac{2 \rho_{\rm gr} a^{2} g}{3 \rho v_{\rm th} \ell} \, \, \, \, \, \, \, \ \ \ \tau_{\rm set,S}=\frac{3 H \rho v_{\rm th} \ell}{2 \rho_{\rm gr} a^{2} g}.
 \eeq
The dust-to-gas ratio finally is given as
\beq\label{eq:fpgstokes}
f_{\rm D/G,S}=\frac{3 \dot{M}_{\rm gr} v_{th} \ell}{8 \pi a^{2}  g R^2 \rho_{\rm gr}}.
\eeq

\subsection{Brownian coagulation}
As \citetalias{movshovitzpodolak2008},  we consider two processes that  lead to grain growth in protoplanetary atmospheres: first, coagulation due to the Brownian motion of the grains, and second, growth due to the differential settling speed of grains of different sizes. The effects of convection that were included by \citet{podolak2003} and \citetalias{movshovitzbodenheimer2010} (and of turbulence in the atmosphere in general) are neglected since \citetalias{movshovitzbodenheimer2010} find that including or neglecting  convection in the grain calculations has a negligible effect. 

The growth timescale due to Brownian motion coagulation\footnote{We find below that growth by differential settling is the dominant  process. We  include Brownian motion mainly for completeness.} can be derived  (e.g., \citealt{helledbodenheimer2011}) by writing the coagulation timescale $\tau_{\rm coag}$ as  $\tau_{\rm coag}=\ell_{\rm gr}/V_{\rm th, gr}$ where $\ell_{\rm gr}$ is the mean free path of a grain between collisions with another grain, while $V_{\rm th, gr}$ is its mean thermal velocity. The mean free path is calculated as
\beq
\ell_{\rm gr}=\frac{1}{\sqrt{2} n_{\rm gr} \sigma_{\rm gr}}
\eeq
where $\sigma_{\rm gr}=4 \pi a^{2}$ is the collisional cross section. Assuming that the temperature of the grains and gas is identical, we further have 
 \beq
 V_{\rm th, gr}=\sqrt{\frac{8 k_{\rm B} T}{\pi  m_{\rm gr}}}.
 \eeq
Combining these equations gives the coagulation timescale for Brownian motion
 \beq\label{eq:taucoagbrown}
\tau_{\rm coag}=\frac{\pi}{6 \fpg \rho}\sqrt{\frac{\rho_{\rm gr}^{3}a^{5}}{3 k_{\rm B} T}}
\eeq
 which is identical to the expression in \citet{helledbodenheimer2011}. The coagulation timescale thus decreases with an increasing dust-to-gas ratio as $1/\fpg$ and increases with grain size as $a^{5/2}$. We also see that fluffy grains with a low material density $\rho_{\rm gr}$ coagulate faster, as expected.

\subsection{Calculation of the typical grain size and opacity}\label{sect:calctypsizeopa}
With the expressions for the settling  and growth timescale at hand, we can  proceed to the central part of the analytical model which is the calculation of the typical grain size $a$. We consider five different regimes separately where we always equate the appropriate growth and settling timescales: Brownian coagulation in the Epstein regime, Brownian coagulation in the Stokes regime, differential settling in the Epstein regime, differential settling in the Stokes regime, and, as a special case,  {Brownian coagulation and growth by differential settling}  in the grain advection regime that is due to the contraction of the (gaseous) envelope.  

In the advection regime it is sometimes found that the grain size resulting from equating the growth and   advection timescale is so small that it is smaller than the  radius thought  to be typical for monomers, $a_{\rm mono}$. In such a case (and also if this should occur in any other regime), we replace the grain radius found from the timescale argument by the assumed monomer size. As \citetalias{movshovitzpodolak2008}, we consider monomer size of 1, 10, and 100 $\mu$m (see Sect. \ref{sect:mp2008amonomer}). It is however found  that this condition rarely happens, and that $a_{\rm mono}$ therefore only has a small effect on the global evolution of the planets (Sect. \ref{sect:mbpl2010fdgamono}), provided that the monomer size is not very large (more than $\sim$100 $\mu$m). Very large grains could in contrast lead to a significant reduction of the formation timescale.

\subsubsection{Brownian coagulation: Epstein regime}
 In the first regime, we insert Eq. \ref{eq:fpgepstein} into \ref{eq:taucoagbrown} to find a coagulation timescale of  
\beq\label{eq:taucoagE}
\tau_{\rm coag,E}=\frac{2 a^{7/2} g  \pi^{2} R^{2} \rho_{\rm gr}^{5/2}}{3 \sqrt{3 k_{\rm B} T} \dot{M}_{\rm gr} v_{\rm th} \rho}.
\eeq
Setting this timescale equal to the settling timescale in the Epstein regime, Eq. \ref{eq:vsetEtausetE},  $\tau_{\rm set,E}=\tau_{\rm coag,E}$, and solving for the grain radius $a_{\rm C,E}$ (coagulation, Epstein) yields
\beq\label{eq:aCEbrownepstein}
a_{\rm C,E}=\left(\frac{27 \dot{M}_{\rm gr}^{2} H^{2} k_{\rm B} T v_{\rm th}^{4} \rho^{4}}{4 g^{4} \pi^{4} R^{4} \rho_{\rm gr}^{7}}\right)^{1/9}.
\eeq
This expression is identical to the one derived in \citetalias{movshovitzbodenheimer2010} if their typical length scale is associated with the atmospheric scale height. Inserting into Eq. \ref{eq:fpgepstein} gives the dust-to-gas ratio
\beq
f_{\rm D/G, C,E}=\frac{1}{2^{16/9} 3^{1/3}}\left(\frac{\dot{M}_{\rm gr}^{7} v_{\rm th}^{5}}{\pi^{5} g^{5} H^{2} k_{\rm B} R^{14} T \rho^{4} \rho_{\rm gr}^{2}}\right)^{1/9} 
\eeq
The opacity is finally found by inserting $a_{\rm C,E}$ and $f_{\rm D/G, C,E}$ into Eq. \ref{eq:kappaxfpga}, yielding
\beq\label{eq:kappabrownepstein}
\kappa_{\rm C,E}=\frac{3^{1/3}}{2^{32/9}}\left(\frac{\dot{M}_{\rm gr}^{5} Q^{9} v_{\rm th} }{\pi k_{\rm B}^{2} T^{2} g H^{4} R^{10} \rho^{8} \rho_{\rm gr}^{4}}\right)^{1/9}
\eeq
or, in terms of the fundamental local properties of the atmospheric layer (and $\dot{M}_{\rm gr}$)
\beq
\kappa_{\rm C,E}=\frac{3^{1/3}}{2^{61/18}}\left(\frac{\dot{M}_{\rm gr}^{5} G^{3} M^{3} (\mu m_{\rm H})^{7/2} Q^{9}}{\pi^{3/2} (k_{\rm B} T)^{11/2} R^{16} \rho^{8} \rho_{\rm gr}^{4}}\right)^{1/9}.
\eeq
We see that $\kappa_{\rm C,E}$ increases as $\dot{M}_{\rm gr}^{5/9}$. The  dependency on the position inside the atmosphere $R$ and on $M$, and therefore the planet's core or total mass, is not straightforward to see because $T$ and $\rho$ depend on these quantities, too. 

\subsubsection{Brownian coagulation: Stokes regime}
 In the second regime, we insert Eq. \ref{eq:fpgstokes} into \ref{eq:taucoagbrown} to find a coagulation timescale of  
\beq\label{eq:taucoagS}
\tau_{\rm coag,S}=\frac{4 a^{9/2} g  \pi^{2} R^{2} \rho_{\rm gr}^{5/2}}{9 \sqrt{3 k_{\rm B} T} \dot{M}_{\rm gr} v_{\rm th} \ell \rho}.
\eeq
Setting this equal to the settling timescale in the Stokes regime (Eq. \ref{eq:vsettausetstokes}),  $\tau_{\rm set,S}=\tau_{\rm coag,S}$, and solving for the grain radius, we find
\beq\label{eq:aCSbrownstokes}
a_{\rm C,S}=\left(\frac{2187 \dot{M}_{\rm gr}^{2} H^{2} k_{\rm B} T v_{\rm th}^{4}\ell^{4}\rho^{4}}{64 \pi^{4} g^{4} R^{4} \rho_{\rm gr}^{7}}\right)^{1/13}.
\eeq 
Inserting back into Eq. \ref{eq:fpgstokes} yields the dust-to-gas ratio
\beq
f_{\rm D/G, C,S}=\frac{1}{2^{2} 6^{1/13}}\left(\frac{\dot{M}_{\rm gr}^{9} v_{\rm th}^{5} \ell^{5} \rho_{\rm gr}}{\pi^{5} g^{5} H^{4} k_{\rm B}^{2} T^{2} \rho^{8} R^{18}}\right)^{1/13}
\eeq
and the opacity 
\beq\label{eq:kappabrownstokes}
\kappa_{\rm C,S}=\frac{3^{5/13}}{2^{47/13}}\left(\frac{\dot{M}_{\rm gr}^{7} v_{\rm th} \ell Q^{13}}{\pi g H^{6} k_{\rm B}^{3} T^{3}\rho^{12}\rho_{\rm gr}^{5} R^{14}}\right)^{1/13}
\eeq
which we can also write as
\beq
\kappa_{\rm C,S}=\frac{3^{5/13}}{2^{46/13}}\left(\frac{\dot{M}_{\rm gr}^{7} G^{5} M^{5} (\mu m_{\rm H})^{13/2}Q^{13}}{\pi^{3/2} (k_{\rm B} T)^{17/2} \rho^{13} \rho_{\rm gr}^{5} \sigma_{\rm m} R^{24}}\right)^{1/13}.
\eeq
In this regime we have $\kappa_{\rm C,E}$ increasing as $\dot{M}_{\rm gr}^{7/13}$. The dependency on $R$ and $M$ is again not obvious because of the interdependency with  $T$ and $\rho$.

\subsubsection{Differential settling (coalescence): Epstein regime}\label{sect:diffsetepstein}
Larger grains can sweep up smaller ones due to their higher sedimentation speed, leading to growth by differential settling. This  process is called coalescence in \citet{rossow1978}. In protoplanetary disk, differential settling is the dominant growth mode for larger grains, and makes a rapid growth to much larger grains possible than  Brownian coagulation alone (e.g., \citealt{brauerdullemond2008}). We will find below that in protoplanetary atmospheres, differential settling is also the dominant growth process. To model this growth mode, we use as \citet{coopersudarsky2003} the growth timescales determined by \citet{rossow1978} who generalizes the results for grains of different sizes to determine the growth timescale when (formally) only one grain size is present per layer.

For small grains in the Epstein drag regime (Kn$>$4/9), the timescale for growth by differential settling is 
\beq\label{eq:taucoalEgeneral}
\tau_{\rm coal,E}=\left(\frac{ 27 \rho}{4 \pi \rho_{gr} g}\right)\sqrt{\frac{2 k_{\rm B} T}{\pi \mu m_{\rm H}}}\left(\frac{1}{a^{3} n_{\rm gr}}\right).
\eeq
When using the expression for $v_{\rm th}$, $n_{\rm gr}$, $\fpg$, and the expression for the settling velocity in the Epstein regime,  this corresponds to
\beq\label{eq:taucoalE}
\tau_{\rm coal,E}=\frac{9 v_{\rm th}}{2 \fpg g}=\frac{18 \pi a \rho_{\rm gr} R^{2}}{\dot{M}_{\rm gr}}.
\eeq

One thus finds that $\tau_{\rm coal,E}\propto 1/\fpg \propto a/\dot{M}_{\rm gr}$.  Equating this with the settling timescale in the Epstein regime $\tau_{\rm coal,E}=\tau_{\rm set,E}$ gives a typical grain size
\beq\label{eq:aDE}
a_{\rm D,E}=\left(\frac{\dot{M}_{\rm gr} H v_{\rm th} \rho}{18 \pi g \rho_{\rm gr}^{2} R^{2}}\right)^{1/2}.
\eeq
The dust-to-gas ratio is then
\beq
f_{\rm D/G, D,E}=\left(\frac{9 \dot{M}_{\rm gr} v_{\rm th} }{8 \pi g H \rho R^{2}}\right)^{1/2}.
\eeq
Finally, the opacity is
\beq\label{eq:kappaDiffEpstein1}
\kappa_{\rm D,E}=\frac{27 Q}{8 H \rho}.
\eeq
The opacity in this regime has thus a simple functional form and the remarkable property that it is independent of $\dot{M}_{\rm gr}$. The result for $\kappa_{\rm gr}$ is therefore in particular also the same in the ``deep'' or ``shallow'' deposition regime for planetesimal impacts (Sect. \ref{sect:grainaccrrateplanetesimal}). The lack of a dependence on $\dot{M}_{\rm gr}$ is a consequence of the fact that $\kappa_{\rm gr}\propto \fpg/a$, and that in this regime, both $\fpg$ and $a$ scale in the same way with $\dot{M}_{\rm gr}^{1/2}$.  Physically it means that if we increase $\dot{M}_{\rm gr}$, we increase the dust-to-gas ratio which would in principle increase the opacity, but the increased dust-to-gas ratio also leads to the formation of larger grains, and this decreases the opacity. In the current regime, the two effects compensate each other. It is found that this regime (in the upper layers) and  differential settling in the Stokes regime (in the lower layers) are the dominant ones for the opacity in protoplanetary atmospheres (see Fig. \ref{fig:mp2008taucompcoagcoal}).  At the same time, the grains are typically large in comparison to the wavelength, so that $Q\approx2$. Therefore, in the outer parts of a protoplanetary atmosphere, the opacity  is simply  $27/(4 H \rho)$.  

The finding that $\kappa_{\rm gr}$ is independent of $\dot{M}_{\rm gr}$ can of course not hold ad infinitum. For example, if $\dot{M}_{\rm gr}$ approaches zero, then also the grain opacity must approach zero, at least in our limit that the steady state is  assumed instantaneously. What actually happens is the following: if  $\dot{M}_{\rm gr}$ becomes very small, then $\tau_{\rm coal,E}$ becomes  long, to an extent that it becomes longer than the advection timescale of the gas. Therefore, the growth regimes changes from the differential settling regime into the advection regime, where there is a dependency on $\dot{M}_{\rm gr}$. We will study the impact of  $\dot{M}_{\rm gr}$  (which can be studied by varying  $f_{\rm D/G, disk}$, because of Eq. \ref{eq:grainaccrationrate}) below in Sect. \ref{sect:effectfpg}.  

With the definition of $H$ and $g$, and the ideal gas law\footnote{One can also use $H^{-1}=(\mathrm{d} P/\mathrm{d} R)(1/P)$ and $\mathrm{d}P/\mathrm{d}r =-\rho g$ from the hydrostatic equilibrium.}, we can also write 
\beq\label{eq:kappaDiffEpstein2}
\kappa_{\rm D,E}=\frac{27 Q g}{8 P}=\frac{27 G M  Q}{8  R^{2} P}.
\eeq
where $P$ is the atmospheric pressure. This form is reminiscent of Eddington's photospheric boundary condition, where $\kappa=2 g/(3 P)$.

\subsubsection{Differential settling (coalescence): Stokes regime}\label{sect:diffsetstokes}
For large particles with Kn$<$4/9 at low velocities (Re$\lesssim$70), the  timescale  for growth by differential settling is given  as \citep{rossow1978}
\beq
\tau_{\rm coal,S}=\frac{9 \eta}{\pi \rho_{\rm gr} g}\frac{1}{n_{\rm gr} a^{4}}.
\eeq
With the definitions of $n_{\rm gr}$ and $\fpg$, this can be written as
\beq\label{eq:taucoalS}
\tau_{\rm coal,S}=\frac{12 \eta}{a \fpg g \rho}=\frac{32 \pi a \rho_{\rm gr} R^{2}}{ 3 \dot{M}_{\rm gr}},
\eeq
which has the same dependency on $\fpg$, $a$ and $\dot{M}_{\rm gr}$ as in the  last regime. Equating this  timescale with the settling timescale for Stokes drag, $\tau_{\rm set,S}=\tau_{\rm coal,S}$, gives a typical grain radius
\beq\label{eq:aDS}
a_{\rm D,S}=\left(\frac{27 \dot{M}_{\rm gr} H \eta }{64 \pi g \rho_{\rm gr}^{2} R^{2}}\right)^{1/3}.
\eeq
The dust-to-gas ratio is 
\beq
f_{\rm D/G, D,S}=\left(\frac{8 \dot{M}_{\rm gr} \eta \rho_{\rm gr}}{\pi g H^{2} \rho^{3}  R^{2}}\right)^{1/3}
\eeq
which finally leads with Eq. \ref{eq:kappaxfpga} to an opacity 
\beq\label{eq:kappaDiffStokes1}
\kappa_{\rm D,S}=\frac{2 Q}{H \rho}
\eeq
which is, except for a factor of order unity (1.69), the same result as for differential settling in the Epstein regime. It is therefore again a very simple expression that is in particular independent of the grain accretion rate, $\dot{M_{\rm gr}}$. The reason is the same as before, except that in this regime, $\fpg$ and $a$ both scale as $\dot{M}_{\rm gr}^{1/3}$ instead of $\dot{M}_{\rm gr}^{1/2}$. Using again the fundamental variables describing the atmosphere, we can  write the opacity also as
\beq\label{eq:kappaDiffStokes2}
\kappa_{\rm D,S}=\frac{2 Q g}{P}=\frac{2 G M  Q}{  R^{2} P}.
\eeq
This equation and the corresponding one for the Epstein regime (Eq. \ref{eq:kappaDiffEpstein2}) are the most important results of this study.

\subsubsection{Advection regime: effect of envelope contraction}\label{sect:advect}
Up to this point, we have completely neglected that the atmosphere is itself not static. In general, this is a good approximation, because the timescale on which the atmosphere changes is much longer than the time on which the grains evolve.

However, under some circumstances, namely at or after crossover, it is necessary to take into account that in reality, the gaseous envelope of the planet is continuously contracting. This means in particular that the  gas itself will also have a (small) downward motion through the atmosphere because the gas that is newly accreted at the outer radius settles down into the deeper parts of the envelope where the mass accumulates (in the convective zone). This leads to a velocity of the gas $v_{\rm gas}$ that must be taken into account to describe the net velocity of the grains relative to the Eulerian atmospheric $p$-$T$ structure at a fixed radius \citepalias{movshovitzpodolak2008}.

Typically, only a very small gas mass is contained in the radiative zone, so that we can assume that the gas accretion rate, $\dot{M}_{\rm XY}$ is radially constant in the atmosphere (this does of course not hold in the inner parts of the envelope where the gas accumulates). The vertical velocity of the gas can then be estimated as 
\beq
v_{\rm gas}=\frac{\dot{M}_{\rm XY}}{4 \pi R^{2} \rho}
\eeq
and the associated timescale is simply $\tau_{\rm adv}={H}/v_{\rm gas}$. We have called this timescale the advection timescale, due to the following: In the limit that $v_{\rm gas}$ becomes large, the gas starts to entrain the grains, so that the grains get externally advected through a scale height by the gas flow. This holds  because the grains are  well coupled to the gas as made clear by the small stopping times (see, e.g., Fig. \ref{fig:mp2008taucompcoagcoal}). 

In principle it is possible to take the effect of the gas flow into account exactly by writing the drag force proportional to the relative velocity. At equilibrium, when the gravitational force $F_{\rm G}$ and drag force $F_{\rm D}$ compensate each other, we  then again have $F_{\rm G} + F_{\rm D}=0$, but  $F_{\rm D}$ has now the form $C_{\rm couple}(v-v_{\rm gas})$, where the coupling constant $C_{\rm couple}$ depends on the drag regime (Eqs. \ref{eq:dragepstein}  and \ref{eq:dragstokes}). Solving for the velocity of the grains, we have $v=(F_{\rm G}+C_{\rm couple}v_{\rm gas})/C_{\rm couple}$. The first  {regime} occurs when the second term is negligibly small (this is the previously treated static case), while the second  {regime} occurs when the first term is negligible, so that $v=v_{\rm gas}$, i.e., when the grains are advected by the gas flow. In this first analytical model, we take the effect of envelope contraction into account only in the case  {of complete advection} where the grains settle exactly at $v_{\rm gas}$ (but we  {show} below  {in Appendix \ref{sect:numsolve} how the relative velocity can be included self-consistently. This has, however, the consequence that the grain size and thus opacity can no longer be found analytically. Instead, a root must be determined numerically in each layer.}) 

 {When the grains settle at $v_{\rm gas}$, the gas accretion rate and scale height alone} define the advection timescale $\tau_{\rm adv}$. To determine the typical grain size, we can ask how big the grains can grow while they are advected through one scale height. This can again be estimated by setting the advection timescale equal to the growth timescale. Therefore, the basic approach to determine the grain size remains the same as in the settling regime.   {As in the previous regimes, we consider grain growth by Brownian motion coagulation and differential settling separately.} 

\subsubsection{ {Advection regime: Brownian motion coagulation}}\label{sect:advectbrown}
 {In contrast to settling regime, we here do not have to consider the Epstein and Stokes case separately}.  {This is due to the important} difference that (in the case  {of complete advection} we consider)  $\tau_{\rm adv}$ is  independent of grain size.  The equation that gives the grain size in  {this regime is thus $\tau_{\rm adv}=\tau_{\rm coag, A}$ where }
\beq\label{eq:taucoagA}
\tau_{\rm coag, A}=\frac{2 \pi^{2} R^{2} v_{\rm gas}}{3 \dot{M}_{\rm gr}}\sqrt{\frac{\rho_{\rm gr}^{3} a^{5}}{3 k_{\rm B} T}}
\eeq
 {from Eqs. \ref{eq:taucoagbrown} and \ref{eq:fpggeneral}. This yields a grain size}
\beq
a_{\rm coag, A}=\left(\frac{1728 H^{2} \dot{M}_{\rm gr}^{2} k_{\rm B} T R^{4} \rho^{4}}{\dot{M}_{\rm XY}^{4}\rho_{\rm gr}^{3}}\right)^{1/5}
\eeq
while the dust-to-gas ratio is simply $f_{\rm D/G, A}=\dot{M}_{\rm gr}/\dot{M}_{\rm XY}$ which is, in the case of deep deposition (Eq. \ref{eq:grainaccrationrate}), simply equal to $f_{\rm D/G, disk}$ as it must be from mass conservation. The opacity is finally given as
\beq
\kappa_{\rm coag, A}=\frac{3^{2/5}}{2^{16/5}}\left(\frac{\dot{M}_{\rm gr}^{3} Q^{5}}{H^{2} \dot{M}_{\rm XY} k_{\rm B} T \rho^{4} \rho_{\rm gr}^{2} R^{4}}\right)^{1/5}
\eeq
which is for deep deposition
\beq
\kappa_{\rm coag, A, deep}=\frac{3^{2/5}}{2^{16/5}}\left(\frac{f_{\rm D/G,disk}^{3} \dot{M}_{\rm XY}^{2} Q^{5}}{H^{2} k_{\rm B} T \rho^{4} \rho_{\rm gr}^{2} R^{4}}\right)^{1/5}.
\eeq
This shows that the opacity increases with both the gas accretion rate and the dust-to-gas ratio in the disk,  {in contrast to growth by differential settling in the settling regime.}

\subsubsection{ {Advection regime: Differential settling}}\label{sect:advectdiff}
 {For differential settling we have to consider the Epstein and Stokes regimes separately as the expressions for the growth timescales differ. The later regime is however not important, as advection only occurs in the upper atmospheric layers, and there the grains are in the Epstein regime. }

 {When the grains are moving at $v_{\rm gas}$ the growth timescale (Eq. \ref{eq:taucoalEgeneral}) becomes}
\beq\label{eq:taucoalAE}
\tau_{\rm coal, A, E}=\frac{18 \pi R^{2} \rho v_{\rm th} v_{\rm gas} }{\dot{M}_{\rm gr} g}.
\eeq
 {This growth timescale is independent of the grain size $a$. As the advection timescale is also independent of the grain size, this means that our normal approach to determine the typical grain size by equating $\tau_{\rm adv}=\tau_{\rm coal, A, E}$ and solving for $a$ is not applicable here. This is a sign that in this regime, the grain size can not be determined locally in one layer. Instead one needs to follow the growth as grains sinks from the top of the atmosphere down to the layer in question. }

 {To model this growth, we can use the definition of the growth timescale $a/\dot{a}$ and fact that the grains travel in this regime at $v_{\rm gas}$. This yields the equation that links the change of the grain size with the change of the radius $R$, }  
\beq\label{eq:daa}
\frac{d a}{a}=-\frac{1}{v_{\rm gas}}\frac{1}{\tau_{\rm coal, A, E}} d R.
\eeq
 {As shown by \citet{ormel2014}, a related approach can be used to derive a general framework where the grain size is found by integrating numerically the expression for $da/dR$. Here we only search for an approximate solution to Eq. \ref{eq:daa}. } {For this, the following finding is useful: as a general result, the advection regime is found to occur first at or after the crossover point (Sect. \ref{sect:tempevoSigma10}) when the gas accretion rate starts to increase. It then first occurs in the very top layers at the outer boundary of the atmosphere. After crossover, the thickness of the advection layer then grows inwards as the gas accretion rate further increases.} 

 {We can therefore estimate the growth of a grain as it traverses the layer where the advection regime occurs by integrating Eq. \ref{eq:daa} from the outer radius of the atmosphere at $R_{0}$ down to the current radius $R$. At $R_{0}$, the grains have some initial size $a_{0}$ (this will typically be the same as the assumed monomer size $a_{\rm mono}$). Inserting the equation for the growth timescale  and $v_{\rm gas}$ and integrating then yields the grain size }
 \beq\label{eq:acoagAEnum}
\ln\left(\frac{a_{\rm coal, A, E}}{a_{0}}\right)=\frac{4 \pi G \dot{M}_{\rm gr} M}{9 \dot{M}_{\rm XY}^{2}}\sqrt{\frac{\pi \mu m_{\rm H}}{k_{\rm B}}}\int_{R}^{R_{\rm 0}}\frac{\rho}{\sqrt{T}}d R
\eeq
 {where we have assumed that the (contained) mass $M$ is constant in the atmosphere. An analytical approximation of the integral is discussed in  Appendix \ref{sect:diffsetbrownadvection}. From the grain size, the opacity can then be calculate in an analogous way as in the other regimes. In Appendix \ref{sect:diffsetbrownadvection} we compare Brownian motion and differential settling in the advection regime. It is found that the two mechanisms occur on comparable timescales leading therefore to similar grain sizes and opacities, in contrast to the normal case where growth by differential settling is much more efficient. In Sect. \ref{sect:combsims} we therefore only consider Brownian motion in the advection regime. It is, however, clear that this regime should be further investigated with a (numerical) approach that is more suitable than the local analytical description used here.}

 {For completeness we finally give the grain size for advection and growth by differential settling  in the Stokes regime. As mentioned, this regime should not usually have a practical meaning. It would occur when the gas accretion rate is very high, so that advection also occurs in the deeper atmospheric layers. This should only occur well after the crossover point, when the planet approaches the disk limited gas accretion rate. This phase is extremely short, and it is questionable whether the 1D radially symmetric approximation still holds. In any case, one finds a growth timescale} 
\beq\label{eq:taucoalAS}
\tau_{\rm coal, A, S}=\frac{16 \pi R^{2} \rho v_{\rm th} v_{\rm gas}  \ell}{\dot{M}_{\rm gr}}.
\eeq
 {This timescale depends on the grain size, therefore it is possible to estimate the typical grain size by equating $\tau_{\rm coal, A, S}$ with $\tau_{\rm adv}$. This yields a grain size}
\beq
a_{\rm coal, A, S}=\frac{48 \pi \eta v_{\rm gas}^{2} R^{2}}{\dot{M}_{\rm gr} H g}.
\eeq

\subsubsection{ {Relevance of the advection regime}}
The timescale arguments indicate that is in principle  plausible that at high gas accretion rates, the advection of grains becomes important,  {leading to high opacities} in the outer layers. Indications of this are also seen in the numerical simulation of \citetalias{movshovitzpodolak2008} (see discussion in  Sect. \ref{sect:effectfpg}). In the current analytical model, only the limiting cases are considered, i.e., that either growth and settling occurs as if the atmosphere would be completely static, or the other extreme where all grains move at  $v_{\rm gas}$.

We find below that  the model therefore predicts a relatively sudden increase of $\kappa$ in the advection regime (\ref{fig:evoM10}). This is understood from the fact (Fig. \ref{fig:mp2008taucompcoagcoal}) that the growth timescale in the  differential settling  and  advection regime can be quite different. In the first regime, the grains therefore  quickly reach sizes of  $\sim$0.01 cm, while in the advection regime, they can stay very small at $a=a_{\rm mono}$ (Fig. \ref{fig:compNumAnM2010V3}). While the decrease of $Q$ to values much smaller than 1 partially compensates for the increase of the absorbing surface for such small grains (Eq. \ref{eq:kappaxfpga}),  we still find that  significant jumps in $\kappa_{\rm gr}$ occur of up to one order of magnitude at the transition between the settling and the advection regime.  {Solving numerically for the grain size to include intermediate states for the relative velocity of grains and gas (Appendix \ref{sect:numsolve}), as well as including growth by differential settling in the advection regime (Appendix \ref{sect:diffsetbrownadvection}) partially changes this, but the basic prediction of a high opacity is found to remain\footnote{During the preparation of this work we became aware of the results of \citet{dangeloweidenschilling2014}. In their Fig. 11, such an effect can be seen for the most massive core they consider. }. On the other hand, the analytical model is built on a number of idealization which  in reality probably do not apply  in the top layers. First, the outer layers partially participate in the general gas flow in the protoplaentary disk \citep{lissauerhubickyj2009,ormel2013}. Therefore, the layers could be turbulent and/or show patterns of atmospheric circulation, in contrast to the simplification made here of a purely radial laminar gas motion. Second, the grains that are accreted from the nebula into the planetary atmosphere  are likely not monodisperse. Potentially this would both make a faster growth due to differential settling possible, in particular by gas motions driving small grains against the large ones \citep{weidenschilling1984}. Such effects related to a bimodal size distribution are difficult to study in the context of the analytical model here, but should be addressed in future work with numerical models \citep{podolak2003,ormel2014}.} 

On the other hand, from a point of view that is mostly interested in the final outcome of the formation process (i.e., in knowing the final mass of a planet) rather than the  atmospheric structure during all phases of the formation, the  details of the advection regime are of secondary importance. This is due to the following (Sect. \ref{sect:combsims}): the advection regime occurs (at least in the cases we studied) only at and (with increasing importance) after crossover, where it influences the time between crossover and the moment the disk-limited maximal gas accretion rate is reached. Including or completely neglecting advection is found to lead to variations of this time of about $10^{5}$ years (Fig. \ref{fig:menvemcoreM1064}). This is short in comparison with typical disk lifetimes which in turn means that the impact on the final properties of a planet will in general only be small.   

\subsection{Grain evaporation}\label{sect:evap}
A last effect we need to take into account is that grains evaporate at sufficiently high temperatures, so that the grain opacity disappears \citep{lenzunigail1995}. The mass loss rate of a grain due to evaporation can be calculated with the Knudsen-Langmuir formula (e.g., \citealt{bronsthen1983}) as
\beq
\frac{d m_{\rm gr}}{dt}= 4 \pi a^{2} p_{\rm vap }(T)\sqrt{\frac{ \mu_{\rm gr} m_{\rm H}}{2 \pi k_{\rm B} T}}.
\eeq
In this equation, $\mu_{\rm gr}$ is the mean molecular weight of an evaporated grain molecule which was set to 33  \citep{pollackpodolak1986}. We have assumed that the grains evaporate from the entire $4 \pi a^{2}$ surface, that the partial pressure of the rock vapor in the atmosphere is small, and that the surface of the grains has the same temperature as the gas. The vapor pressure $p_{\rm vap }$ is approximately given as $p_{\rm vap}=p_{\rm vap,0}e^{-C_{\rm vap}/T}$ where $p_{\rm vap,0}$ and $C_{\rm vap}$ are material properties. We take the values suggested by \citet{podolakpollack1988} for rocky material, so that $p_{\rm vap,0}=1.5\times10^{13}$ dyn/cm$^{2}$ and $C_{\rm vap}$=56\,655 K.  These parameters mean that grains evaporate at a temperature of roughly 1300 K.  {It is worth noting that instead of the pure stoichiometric evaporation considered here, minerals can also get decomposed by reacting with the surrounding H$_{2}$ gas. This can increase the mass loss rate \citep[][]{bossalexander2012}.}

The timescale of grain evaporation is 
\beq
\tau_{\rm evap}=\frac{ \rho_{\rm gr} a}{3 p_{\rm vap }(T)}\sqrt{\frac{2 \pi k_{\rm B} T }{\mu_{\rm gr} m_{\rm H}}}.
\eeq
In principle, one could imagine that in certain parts of the atmosphere, grain growth (by Brownian coagulation or differential settling) is balanced by evaporation, so that the two processes must be combined to obtain the actual timescale on which the grain size changes. However, due to $p_{\rm vap}$ there is a very strong exponential dependency of $\tau_{\rm evap}$ on the temperature. In practice it means that either  the evaporation timescale is  many orders of magnitudes longer than the growth timescale at low temperatures, or many orders of magnitudes shorter as soon as the temperature is sufficiently high. This means that the radial domain in the envelope where evaporation changes from  completely a negligible to  completely  the dominating process is very small. This in turn means that there is a well defined evaporation temperature and evaporation radius $R_{\rm evap}$.   

Therefore, it is possible to simplify the treatment of evaporation. If $\tau_{\rm evap}$ is longer than the relevant growth timescale $\tau_{\rm growth}$ (given by one of the regimes of Sect. \ref{sect:calctypsizeopa}, typically  differential settling regime with Stokes drag), it is possible to completely neglect evaporation. We need to take it into account as soon as  $\tau_{\rm evap}<\tau_{\rm growth}$, in a way that the grain opacity decreases rapidly towards the interior. Operationally, we therefore multiply in the latter case $\kappa_{\rm gr}$ calculated without evaporation by a factor $\tau_{\rm growth}/\tau_{\rm evap}$. This  factor very quickly approaches zero with decreasing radius inside of the evaporation radius. Clearly, this particular setting is used for convenience rather than  physical reasons. However, the particular setting for the reduction of $\kappa_{\rm gr}$ due to evaporation remains  in any case without consequences because of the following interesting result: 

Due to grain growth, it is found (e.g., Fig. \ref{fig:crossM10}) that the grain opacity decreases with decreasing height in the atmosphere by about three orders of magnitude (as already found by \citetalias{movshovitzpodolak2008} or \citetalias{movshovitzbodenheimer2010}). For this reason the opacity of the grain-free gas (molecular and atomic opacities) becomes larger than the grain opacity at some radius. This radius  ($R_{\kappa\rm{-DG}}$) is found to be larger than the radius where the grains evaporate $R_{\rm evap}$ (Fig. \ref{fig:massdepo}). This means that grain evaporation remains without consequences for the actual total opacity.  In other words,  the grains are so large in the deeper layers that before they evaporate, their contribution to the total opacity becomes anyway smaller than the one of the grain-free gas. Our simplified treatment of evaporation is therefore justified. 

 {It is clear that besides the destruction by evaporation, grains can also get fragmented in mutual collisions. We follow in this paper earlier works on grain growth in protoplanetary atmospheres (\citealt{podolak2003}, \citetalias{movshovitzpodolak2008}, and \citetalias{movshovitzbodenheimer2010})  and currently neglect this effect. Instead, we simply assume perfect sticking. For the cases studied in detail here  (see Sect. \ref{sect:mp08vsetdensqbounce} and \ref{sect:perfectsticking}) it is found that the relative velocities of the grains are likely too small for fragmentation. Bouncing could in contrast be important. Future work should  explore the importance of non-perfect sticking.}

\subsection{Extinction coefficient $Q$}\label{sect:extinctioncoeff}
Once the grain size and dust-to-gas ratio is calculated, in order to get the opacity, we need to calculate the extinction coefficient $Q$ that gives the ratio of the extinction cross section of a grain for radiation of wavelength $\lambda$ to its geometric cross section $\pi a^{2}$. 

The extinction coefficient is the sum of the absorption and scattering efficiency. For spherical grains, it can be calculated with Mie theory. It is then a function of the real $n_{\rm r}$ and imaginary $n_{\rm i}$ refractive indices of the grain material (e.g., \citealt{dorschnerbegemann1995}) and of the size  parameter $x={2 \pi a}/{\lambda}$. As discussed in \citet{podolak2003} and \citetalias{movshovitzpodolak2008}, instead of full Mie theory, it is possible to use approximative expressions for $Q$ as a function of $x$, $n_{\rm r}$, and $n_{\rm i}$, derived by Dr. J. Cuzzi\footnote{During the preparation of this paper, we became aware of the work of \cite{cuzziestrada2013}. In future work it will be possible to use the results for $Q$ from this model instead of the less general approximations presented here.}.

In the context of protoplanetary atmospheres, it is first of interest to see which  {values of} $x$ occur. The  typical wavelength of the radiation in the atmosphere at a temperature $T$ can be estimated with Wien's displacement law as ${C_{\rm Wien}}/{T}$.
The temperature in the envelope  can vary from $\sim100$ K in the outer parts of the atmosphere to about 1500 K, the temperature  where the grain evaporate. The grain size will vary from about 1 $\mu$m, the monomer size, to about 0.1 cm in the deep layers (\citetalias{movshovitzpodolak2008}, Fig. \ref{fig:mp08fpg}). This means that $x$ roughly runs from 0.2 to 3000. 

\begin{figure}
\begin{center}
\includegraphics[width=\columnwidth]{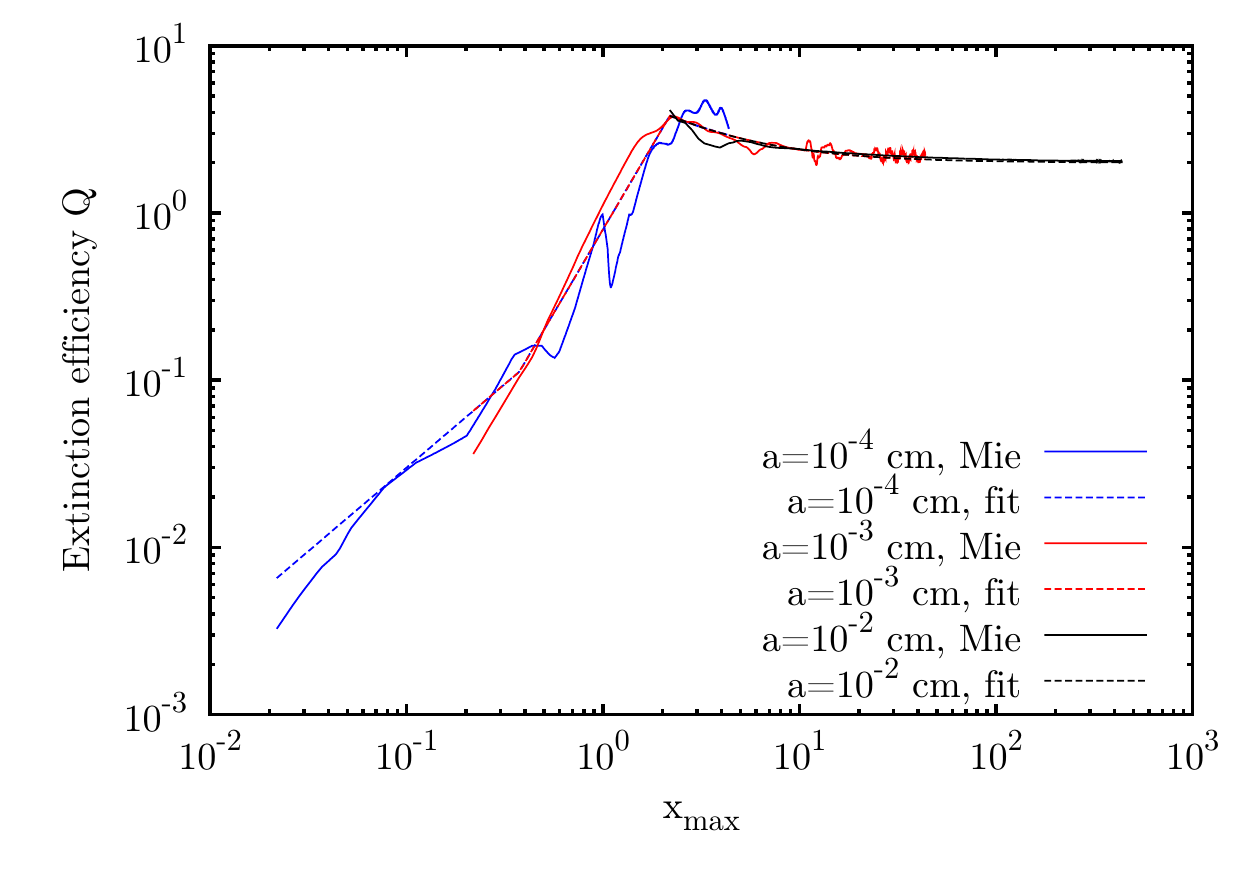}
\caption{Extinction efficiency $Q$ as a function of $x_{\rm max}= 2 \pi a/\lambda_{\rm max}$. The coefficient is shown for grain sizes of $a$=1, 10, and 100 $\mu$m,  computed with Mie theory (solid lines) and with the fit of Eq. \ref{eq:figQ}  (dashed lines). In all three cases, $\lambda_{\rm max}$  {corresponds to the wavelength of the maximum in the blackbody spectrum.}}\label{fig:qfit}
\end{center}
\end{figure}

\citetalias{movshovitzpodolak2008} investigated the impact of different grain materials (tholine, olivine, and iron) that have different refractive indices. They found that the resulting opacity in the atmosphere is not very sensitive to the specific material type, i.e., to $n_{\rm r}$ and $n_{\rm i}$. We therefore restrict ourselves in this study to tholins, the nominal case in \citetalias{movshovitzpodolak2008}. Note that this is strictly speaking inconsistent with the assumption of silicate grains for the evaporation. In Figure \ref{fig:qfit}, the solid lines show $Q(x_{\rm max})$ for grains with $a$=1, 10 and 100 $\mu$m. The size parameter was calculated as $x_{\rm max}=2 \pi a/\lambda_{\rm max}(T)$, and $T$ was running in each case from 10 to 2000 K.  

In Figure \ref{fig:qfit}, the extinction efficiency was calculated using a table of wavelength dependent refractive indices for tholins given by \citet{kharesagan1984} and the Mie code of \citet{bohrenhuffman1983}. We derived then a simple empirical fit by comparison with the actual $Q$ in Figure \ref{fig:qfit}, finding

\beq\label{eq:figQ}
 Q(x) \approx
  \begin{cases}
  0.3 x & \textrm{if }  x < 0.375  \\
  0.8 x^{2}   & \text{if } 0.375\leq x < 2.188   \\ 
  2+\frac{4}{x}   & \text{if } 2.188\leq x < 1000 \\ 
  2 & \text{if } x \geq 1000
  \end{cases}
\eeq
This fit is shown with dashed lines in Fig. \ref{fig:qfit}. For small $x$ the fit approaches the absorption efficiency for small particles that is linear in $x$ as found for isotropic Rayleigh spheres. It dominates over scattering (that would scale as $x^{4}$) because of the complex refractory index  \citep[e.g.,][]{hansentravis1974}. Expressed with $n_{\rm r}$, and $n_{\rm i}$, the absorption efficiency can be approximated as \citepalias{movshovitzpodolak2008}
\beq
Q_{\rm abs}=\frac{24 n_{\rm r} n_{\rm i}}{(n_{\rm r}^{2}+2)^{2}}x.
\eeq
For $n_{\rm r}=1.5$ and $n_{\rm i}=0.1$, the typical values mentioned by \citetalias{movshovitzbodenheimer2010}, one finds $Q_{\rm abs}=0.20 x$, while using $n_{\rm r}=1.6$ and $n_{\rm i}=0.16$ (which we find to provide a somewhat better fit to the tholine data of  \citealt{kharesagan1984}) yields $Q_{\rm abs}=0.28 x$. This is  close to the value given by the fitting function that was derived by comparing visually a fitting function and the Mie results.  On the other hand, for large very large $x$, the fit approaches the  geometrical limit of $Q=2$.

Equation \ref{eq:kappaxfpga} gives the wavelength-dependent opacity of a grain of size $a$ for radiation of wavelength $\lambda$. For the calculation of the atmospheric structure, one needs in contrast the  Rosseland mean opacity. In the light of an absence of a grain size distribution in the simple analytical model, and, most importantly, the fact that under most circumstances $a\gg \lambda$, so that $Q\approx 2$  (i.e., without a significant wavelength dependency for the contributing part of the Planck function), we here use the following simplification: instead of calculation the actual Rosseland mean, we use the opacity as found with $Q(x)$ evaluated at $x_{\rm max}$, i.e., at the wavelength $\lambda_{\rm max}(T)$ of peak flux in the Planck distribution,  where $T$ is the temperature of the gas in the atmospheric layer. 

The comparison with the results of \citetalias{movshovitzpodolak2008} and \citetalias{movshovitzbodenheimer2010} (where the actual Rosseland mean opacity is calculated) indicates that this simplification does not have a significant impact on the predicted  grain opacity, at least for the cases studied, because grain growth is so efficient that we are in the geometrical limit. This is confirmed by Figure 10 in \citet{cuzziestrada2013}: from the figure it becomes clear that the wavelength-independent geometrical limit for $Q$ (and therefore our approximation) is applicable for grains of $10$ $\mu$m if the temperature is larger than 300 K, and for grains of  100 $\mu$m (or larger) if $T\gtrsim50$ K. These conditions are usually met in protoplanetary atmospheres except for the advection regime where the grains remain small in the outer layers of the atmosphere. Fortunately, this regime occurs only when the planet is already at or close to the critical core mass, so that the impact on the total formation history of a planet remains small (Sect. \ref{sect:combsims}). Nevertheless, this is a point that will be further addressed in future work.

\subsection{Combination of the different regimes, gas opacity and final expression for $\kappa$}
In the previous sections, we have derived the timescales of several processes that govern the evolution of grains and the associated opacity. To get the final expression of $\kappa$, we combine them, following the general approach that the process with the shortest timescale is taken as the dominant one  \citep{coopersudarsky2003}. 

Operationally, we first calculate all four growth (and by definition equal settling) timescales for Brownian coagulation and for differential settling in both drag regimes. The appropriate drag regime is then chosen by checking which regime leads to a self-consistent Knudsen number (see Sect. \ref{sect:aeroregimes}). Next, the timescale of growth by Brownian motion and differential settling are compared, and the one with the shorter timescale  is retained. This timescale is then compared with the advection timescale, and again the shorter one is retained. Finally, this timescale is compared with the evaporation timescale. If the latter is shorter, the opacity is reduced as described in Sect. \ref{sect:evap}. 

Up to this point, we have only considered the opacity due to grains $\kappa_{\rm gr}$.  To get the total opacity, we also need to take into account the molecular and atomic opacity of the grain-free gas, $\kappa_{\rm gas}$. This gives the opacity in the deep convective interior of the envelope, but also in the deeper parts of the atmosphere because the grain opacity becomes so small there due to grain growth (see Fig. \ref{fig:mp08kappas}). The gas opacity is obtained from the combination of the analytical expressions of \citet{belllin1994} for $T>$4000 K and the tables of \citet{freedmanmarley2008} at lower temperatures, assuming a solar-composition gas. As already noted by \citetalias{movshovitzbodenheimer2010}, the details of the opacities at high temperatures are unimportant, because the energy transport is due  to convection.  

For simplicity, we set the resulting, total opacity equal to $\kappa={\rm MAX}(\kappa_{\rm gr},\kappa_{\rm gas})$. It is found that the specific way the gas and grain opacity are combined is not very important, because at the radius where the two become equal,  $\kappa_{\rm gr}$ rapidly decreases towards the interior, whereas $\kappa_{\rm gas}$ rapidly increases (Fig. \ref{fig:mp08kappas}). This means that the radial domain where the two are of the same magnitude is small.

\subsection{Limitations of the model}\label{sect:limitations}
This first simple analytical model has a number of  limitations. An obvious one is the absence of a grain size distribution in each  atmospheric layer. As explained in \citetalias{movshovitzpodolak2008}, it is not assured that the opacity in a layer is always dominated by the grains with the most frequent size. For example, the analytical model predicts that in deep layers, there are only large grains because it assumes that small grains are brought into the atmosphere only at its outer boundary. If in reality, planetesimal impacts deposit many  small grains  in the lower layers, these could dominate the opacity, even if they do not dominate the size distribution. On the other hand, in our comparisons with the results of \citetalias{movshovitzpodolak2008} and \citetalias{movshovitzbodenheimer2010}, the consequences of a grain size distribution instead of one single size becomes apparent only in one rather minor feature, which turns out to remain without consequences for the overall opacity (Sect. \ref{sect:mp08structopa}). Nevertheless, more work is needed to further investigate and improve this point. It could be possible to extend the model by including size distributions centered on the typical size, with a shape and width typical for either Brownian coagulation or growth by differential settling. 

A second, related important limitation is that the analytical models assumes that the total flux of grains remains radially constant. This makes it impossible to capture the consequences for $\kappa_{\rm gr}$ if planetesimal impacts locally increase the dust input. Our results indicate that this should  typically  not have very strong consequences for the overall opacity (and therefore optical depth) due to two reasons: First, the opacity in the most common regime (Eqs. \ref{eq:kappaDiffEpstein2} and \ref{eq:kappaDiffStokes2}) is independent of $\dot{M}_{\rm gr}$. This picture is, roughly speaking, confirmed by the results of \citetalias{movshovitzpodolak2008} (their Fig. 11): a very strong increase of  the planetesimal input (by a factor 100) in a certain layer leads locally to a  higher opacity, but the total optical depth only changes by a factor of a few (except in the special case that the increased deposition occurs just at the bottom of the atmosphere). Second, at least for km sized planetesimals, most mass deposition should occur in any case sufficiently deep not to influence much the grain opacity (see Fig. \ref{fig:massdepo}). 

A third limitation is that we assume that the opacity of the grains for radiation at $\lambda_{\rm max}$ can directly be used as the Rosseland mean opacity. In the most frequent growth regime (differential settling) and in most parts of the envelope, $a\gg \lambda_{\rm max}$, so that the extinction efficiency is largely independent of the wavelength for the relevant part of the Planck function. In this regime, the difference between $\kappa(\lambda_{\rm max})$ and the Rosseland mean is therefore small. This is confirmed in tests where we calculated the Rosseland mean. On the other hand, in the advection regime, there are small grains, and the wavelength dependency becomes more important. Tests show that the Rosseland mean and  $\kappa(\lambda_{\rm max})$ can differ in this regime by up to one order of magnitude. On the other hand, the advection regime only occurs at or after crossover, so that it is less important (Sect. \ref{sect:combsims}).

 {A fourth important limitation is the assumption of perfect sticking despite the fact that bouncing may be important  (Sects. \ref{sect:mp08vsetdensqbounce} and \ref{sect:perfectsticking}). This is a limitation that is shared with the numerical models of \citetalias{movshovitzpodolak2008} and \citetalias{movshovitzbodenheimer2010}. Additional limitations that are shared with these numerical models are the assumptions that} the grains are spherical, that no (strong) turbulence occurs in the envelope, and that only silicate grains are considered. The effects of convection are also ignored in the analytical model because \citetalias{movshovitzbodenheimer2010} showed that convection  hardly influences their results.

\section{Comparison with numerical models}\label{sect:compMPforgivenrhoT}
We now compare the analytical expression for the opacity derived in the previous sections with the results of  \citetalias{movshovitzpodolak2008}, and then of   \citetalias{movshovitzbodenheimer2010}  (Sect. \ref{sect:combsims}). Building on the initial work of \citet{podolak2003}, \citetalias{movshovitzpodolak2008} and \citetalias{movshovitzbodenheimer2010} use a complex (and computationally very heavy) numerically model to simulate the temporal evolution of grains under the influence of vertical settling and growth by Brownian coagulation and differential settling. In contrast to the analytical model, at each layer and moment in time they have a full distribution of grain sizes, expressed in the number of grains of a given size per unit volume. The temporal evolution of this quantity is obtained by solving numerically the Smoluchowski equation. The  Smoluchowski equation contains the terms representing coagulation and settling, plus a source term. The source term has two contributions: first, grains that are deposited by impacting planetesimals, and second, grains that are accreted together with the gas. The radial distribution of the grain input due to planetesimals is  given directly by the radial mass deposition profiles of the impacting planetesimals obtained with the code of \citet{podolakpollack1988}. This is in contrast to the analytical model where $\dot{M}_{\rm gr}$ has to be radially constant. The grains accreted with the gas are injected into the envelope at the top layer (which is the same as in the analytical model). In their nominal model, \citetalias{movshovitzpodolak2008} assume that the monomer size both of grains accreting with the gas and resulting from planetesimal deposition is approximately 1 $\mu$m.

The effects of envelope contraction which advects small grains (Sect. \ref{sect:advect}) and of convection are also included in some simulations. Once the grain size distribution is calculated, the wavelength-dependent opacity is computed using an approximation to Mie theory \citep{cuzziestrada2013}. Finally, the Rosseland mean is calculated using the  Planck function at the local temperature of the gas. It is clear that the numerical model is itself  simplified in a number of points, the most important ones of which were mentioned in the last section.  This means that  the results of the numerical model must be regarded as an approximation, too. 

Nevertheless, the results of the numerical model are crucial in this study, as they represent the best currently available information on how grains evolve in protoplanetary atmospheres. Therefore, they serve as the benchmark case. Ideally, the analytical model would perfectly reproduce the results of the numerical model. This is of course impossible due to the numerous  simplifications. Still, we find below that the numerical model and the analytical approximation agree almost surprisingly well and share may common key features, also in a quantitative sense. An litmus test for the analytical model is its prediction for the opacity if the dust-to-gas ratio in the accreted gas $f_{\rm D/G, disk}$ is varied. We will see that the analytical model passes this test very well, and that this is a very telling result (Sects. \ref{sect:effectfpg} and \ref{sect:implicationsfehpebble}).

The analytical model yields in particular a radial opacity structure that is much closer to the numerical result than the one obtained by reducing the ISM grain opacity by some constant factor (Fig. \ref{fig:mp08kappas}), as was often done in many previous works of different groups  as mentioned in the introduction. This is one of the goals of this work. The most important input parameters for the  comparison cases are given in Table \ref{tab:compcalcs}.

\begin{table}
\caption{Settings for the comparison with \citetalias{movshovitzpodolak2008}.}\label{tab:compcalcs}
\begin{center}
\begin{tabular}{lcc}
\hline\hline
Quantity & Nominal & Non-nominal \\ \hline
Semi-major axis a [AU] & 5.2 & 5.2\\
Stage & mid phase II &  mid phase II\\
$M$ [$\mearth$]    & 15.34 &  15.34 \\                                              
$\dot{M}_{\rm XY}$ [$\mearth$/yr]& $8\times10^{-6}$ &  $8\times10^{-6}$\\
Disk dust-to-gas ratio  $f_{\rm D/G,disk}$&0.01 &  10$^{-4}$ - 0.5\\
Monomer size [$\mu$m] & 1 & 1, 10, 100 \\
Planetesimal mass deposition & deep & deep, shallow\\ \hline
\end{tabular}
\end{center}
\end{table}

\subsection{Comparison with \citetalias{movshovitzpodolak2008}}
The first numerical simulation to which we compare our analytical model is the calculation of \citetalias{movshovitzpodolak2008} for a giant planet forming at 5.2 AU. This simulation is very important for the validation of the analytical model because it is the only published numerical simulation with a fairly complete information on the grain dynamics. This means that not only the radial profile of the opacity is known  \citepalias[as in][]{movshovitzbodenheimer2010}, but also the shape of the grain size distribution at different heights in the atmosphere, the mass density and settling velocity of grains as a function of height, the impact of different monomer sizes, and, very importantly, the effect of different dust-to-gas ratios in the newly accreted gas. The latter information is important as it helps to understand which growth mechanism (Brownian motion or differential settling) is dominant.

\begin{figure*}
\begin{minipage}{0.53\textwidth}
	      \centering
       \includegraphics[width=0.95\textwidth]{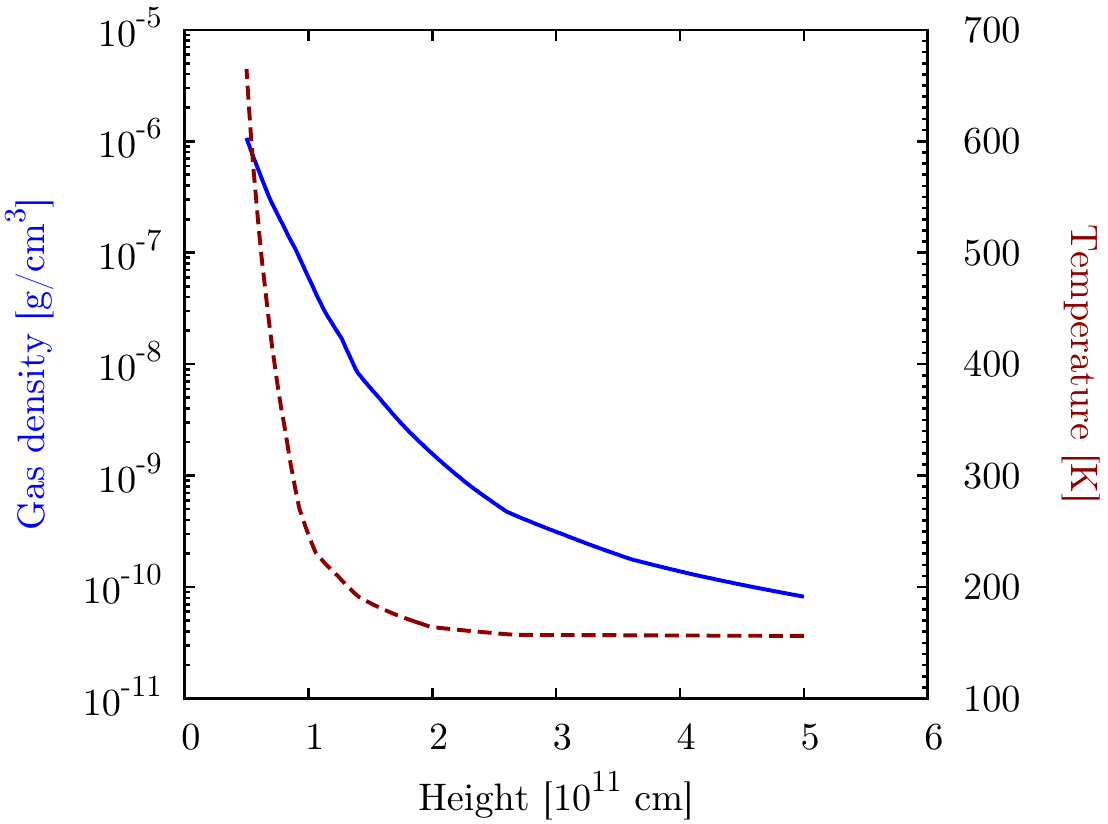}
     \end{minipage}
          \hfill
     \begin{minipage}{0.45\textwidth}
      \centering
                    \includegraphics[width=0.95\textwidth]{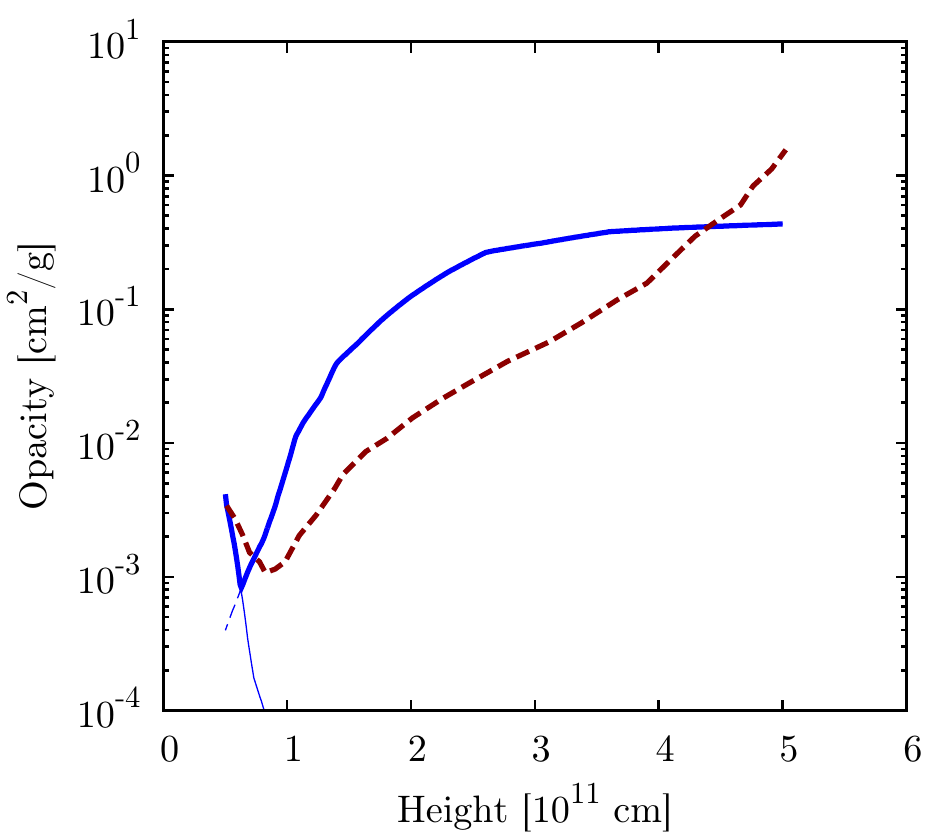}
     \end{minipage}
                  \caption{\citetalias{movshovitzpodolak2008} comparison case. \textit{Left panel:} Atmospheric gas density (solid line) and temperature (dashed) as a function of height (distance from the planet's center) in the protoplanet's atmosphere (outer radiative zone). \textit{Right panel:} Opacity as a function of height in the numerical model of \citetalias{movshovitzpodolak2008} (dashed line: grain opacity only) and in the analytical model of this work (thick solid line:  combined grain and gas opacity). The thin blue solid (dashed) line is the gas (grain) opacity alone.}
             \label{fig:2008rt} 
\end{figure*}

The atmospheric temperature and density structure used by \citetalias{movshovitzpodolak2008} was taken from \citet{hubickyjbodenheimer2005} and correspond to a moment in time when the planet is in the middle of phase II (see \citealt{pollackhubickyj1996} for the different phases occurring during classical in-situ giant planet formation simulations). At this moment in time, the mass of the core is 12.61 $\mearth$, and the envelope mass is 2.73 $\mearth$. The outer radiative zone/atmosphere of the planet only contains very little mass, therefore we assume that the mass interior of the atmosphere driving the settling is equal to the sum of the two.  \citetalias{movshovitzpodolak2008} do not explicitly state the gas accretion rate at this moment (which we need to calculate the grain accretion rate, Eq. \ref{eq:grainaccrationrate}), but the simulations of \citet{hubickyjbodenheimer2005} indicate that it should be approximately $8\times10^{-6}$ $\mearth$/yr. We assume that the planetesimals deposit their mass deep into the convective zone so that the advection of grains with the newly accreted gas is the only source of grains.  Our results are neither sensitive to the exact value of $\dot{M}_{\rm XY}$ nor to the assumption of deep/shallow deposition (Sect. \ref{sect:effectfpg}). In \citetalias{movshovitzpodolak2008}, grains are brought into the atmosphere both by gas accretion in the top layer as in the analytical model, but also at other heights due to planetesimals ablation (which is not included in the analytical model). It is therefore not a priori clear if a useful comparison can still be made. However, as discussed by \citetalias{movshovitzpodolak2008}, the planetesimal source is important only in the deepest layers because these grains are released mostly in the deep atmosphere (below $\sim10^{11}$ cm, see their Fig. 9). Therefore, it is possible to compare the analytical and numerical result. 

Note that the calculation of  $\kappa_{\rm gr}$ in \citetalias{movshovitzpodolak2008} (and of course also in this work) is not self-consistent in the sense that the  temperature and density structures were obtained by \citet{hubickyjbodenheimer2005} with a pre-specified (and different) opacity, and not the one obtained in the grain  calculations. The calculations of \citetalias{movshovitzbodenheimer2010} discussed in Sect. \ref{sect:combsims} are in contrast self-consistent. For our purpose of comparison of analytical and numerical model this has, however, no negative impact. 

\subsubsection{Atmospheric structure and opacity}\label{sect:mp08structopa}
The left panel of Figure \ref{fig:2008rt} shows the atmospheric gas density (blue solid line, left y-axis) and the temperature (dashed brown line, right y-axis) as a function of height (distance from the planetary center). These lines are simply taken from Fig. 1 in \citetalias{movshovitzpodolak2008}. As in \citetalias{movshovitzpodolak2008}, only the outer radiative zone of the gaseous envelope  is considered. The density increases from about $10^{-10}$ g/cm$^{3}$ at the outer radius (which is comparable to typical nebular densities)  to about  $10^{-6}$ g/cm$^{3}$ at the inner boundary. The temperature structure consists of an approximately isothermal outer part at 150 K (again the nebular temperature), followed by an increase to about 670 K at the inner boundary. This temperature is so low that silicate grains do not evaporate in the atmosphere. 

The right panel of Fig. \ref{fig:2008rt}  directly shows  the main result of the calculations  which is the  opacity as a function of height. Both the results of \citetalias{movshovitzpodolak2008} and of the analytical model are shown. One sees a radial opacity structure that is characteristic for all calculations, at least during phase II: At the outer boundary, the opacity is high, of order 1 cm$^{2}$/g, and therefore comparable to  ISM opacities. Towards the interior, it decreases strongly, going down to about $10^{-3}$ cm$^{2}$/g at the inner boundary. It is the consequence of the increase of the typical grain size $a$ and of the decrease of $\fpg$ with decreasing height (see Fig. \ref{fig:mp08fpg} below). Clearly, this is a consequence of grain growth.

This general  structure is found both in the numerical and analytical model. Note that for the former, only the opacity due to grains  $\kappa_{\rm gr}$ is shown, while for the latter, the thick line shows the total opacity, including the contributions of gas and grains. The thin dashed line shows the grain opacity alone in the analytical model. We see that in the analytical model, $\kappa_{\rm gr}$ is a strictly monotonically increasing function of  the height. In the numerical model, $\kappa_{\rm gr}$ has in contrast a local minimum at about $10^{11}$ cm. Inside of this point, it increases again. The reason for this is explained in \citetalias{movshovitzpodolak2008}: At   the minimum, the drag law for the large grains changes from Epstein into the Stokes regime, leading to an increase in the settling velocity (we find the same behavior in the analytical model, see Fig. \ref{fig:mp08vset}). The small grains (which are brought deep into the atmosphere by local planetesimal mass deposition, see \citetalias{movshovitzpodolak2008}) in contrast still remain in the Epstein regime. This causes a shift in the grain size distribution, resulting in an increase of $\kappa_{\rm gr}$. Because we do not consider planetesimal ablation as a local source of small grains deep in the envelope (as mentioned, $\dot{M}_{\rm gr}$ needs to be radially constant in the analytical model), and because we only have one grain size (which corresponds to the big grains in \citetalias{movshovitzpodolak2008}), we cannot have such a behavior in the analytical model. This is the only  point where a difference between a grain size distribution, and only one representative size becomes apparent, at least in the comparisons made here.  

\begin{figure*}
\begin{minipage}{0.48\textwidth}
	      \centering
                    \includegraphics[width=0.96\textwidth]{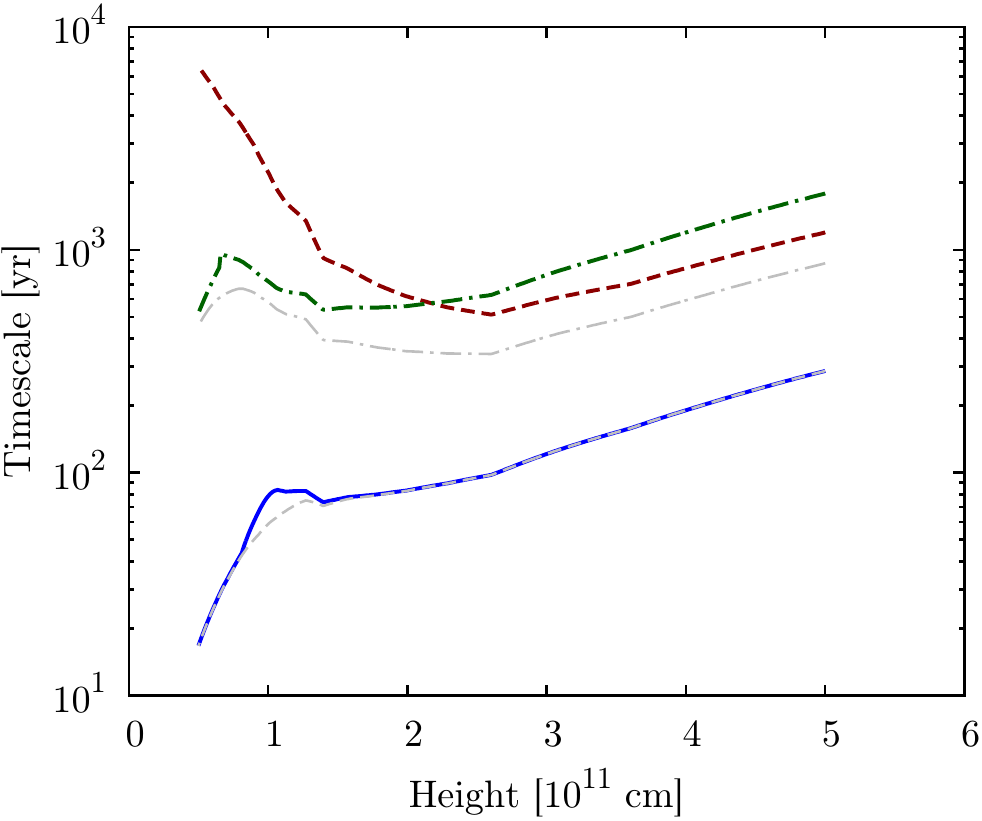}
     \end{minipage}
     \hfill
     \begin{minipage}{0.5 \textwidth}
                         \includegraphics[width=0.97\textwidth]{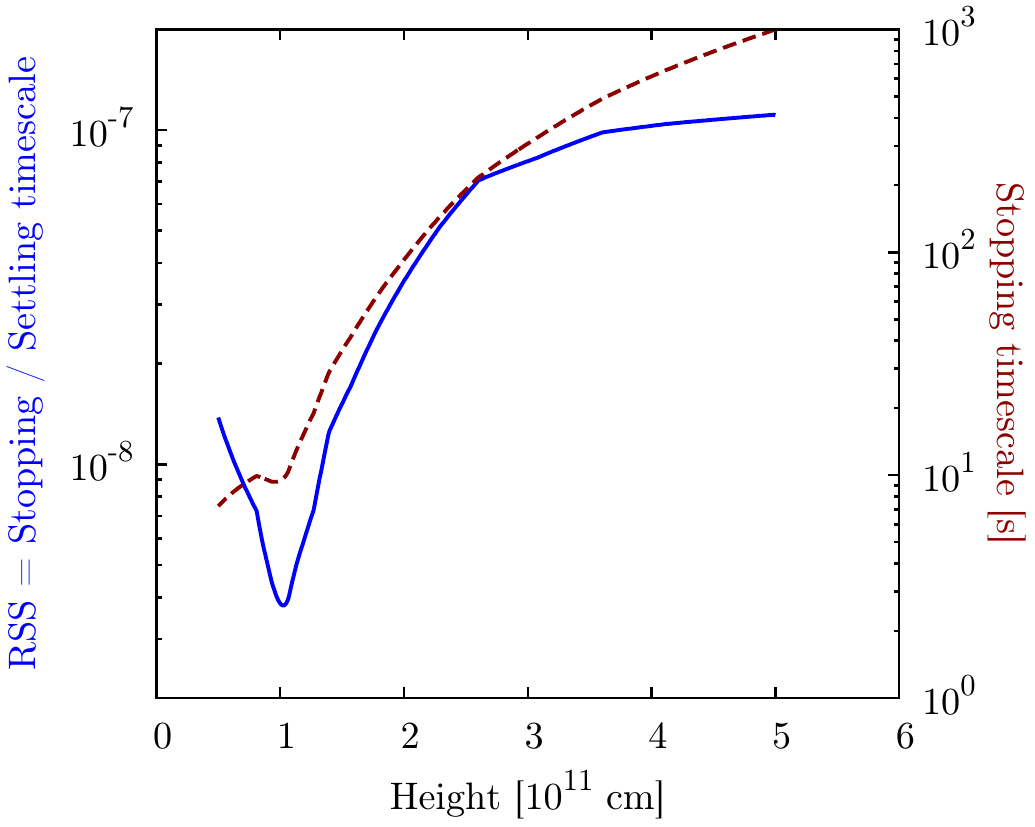}
      \centering
     \end{minipage}
                  \caption{\citetalias{movshovitzpodolak2008} comparison case. \textit{Left panel:}  Timescale of growth by differential settling/coalescence (blue solid line) and Brownian coagulation (green dashed-dotted). These timescales are by construction equal to the associated settling timescale. The timescale of gas advection is shown by the brown dashed line. In this atmosphere, the evaporation timescale of the grains is always many orders of magnitude longer due to the low temperatures.  {The thin gray dashed and dashed-dotted lines show the result if the velocity of the gas is directly included to calculated the growth timescales of differential settling and Brownian coagulation, respectively (Appendix \ref{sect:numsolve}).}   \textit{Right panel:} Ratio of the stopping to the settling timescale RSS  (solid) and stopping timescale (dashed line) of the grains. }
             \label{fig:mp2008taucompcoagcoal} 
\end{figure*}

Interestingly, the difference between the  numerical and analytical model gets reduced due to the following: if one considers the total opacity as predicted by the analytical model including the opacity due to the grain-free gas which is the physically relevant quantity, then we see that also the total opacity in the analytical model has a local minimum at a radius $\approx6\times10^{10}$ cm. Inside of this distance, the total opacity also increases rapidly, but now due to the molecular opacities which were not considered in \citetalias{movshovitzpodolak2008}. Clearly, this is a priori a chance result. But we find below also in the other comparisons (Sect. \ref{sect:Sigma10structcrossover}), that this  {is} actually a general behavior: in the deeper parts of the atmosphere, the grain opacity decreases so much with decreasing altitude that the gas opacities become dominant at some height. Interior of this height, the opacity increases again, but now due to the molecules. This means that there is a local minimum in the opacity. It is found that the analytical model predicts both the location of the local minimum and its value (typically a few $10^{-3}$ cm$^{2}$/g) in a way that agrees well with the results from the numerical calculations (see Figs. \ref{fig:crossM10} and \ref{fig:crossM4}). This makes it possible to use the analytical model also for quantitative calculations. 

We further see that the radial structure of the opacity in the two models can differ by up to one order of magnitude at some heights.  Interestingly, the shape of $\kappa$ agrees clearly better in the simulations where the atmospheric structure is calculated self-consistently (see Figs. \ref{fig:crossM10} and \ref{fig:crossM4}). In the simulation here,  at an intermediate height of about 2.5$\times10^{11}$ cm, the opacity  is in particular about a factor 10 higher in the analytical model.  The total optical depth of the atmosphere on the other hand agrees to a factor 2-3 in the two models (Fig. \ref{fig:mp08kappas}). The reason for the difference could be an overestimation of the efficiency of grain growth in the outer and intermediate layers in the analytical model, leading to grains of $\gtrsim10$ $\mu$m in size which  cause a relatively high opacity (see \citetalias{movshovitzpodolak2008} and the discussion further below). In the nominal numerical model of \citetalias{movshovitzpodolak2008} that assumes a monomer size of 1 $\mu$m,  the  grains in the outer layers are smaller, causing a lower $\kappa_{\rm gr}$ due to their low $Q$. On the other hand, if a monomer size of 10 $\mu$m is used in the numerical model, then the results for $\kappa_{\rm gr}$ agree better with the analytical result (see Fig. \ref{fig:mp08mono}).

\subsubsection{Timescales}\label{timescales}
We now discuss the microphysical grain dynamics that lead to the opacity. The analytical model is based on timescales, therefore we first show in  the left panels of Figure \ref{fig:mp2008taucompcoagcoal}  the timescales of growth by differential settling, Brownian motion coagulation, and  gas advection. By construction, in all three regimes, the growth  and  settling timescales  are the same.

\citetalias{movshovitzpodolak2008} found that the grain evolution approaches a steady state in only 100 yrs, and that after 1000 yrs, the steady state is fully established. Similarly short timescales are also found in the analytical model. The timescale for Brownian coagulation varies between about 500 and 2000 years. The timescale for gas advection is also about  1000 years in the outer layers and raises to about 6000 years at the bottom which is however irrelevant. The process with the shortest timescale is growth by differential settling.  It occurs on a timescale of about 300 years in the outer layers, and of less than 20 years at the radiative-convective boundary. This is indeed a very short timescale. 

\begin{figure*}
\begin{minipage}{0.49\textwidth}
	      \centering
       \includegraphics[width=0.95\textwidth]{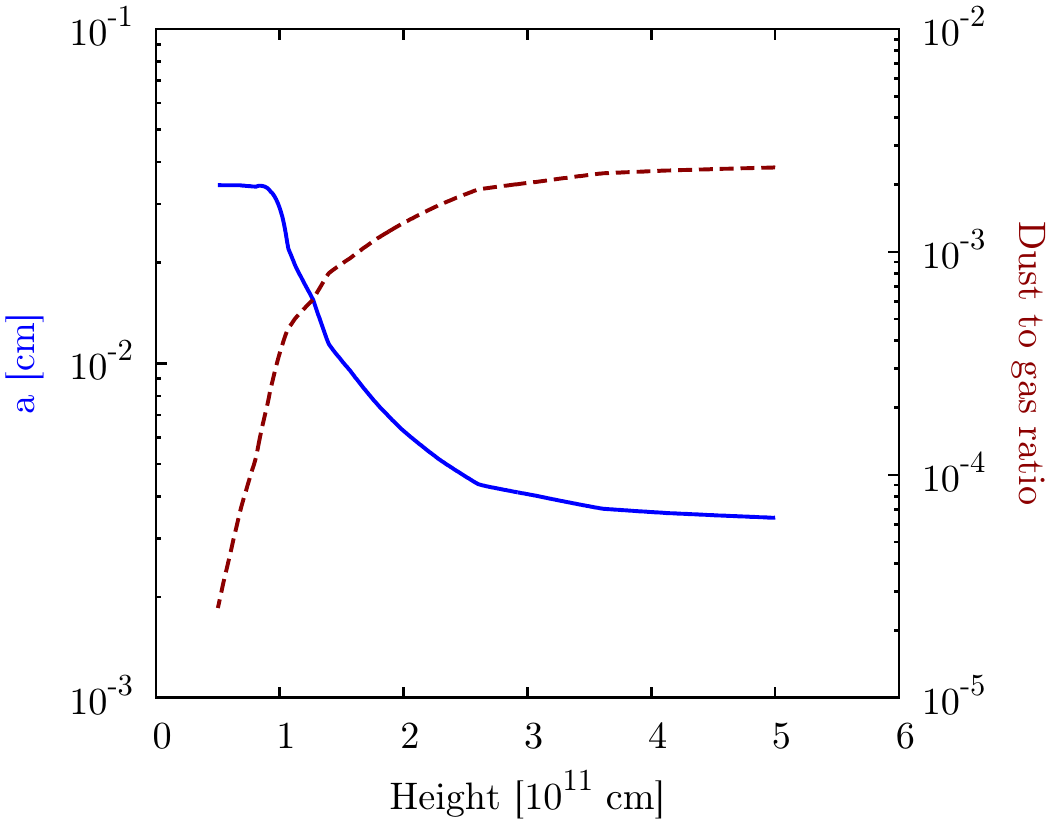}
     \end{minipage}
     \hfill
     \begin{minipage}{0.49\textwidth}
      \centering
             \includegraphics[width=0.95\textwidth]{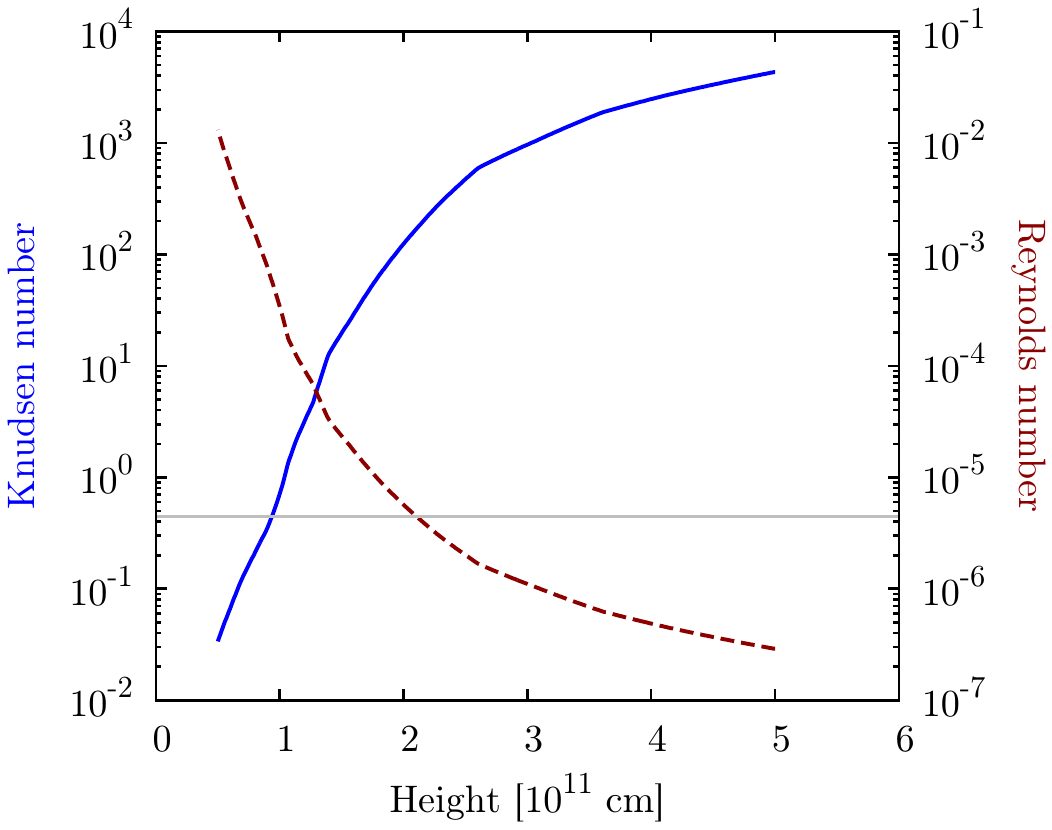}
     \end{minipage}
                  \caption{\citetalias{movshovitzpodolak2008} comparison case. \textit{Left panel:}  Grain radius (solid) and dust-to-gas ratio (dashed line).  \textit{Right panel:}  Knudsen (solid) and Reynolds number (dashed line) of the grains as a function of height. The horizontal line indicates the critical Knudsen number of 4/9 where the drag regime changes. }
             \label{fig:mp2008fpgknre} 
\end{figure*}

As \citet{coopersudarsky2003} we assume that the process with the shortest timescale is the dominant one. The figure shows that therefore, in the entire atmosphere, differential settling is the dominant process, meaning that the opacity is given by either Eq. \ref{eq:kappaDiffEpstein2} or \ref{eq:kappaDiffStokes2}, depending on the Knudsen number.  The dominance of differential settling is a general result, that applies to most atmospheres (see Sect. \ref{sect:tempevoSigma10}). Coagulation is about one order of magnitude slower in the outer layers, and a factor 30 slower at the bottom. Also gas advection is clearly slower, and therefore not important.  {This can be seen from the negligible  change of the timescale for differential settling if the effect of the gas velocity is directly included (gray dashed line), as described in Appendix \ref{sect:numsolve}.} This will be partially different in the simulation discussed in Sect. \ref{sect:Sigma10structcrossover}. There, the core is more massive at the crossover point, so that the gas accretion rate is higher, and therefore the gas advection timescale shorter.

Regarding the importance of Brownian coagulation versus growth by differential settling, one could argue that for differential settling, grains of different sizes must already exist. In the deeper atmosphere, this should always be the case because grains of different sizes rain into it from above. At the outer radius, in the limit that the initial size of the grains is identical (which is probably not realistic as grains already grow in the protoplanetary disk), this is not the case. Therefore, at the outer boundary, it is likely that first the slower process, i.e., Brownian coagulation sets  as the bottleneck the timescale up to a moment when differential settling kicks in. This would mean that Brownian motion needs first to act as the trigger of the  coagulation process as it is the case in disks \citep{weidenschilling1984,brauerdullemond2008}. This is an effect that we do not capture in the analytical model. Therefore, we likely overestimate the efficiency of grain growth in the top layers, and therefore underestimate the opacity. We will indeed see  indications of such a behavior when studying the effect of different dust-to-gas ratios in the disk (Sec. \ref{sect:effectfpg}).

We have assumed that grains always settle at the terminal velocity. The right panel of Fig.  \ref{fig:mp2008taucompcoagcoal}  shows that this is justified. It shows that the stopping time of the grains is only very short,  10-1000 seconds. As the settling timescales are of order 10-100 years, this means that the ratio of stopping to settling timescale RSS is tiny. As shown by the figure, it has values of order $10^{-8}$ to $10^{-7}$, meaning that the grains very quickly  reach the equilibrium between drag and gravitational force.

\subsubsection{Grain size, dust-to-gas ratio, and aerodynamic regime}
Now that differential settling is identified as the process determining the typical grain size,  local dust-to-gas ratio, and opacity, it is interesting to study these quantities.  The left panel of  Fig. \ref{fig:mp2008fpgknre} shows the typical size of the grains $a$ as a function of height. At the outer boundary, the grains are relatively small ( $\approx3\times10^{-3}$ cm). This size is a factor 30 larger than the assumed monomer size. In the analytical model, the size found by equating the timescales is almost always larger than the (typical) monomer size ($\sim 1 \mu$m). Exceptions are the rarely occurring advection regime  or when large monomer sizes are assumed (100 $\mu$m). The results of the analytical model are therefore usually independent of the assumed monomer size. This only holds because we do not inject grains at the monomer size also into the deeper layers. 

Further down into the atmosphere, the grains then grow to a size of about $3\times10^{-2}$ cm at a height of slightly less than $10^{11}$ cm. Inside of this height, the size remains approximately constant. This change at  $10^{11}$ cm comes from a transition in the drag regime, as we will see below in this section. 

It is interesting to compare the typical size predicted by the analytical model  with the grain size distribution found by \citetalias{movshovitzpodolak2008} where the grain size distribution is shown for four layers in the atmosphere (their Fig. 4). At the outer boundary, their size distribution has only one global maximum which is at the assumed monomer size (1 $\mu$m). Inside the atmosphere, the size distribution is bimodal. One (larger) maximum is always found at the  monomer size (1 $\mu$m). This peak is likely due to locally injected monomers originating from planetesimal ablation. The second maximum in contrast shifts to larger grain sizes as we move inwards, as in the analytical model. It is likely that this peak corresponds to grains which have already settled through a significant part of the atmosphere while growing. They therefore correspond to the grains studied here. In Table \ref{tab:compsize} we compare the grain size of these maxima with the typical grain size found in the analytical model.

\begin{table}
\caption{Grain size $a$ in cm in \citetalias{movshovitzpodolak2008} and the analytical model as a function of height in the atmosphere.}\label{tab:compsize}
\begin{center}
\begin{tabular}{lcc}
\hline\hline
Height [cm] & \citetalias{movshovitzpodolak2008} & this work \\ \hline
$5\times10^{11}$ &  $1\times10^{-4}$ & $3.4\times10^{-3}$\\
$2\times10^{11}$ &  $5\times10^{-3}$  &$6.3\times10^{-3}$\\
$1\times10^{11}$ &  $2\times10^{-2}$  &  $3.0\times10^{-2}$ \\                                              
$5\times10^{10}$ &  $6\times10^{-2}$    &  $3.3\times10^{-2}$\\ \hline
\end{tabular}
\end{center}
\end{table}

The table shows that except for the top layer, the maximum of the grain size distribution in the numerical model and the typical size found here agree relatively well, namely to a factor of two or less. This indicates that despite its simplicity, the analytical model is able to describe the essence of the grain evolution quite well, also quantitatively. 

\begin{figure*}
\begin{minipage}{0.492\textwidth}
	      \centering
            \includegraphics[width=0.95\textwidth]{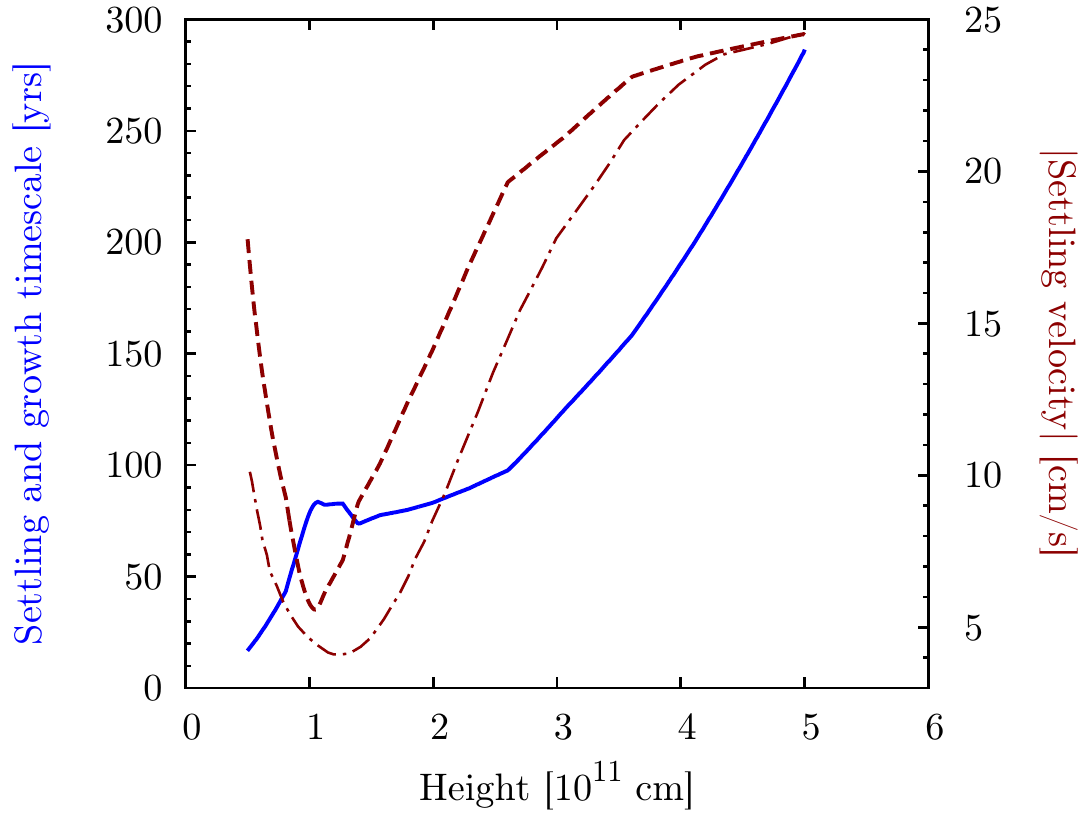}
     \end{minipage}
     \hfill
     \begin{minipage}{0.49\textwidth}
	                          \includegraphics[width=0.99\textwidth]{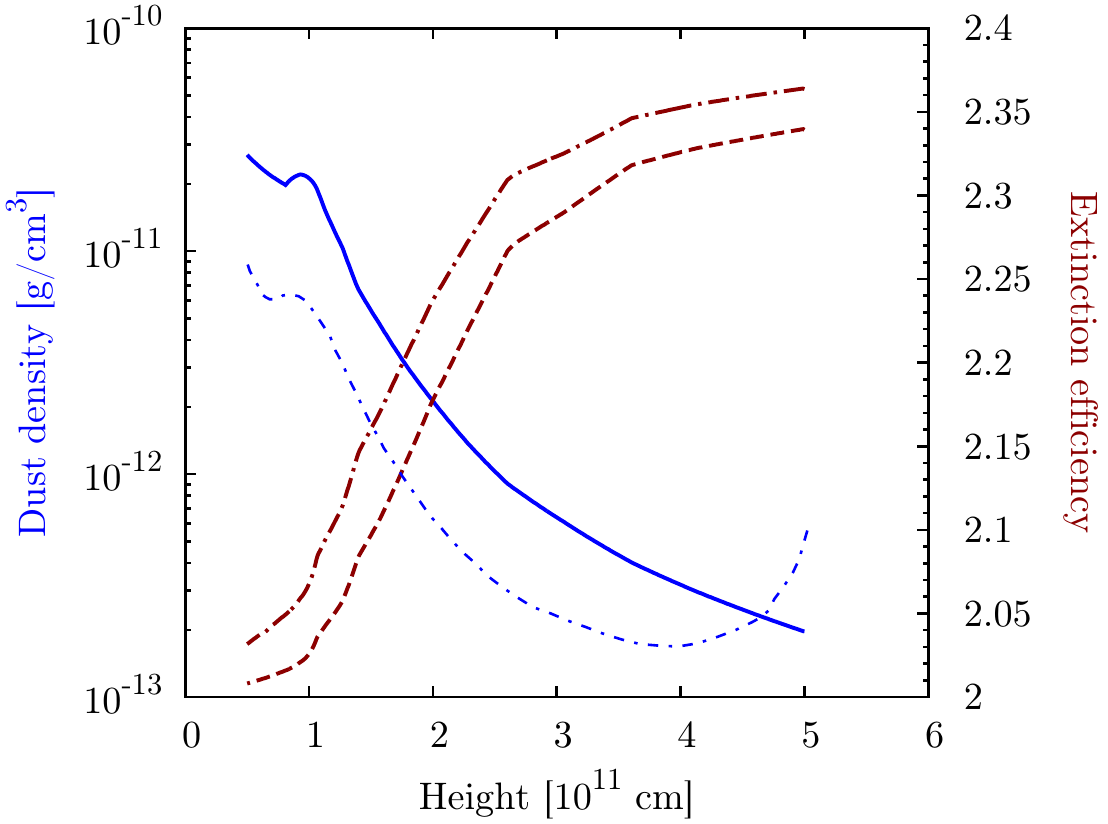}
      \centering
     \end{minipage}
                  \caption{\citetalias{movshovitzpodolak2008} comparison case. \textit{Left panel:} Settling and by construction identical growth timescale in years (solid line) and settling velocity (dashed line) as a function of height. The thin dashed-dotted line is the settling velocity found by \citetalias{movshovitzpodolak2008}. As they only give the relative velocity $v_{\rm set}(R)/v_{\rm set}(R_{\rm out})$, we have normalized their curve to the value predicted by the analytical model at  $R_{\rm out}$.  \textit{Right panel:} dust mass density ($\fpg\times\rho$) found by the analytical model (blue solid line) and by \citetalias{movshovitzpodolak2008} (thin dashed-dotted line). The two brown lines are the extinction efficiency, calculated with the fit of Eq. \ref{eq:figQ} (dashed) and directly with Mie theory (dashed dotted line). }
             \label{fig:mp08vset} 
\end{figure*}

The left panel of Fig.  \ref{fig:mp2008fpgknre} also shows the local dust-to-gas ratio $\fpg$ as a function of height. It scales as $1/(R^{2}\rho v_{\rm set})$ (Eq. \ref{eq:fpggeneral}). The plot shows that $\fpg$ decreases by two orders of magnitude across the atmosphere, showing that the effect of an increasing  gas density and settling velocity of the grains (due to grain growth) dominates over the decrease of $R$. This decrease of $\fpg$ and the increase of the grain size $a$ both conspire to the strong decrease of $\kappa_{\rm gr}\propto \fpg/a$ in the deeper parts of the atmosphere.  

When calculating $\dot{M}_{\rm gr}$, it was assumed that $f_{\rm D/G,disk}=0.01$ as in \citetalias{movshovitzpodolak2008}. The plot shows that the model predicts a sudden jump down to about 0.003 in the outermost layer. Such a non-continuous behavior is in principle not physical. It is a consequence of the simplification that the growth by differential settling sets in instantaneously already in the outermost shell, which overestimates the growth as discussed in Sect. \ref{timescales} and therefore artificially reduces $\fpg$.  

The right panel of Fig.  \ref{fig:mp2008fpgknre} shows the Knudsen and Reynolds number as a function of height. In the outer parts, the Knudsen number is much larger than 4/9, therefore the Epstein regime applies. As we move deeper into the atmosphere,  the mean free path $\ell$ of a gas molecule decreases while the grain size increases, so that at a height of about $10^{11}$ cm, the drag changes into the Stokes regime. This affects the settling velocity as we will see in the next section, and therefore also the grain size. 

The Reynolds number is very small in the outer atmospheric layers ($\sim10^{-7}$)  and increases to about $\sim10^{-2}$ at the boundary to the convective zone. This shows that the grains  always settle in a laminar flow,  meaning that the expressions for the drag laws we use are appropriate. 

\subsubsection{Settling velocity, dust density, and extinction efficiency}\label{sect:mp08vsetdensqbounce}
The left panel of Fig. \ref{fig:mp08vset}  shows the settling velocity. We see that the settling velocity is about 25 cm/s in the top layers. It then decreases to about 5 cm/s at a height of $10^{11}$ cm. After this local minimum, it increases again, reaching about 18 cm/s at the boundary to the convective zone. This increase is related to the change of the drag regime, as mentioned by \citetalias{movshovitzpodolak2008}. In the Epstein regime, the settling velocity scales as $a/(\rho \sqrt{T})$, while in the Stokes regime it scales as $a^{2}/\sqrt{T}$ without a dependency on the gas density. This means for the Stokes regime that the settling velocity increases faster with increasing particle size $a$ (or in other words, that it becomes more difficult to grow larger, as visible in Fig. \ref{fig:mp2008fpgknre} ), and that the settling velocity does not decrease with increasing density as we go deeper into the atmosphere.

The settling velocity found in the analytical model can be compared with Figure 8 in \citetalias{movshovitzpodolak2008}. A direct quantitative comparison is difficult because \citetalias{movshovitzpodolak2008} plot only the normalized velocity $v_{\rm set}(R)/v_{\rm set}(R_{\rm out})$ relative to the value in the top layer at $R_{\rm out}$ for two specific grain sizes (0.5 and 0.005 cm). In Fig. \ref{fig:mp08vset}, we also include their result for 0.5 cm grains, normalized to the settling velocity given by the analytical model at  $R_{\rm out}$. Nevertheless, we can see that the shape of the curves is  similar in both models with the maximum value at the top, the local minimum at about $10^{11}$ cm for the large grains, and the re-increase below this height. 

The settling velocity is also important from another point of view. In this simple model presented here, we implicitly assume that all particle collisions lead to perfect sticking. The very high number of studies regarding this subject in the context of grain as well as planetesimal growth in protoplanetary disks (e.g., \citealt{blummunch1993,guettlerblum2010,okuzumitanaka2012, windmarkbirnstiel2012} to name just a few), shows that this is an optimistic assumption, which is justified only for  certain collisional regimes. In other regimes, fragmentation or bouncing occurs. The outcome of a collision depends on several quantities  \citep[see, e.g.,][]{blumwurm2008}, namely the particle sizes, the collisional velocity, the particle inner structure (fractal dimension, compact versus porous), and the material type (ice versus silicates). 
 
Laboratory experiments indicate (see \citealt{guettlerblum2010}) that the critical velocity for collisional fragmentation of similar sized particles made of SiO$_{2}$ monomers is approximately 100 cm/s.  This is more than the settling velocity found in Fig.  \ref{fig:mp08vset} (the collision velocities should be a fraction of settling velocity, depending on the relative size), so that fragmentation should not occur. 

\begin{figure*}
\begin{minipage}{0.5\textwidth}
	      \centering
       \includegraphics[width=0.95\textwidth]{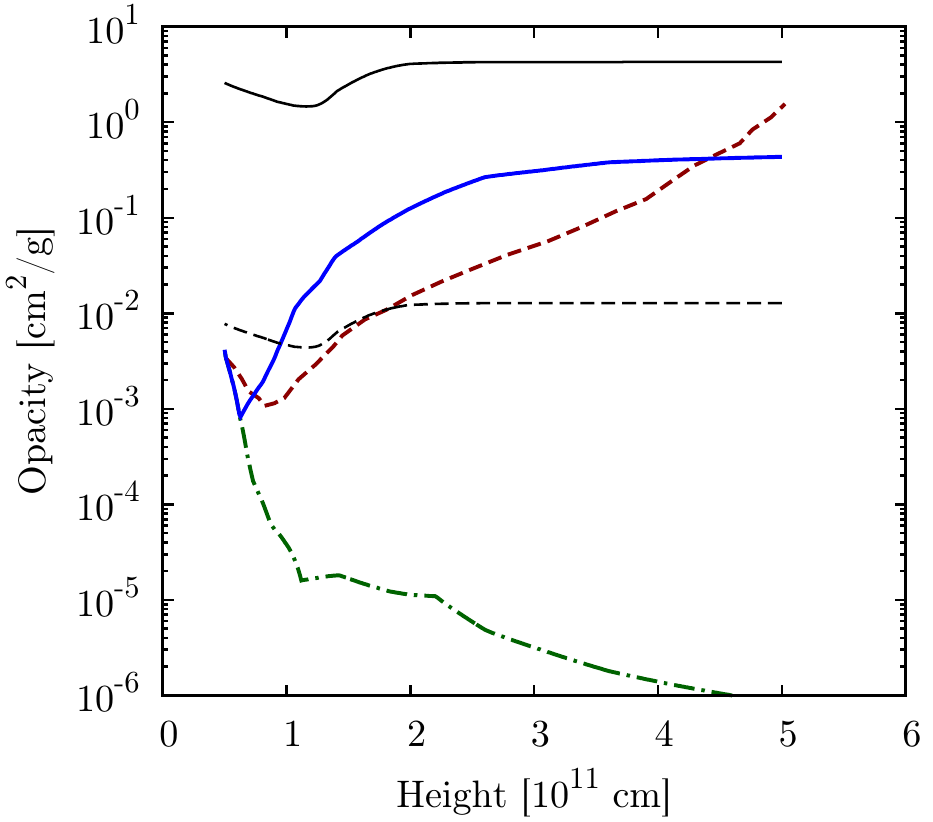}
     \end{minipage}
          \hfill
     \begin{minipage}{0.495\textwidth}
      \centering
                    \includegraphics[width=0.95\textwidth]{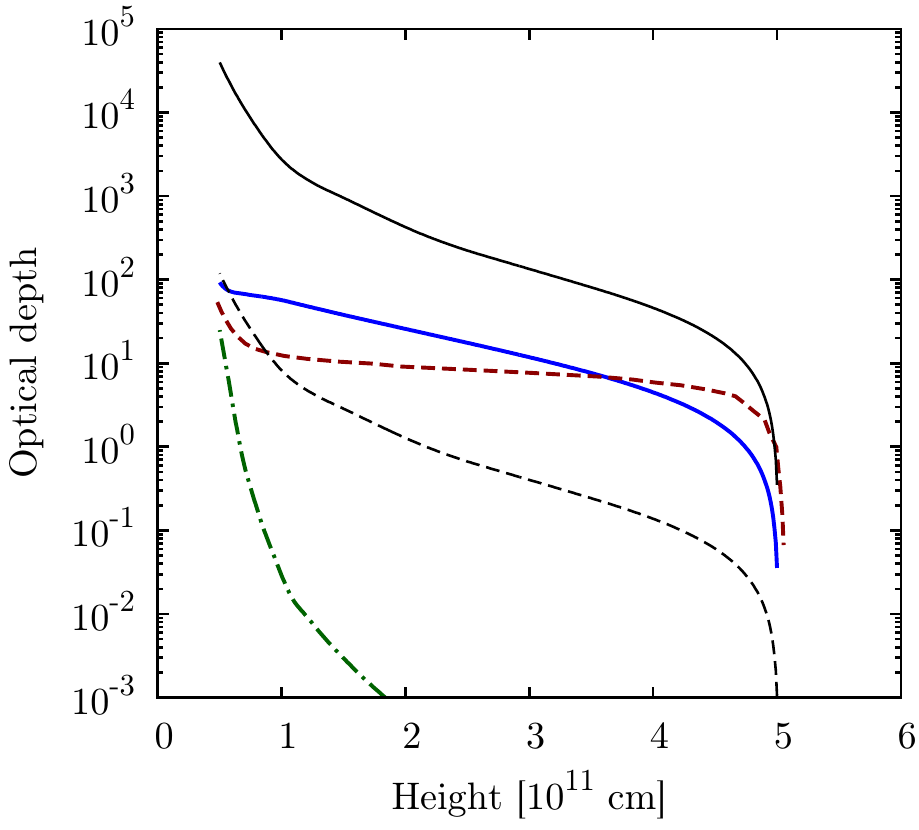}
     \end{minipage}
                  \caption{\citetalias{movshovitzpodolak2008} comparison case. \textit{Left panel:} Opacity as a function of height. The brown dashed line is the result of \citetalias{movshovitzpodolak2008}. Other lines show the total opacity found in this work (blue solid line), the gas opacity only (green dashed-dotted line), the full ISM opacity (thin  solid black line) and the ISM opacity $\times \ 0.003$ (thin dashed black line).  \textit{Right panel:} Corresponding optical depths.  }
             \label{fig:mp08kappas} 
\end{figure*}

However, when comparing the grain mass-settling velocity relation predicted by the analytical model for this atmosphere with Figure 11 in \citet{guettlerblum2010}  {(see also Sect. \ref{sect:perfectsticking} below)}, one sees that the collisions could fall into the bouncing regime, except if the collision velocities are only of order 1\% of the settling velocity, or if mostly particles of significantly different sizes collide.  In this case, the collisions would again lead to growth. Note that there is currently a disagreement between numerical simulations (where little bouncing is observed) and laboratory experiments \citep{windmarkbirnstiel2012}.  The uncertainty associated with the impact of bouncing should be kept in mind when interpreting the result of the present work. To better understand this mechanism, future models of grain growth in protoplanetary atmospheres should build on the  apparatus that has been developed in the past to study grain growth in protoplanetary disks (Ormel \& Mordasini, in prep).

The left panel of Fig. \ref{fig:mp08vset} also shows again the timescale of growth by differential settling (coalescence) which is the relevant regime. It is clear that a sticking efficiency of less than unity would also affect the timescales.  \citetalias{movshovitzpodolak2008} investigated the effect of uniformly reduced sticking efficiencies. At a lower efficiency of 0.4, the opacity increased mildly.  Lower efficiencies also mean that the time until a steady state is reached, increases. At an efficiency of 0.1, the a steady state is established after $\sim3500$ years instead of less than 1000 years. 

The right panel of Fig. \ref{fig:mp08vset} shows the mass density of dust ($\rho \times \fpg$) as a function of height. \citetalias{movshovitzpodolak2008} show the same quantity in their Figure 5, and their curve is also included here. We see that the two lines have a similar shape, but that there is a certain offset. Even relatively fine structures like the small bump close to $10^{11}$ cm are seen in both models. Interestingly, the bump only appears clearly in the product of  $\rho$ and $\fpg$, but not the two quantities alone. 

There are also differences: in \citetalias{movshovitzpodolak2008}, the dust density increases outside of about 4.5$\times10^{11}$ cm to reach at the outer boundary the value that is simply given by the nebular density times the dust-to-gas ratio in the disk, corresponding to about $10^{-12}$ g/cm$^{2}$. In the analytical model, the density only decreases towards the exterior. This is probably a consequence of the mentioned shortcoming of the analytical model that predicts that differential settling dominates everywhere, while in the numerical model, Brownian motion dominates in the outer layers \citepalias{movshovitzpodolak2008}. The comparison allows to deduce that the thickness of the layer where Brownian motion dominates is about $0.5\times10^{11}$ cm. A similar a length scale is also seen in the numerical model when the dust-to-gas ratio in the disk is varied (Sect. \ref{sect:effectfpg}). 

A second difference is that the analytical model predicts in the middle and lower parts a mass density of dust that is about a factor 3 larger than in \citetalias{movshovitzpodolak2008}.  This could be the reason why also the opacity  predicted by the analytical model is for most of the atmosphere higher than in the numerical model (Fig. \ref{fig:2008rt}).

In the right panel of Fig. \ref{fig:mp08vset} also the extinction efficiency  $Q$ is given.  For the dashed line $Q$ was calculated using the fit from Sect. \ref{sect:extinctioncoeff}, while the dashed-dotted line  it was calculated directly with Mie theory. We see that in both cases, $Q$ is approximately equal to 2 in the entire atmosphere. This is the consequence of $a\gg\lambda_{\rm max}$. This is again a general result. We find that in most atmospheres, grain growth is so efficient that $Q$ is to good approximation equal to 2. This means that the expressions derived for $\kappa_{\rm gr}$ in Eqs. \ref{eq:kappaDiffEpstein2} and \ref{eq:kappaDiffStokes2} are indeed useful ones as typically, they can be directly used for the Rosseland mean opacity (except for the outermost layers). Finally, we have tested if the results for $Q$ change  if we calculate the Rosseland mean opacity instead  of evaluating $Q$ only at the wavelength of maximum flux as described in Sect. \ref{sect:extinctioncoeff}. We found that the results are again very similar. This is due to the fact that the grains are so large and the temperature sufficiently high that we are  in the wavelength-independent regime \citep{cuzziestrada2013}.

\subsubsection{Comparison with scaled ISM opacity and optical depth of the atmosphere}\label{sect:compism}

In the past, numerous studies have  used the ISM grain opacity multiplied by some constant reduction factor to mimic the effects of grain evolution \citepalias[see][]{mordasiniklahr2014}. It is therefore interesting to compare the radial structure of the ISM opacity and the opacity found with the grain evolution calculations. Figure \ref{fig:mp08kappas}, left panel, shows the opacity as found by \citetalias{movshovitzpodolak2008} and the analytical model together with the full ISM opacity (from \citealt{belllin1994}) and the ISM opacity multiplied by 0.003. This reduction factor was determined in \citetalias[][]{mordasiniklahr2014}  as the fitting factor that leads to a duration of phase II that is similar as in \citet{movshovitzbodenheimer2010}. The figure also contains the grain-free gas opacity for a solar composition gas from \citet{freedmanmarley2008}.  The right panel shows the corresponding optical depths of the atmosphere, $\tau_{\rm atmo}$. 

\begin{figure*}
\begin{minipage}{0.46\textwidth}
	      \centering
       \includegraphics[width=0.95\textwidth]{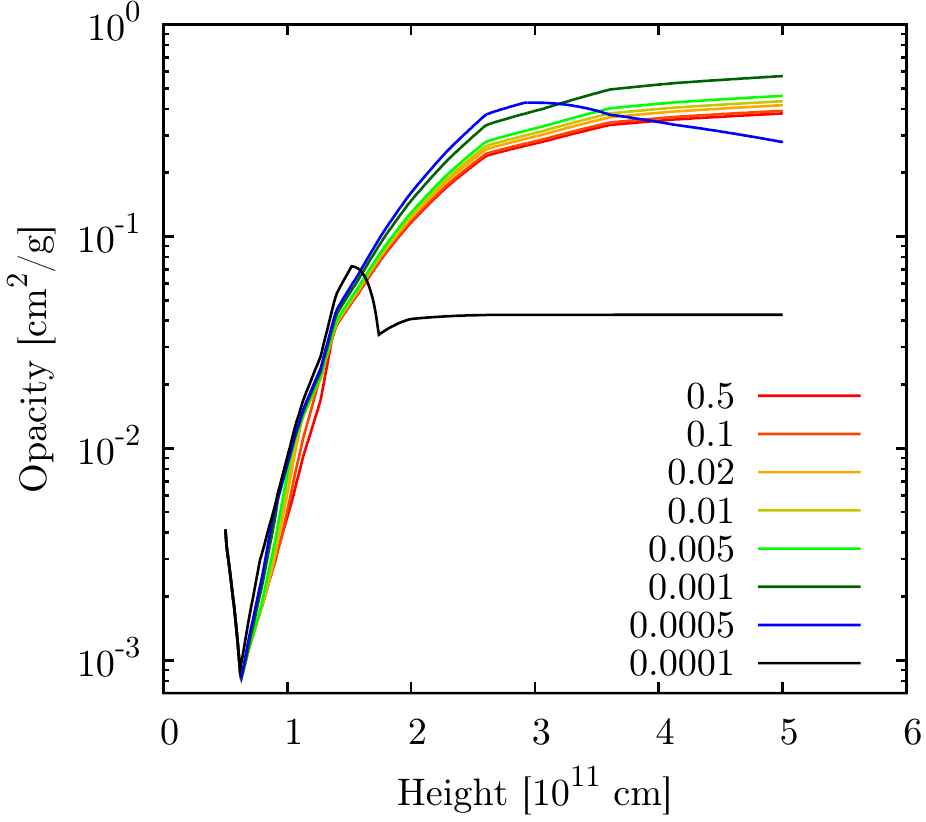}
     \end{minipage}
     \hfill
     \begin{minipage}{0.52\textwidth}
      \centering
                    \includegraphics[width=0.95\textwidth]{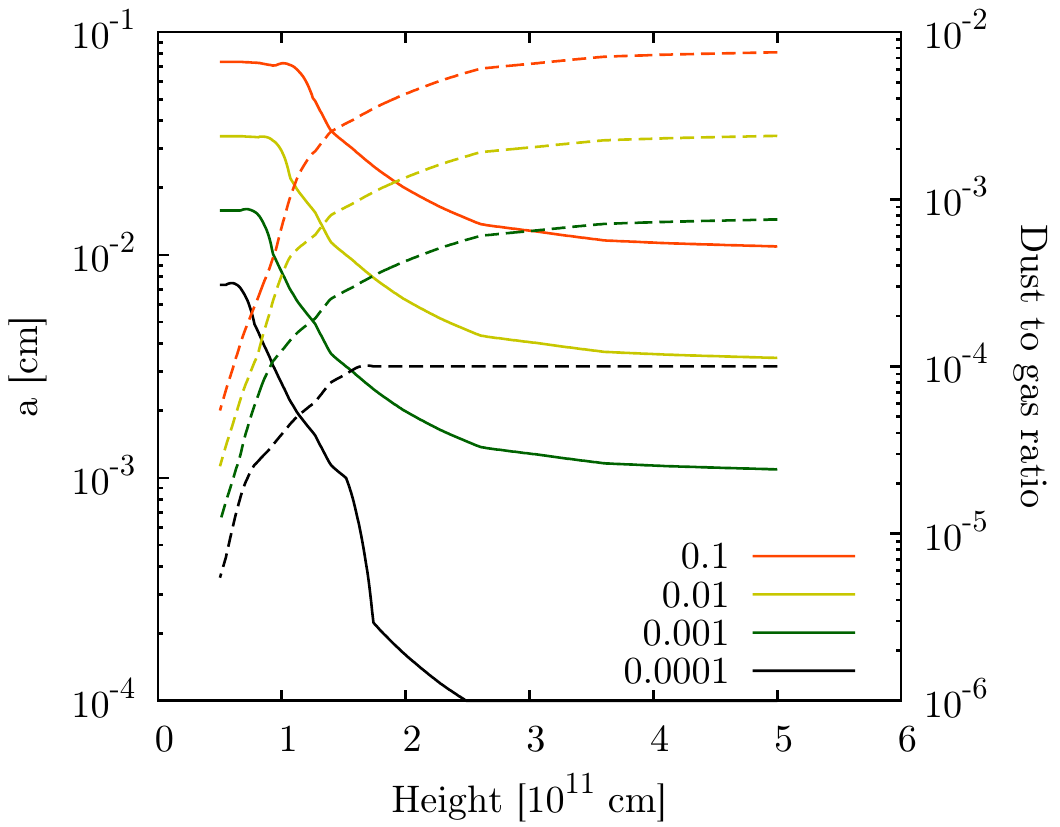}
     \end{minipage}
                  \caption{\citetalias{movshovitzpodolak2008} comparison case. \textit{Left panel:}  Opacity as a function of height for eight different dust-to-gas ratios in the accreted gas $f_{\rm D/G, disk}$ indicated in the plot. \textit{Right panel:}  Corresponding size of the grains (solid lines) and local dust-to-gas ratio (dashed lines) for the four dust-to-gas ratios in the accreted gas indicated in the plot.}
             \label{fig:mp08fpg} 
\end{figure*}

The plot shows that the full ISM opacity is more or less radially constant at a high value  ($\sim$1.3-4 g/cm$^{2}$), in clear contrast to the dynamically calculated opacity. The ISM opacity is therefore up to a factor 2000 higher (in the deeper layers) than the opacity predicted both by the numerical and analytical model. This is a very large difference. The corresponding total optical depth is also about a factor $\sim400-800$ larger than predicted by the grain calculations. The optical depth for full ISM opacity is approximately 4$\times$10$^{4}$, while the numerical and analytical models predict $\tau_{\rm atmo}$ of 90 and 50, respectively. They therefore agree to a factor $\sim$2. 

The ISM opacity multiplied by 0.003 is lower than the result from the microphysical calculations in the outer layers, but higher in the inner layers, meaning that it has a clearly different (roughly speaking constant) radial shape. The differences are up to one and a half orders of magnitude. The total optical depth ($\approx$110) on the other hand agrees well with the numerical result. This is probably not surprising, since the reduction factor was derived by fitting similar simulations \citepalias{mordasiniklahr2014}.

The grain-free opacity is again different. It only decreases towards the outer layers, which is also in contrast to the grain calculations.  {This means that it predicts a vanishing optical depth for the  atmosphere between 1 and $5\times10^{11}$ cm, whereas the calculations including the grain dynamics predict an optical depth of at least 10 for this part, which is not negligible.} In the deeper layers,  {the grain free opacity} dominates over the grains as described earlier. This makes that the total optical depth is about 25  {even without grains}. While this is significantly lower than the result from the grain calculations, it still yields an interesting insight: using grain-free opacities (cf. \citealt{horiikoma2010}) provides an optical depth that is closer to the result from the grain calculations than the full ISM value.  {On the other hand, the substantially different radial structure and also the too low total optical depth mean that also a completely grain-free opacity differs from the result obtained when the grain dynamics are considered.  }

In retrospect it means that classical studies like \cite{pollackhubickyj1996} should have considered the grain-free case as a meaningful alternative to the case that grains exist in the atmosphere. Potentially, this would have meant that the classical timescale problem of the core accretion theory would not have been regarded as an important drawback of this theory for several decennia.

\subsubsection{Impact of the dust-to-gas ratio in the disk}\label{sect:effectfpg}
The impact of different dust-to-gas ratios in the disk $f_{\rm D/G, disk}$ and therefore different $\dot{M}_{\rm gr}=\fpg \dot{M}_{\rm XY}$ on the opacity is the litmus test showing which grain growth mechanism (Brownian coagulation vs. differential settling) dominates. This is because for Brownian coalition, $\kappa_{\rm gr}$ increases for a fixed gas accretion rate $\dot{M}_{\rm XY}$ as $f_{\rm D/G, disk}^{5/9}$ and $f_{\rm D/G, disk}^{7/13}$ in the Epstein and Stokes regime, respectively (Eqs. \ref{eq:kappabrownepstein}, \ref{eq:kappabrownstokes}). While these are not very strong dependencies, they are nevertheless in clear contrast to the growth by differential settling: for differential settling, $\kappa_{\rm gr}$ is independent of $f_{\rm D/G, disk}$  both for the Epstein and Stokes  regime (Eqs. \ref{eq:kappaDiffEpstein2}, \ref{eq:kappaDiffStokes2}). 

\citetalias{movshovitzpodolak2008} investigated the impact of varying $f_{\rm D/G, disk}$ between 0.001 and 0.02, i.e., by a factor 20. For Brownian coagulation, this would lead to a variation of $\kappa_{\rm gr}$ by a factor of approximately five according to the analytical estimates. But \citetalias{movshovitzpodolak2008} actually found  that in the dominant part of the atmosphere (between a height of about 1 and 4.5$\times10^{11}$ cm), the opacity was  nearly independent of $f_{\rm D/G, disk}$. The analytical model now gives an explanation for this behavior. It is also a very clear sign that differential settling is indeed the dominant growth mode and not Brownian coagulation, as indicated by the analytical timescale criteria. This is a result with important wider implications which we discuss in Sect. \ref{sect:implicationsfehpebble}. Inside of 1$\times10^{11}$ cm, \citetalias{movshovitzpodolak2008} found that $f_{\rm D/G, disk}$ is slightly anti-correlated with $\kappa_{\rm gr}$ (variation by a factor 3). This is likely a consequence of the a change in the grain size distribution, an effect that we cannot capture in the analytical model (see Sect. \ref{sect:mp08structopa}). Outside of 4.5$\times10^{11}$ cm, the opacity in contrast increases with $f_{\rm D/G, disk}$ (variation by a factor $\sim$20). This  is again a clear sign that in the outermost layers, Brownian coagulation is important \citep{movshovitzpodolak2008}.

The left panel of  Fig. \ref{fig:mp08fpg}  shows the opacity for $f_{\rm D/G, disk}$ between 0.0001 and 0.5, a very wide (and not necessarily realistic) range.  {One} sees that for all $f_{\rm D/G, disk}$ except 10$^{-4}$, the model predicts a very similar opacity. The variation that still exists is a consequence of the decrease of $Q(x_{\rm max})$ that approaches 2  (Fig. \ref{fig:qfit}) as the grains become larger (see next).  Only for $f_{\rm D/G, disk}=10^{-4}$, the opacity decreases in the outer parts of the atmosphere. The reason is that at such low grain concentrations, the timescale of growth by differential settling becomes longer than the gas advection timescale. The growth regime therefore changes in the outer layers into the advection regime, resulting in low opacities due to the very low $\fpg$. The analytical model  thus agrees with the numerical result that in the dominant part of the atmosphere, the opacity is nearly independent of  $f_{\rm D/G, disk}$. It differs by predicting that this independence holds for the entire atmosphere, including the top layers. This is due to the overestimation of the efficiency of differential settling also in the outermost layers that was mentioned earlier (Sect. \ref{timescales}).  

The right panel illustrates the reason for the independence of $\kappa_{\rm gr}$ on  $f_{\rm D/G, disk}$. The figure shows the corresponding grain size $a$ and the local dust-to-gas-ratio $\fpg$ as a function of height. For an approximately constant $Q$ (because of the always large grains relative to the wavelength), $\kappa_{\rm gr}$ scales simply $\propto \fpg/a$ (Eq. \ref{eq:kappaxfpga}). One sees that high  $f_{\rm D/G, disk}$ indeed lead to a higher $\fpg$ which would increase the opacity, but at the same time these higher grain concentrations also lead to larger grains, reducing the opacity. As $\fpg$ and $a$ depend in the settling regime in the same way on $f_{\rm D/G, disk}$, these two effects cancel. We have checked that even for the largest grain, the Reynolds number stays below unity, meaning that the applied drag laws remain valid.  {We note that for very large grains expected for high $\fpg$, bouncing and fragmentation could become important (Sect. \ref{sect:perfectsticking}). This could limit the efficiency of grain growth in very metal rich envelopes.}

Up to now, we have discussed the effect of varying the grain input in the top layer that is due to a different  $f_{\rm D/G, disk}$ in the accreted gas. \citetalias{movshovitzpodolak2008} also studied the effect of varying the grain input due to the ablation of planetesimals. For an increase of the grain input in a certain layer, the general result is that around this layer, the opacity is also increased, while in the remaining parts of the atmosphere, $\kappa$ stays approximately unaffected. This means that around the layer of increased grain input, Brownian motion dominates, while far away, it is differential settling. This is the same general behavior as for an increase due to a higher $f_{\rm D/G, disk}$. The simulations of \citetalias{movshovitzpodolak2008}  also show that the thickness of the layer with an increased opacity becomes smaller if the grains are injected in deeper layers. This is expected, because the timescale of growth decrease with depth (Fig. \ref{fig:mp2008taucompcoagcoal}). 

\subsubsection{Implications for the ``metallicity effect'' and pebble accretion}\label{sect:implicationsfehpebble}
The independence of the opacity on the grain accretion rate is the result of this study that has the most important wider implications for planet formation theory.  It means that even if  more grains are added  {at the top of the atmosphere}, the $\kappa_{\rm gr}$ does not significantly increase, as one may naively expect. This is important in two contexts:

First, in the context of the ``metallicity effect'', which is the observational finding that giant planets are clearly more frequent around host stars with a high [Fe/H]  \citep{gonzalez1997,santosisraelian2001,fischervalenti2005}. Within the core accretion theory, this can be explained  if a higher stellar [Fe/H] means a higher associated [Fe/H] in the protoplanetary disk. This likely leads to a higher surface density of planetesimals \citep{kornetbodenheimer2005}, which in turn leads to a faster growth of more massive cores that can trigger gas runaway accretion \citep[e.g.,][]{pollackhubickyj1996}. Due to this mechanism, planet population synthesis models built on the core accretion paradigm  reproduce the ``metallicity effect'' \citep{idalin2004,mordasinialibert2012a}. But one could also argue that at a higher stellar [Fe/H], not only the planetesimal surface density, but also the $f_{\rm D/G, disk}$ should be higher, which could (via a higher opacity) be detrimental for giant planet formation. The analytical model (and the numerical simulations of \citetalias{movshovitzpodolak2008}) show this is not the case. A higher stellar [Fe/H] should not strongly increase the grain opacity in the protoplanetary atmospheres. This is an important insight for the core accretion theory. Also, in this light,  the a priori questionable assumption of past population synthesis calculations \citep[e.g.,][]{idalin2004,mordasinialibert2012a} that the opacity is independent of the disk [Fe/H] can be physically justified, at least as a first order approximation. 

Second, the independence is important in the context of the recently proposed mechanism that the accretion of small objects (pebbles instead of 100 km planetesimals) is a key element for the formation of giant planet cores \citep[e.g.,][]{ormelklahr2010,lambrechtsjohansen2012,morbidellinesvorny2012}. Such small objects make a rapid formation of the core possible thanks to their low random velocities and the large, drag-enhanced capture radius of the protoplanet. This  mechanism could, however, a priori suffer from a disadvantage: In contrast to 100 km planetesimals (that deposit most of their mass in the convective zone or even directly on the core), small objects will strongly increase the grain input in the outer layers of the atmosphere (as illustrated by Fig. \ref{fig:massdepo}). This could potentially increase the opacity, making giant planet formation harder. The study here shows that this is at least to first order not the case. An accretion of the core via small bodies should not strongly increase the opacity. We have further tested this by re-calculating the opacity for the case of  ``shallow deposition'' (Eq. \ref{eq:grainaccrationrate}), adding grains in the top layer at a rate of $10^{-6}$ $\mearth$/yr (\citetalias{movshovitzpodolak2008}). We found that the opacity again remains virtually identical. This means that the small bodies would not strongly increase the opacity, but they would increase the mean molecular weight of the envelope gas in the layers that are sufficiently hot for grain evaporation (see \citealt{fortneymordasini2013} for estimates of the mean molecular weight in protoplanetary atmosphere). A high mean molecular weight is known to reduce the critical core mass \citep{stevenson1982,horiikoma2011}.  The low random velocity, the large capture radius, the high mean molecular mass, and the low opacity thus conspire to facilitate giant planet formation.    

We have found an absence of a strong dependency of $\kappa_{\rm gr}$ on the dust-to-gas ratio in the dominant differential settling regime. The quantity that enters the equations is, however, in principle not $f_{\rm D/G,disk}$ alone, but $f_{\rm D/G,disk} \dot{M}_{\rm XY}$. This means that for a fixed dust-to-gas ratio, the opacity is also not explicitly dependent on the gas accretion rate. It is on the other hand clear that the quantities on which $\kappa_{\rm gr}$ does depend (local atmospheric pressure  $P$ and gravity $g$) and $\dot{M}_{\rm XY}$  are not independent. The absence of  an explicit dependency on $\dot{M}_{\rm XY}$ still means that if $P$ and $g$ remain similar when $\dot{M}_{\rm XY}$ changes, then also $\kappa_{\rm gr}$ remains similar. This could be part of the explanation why the global structure of $\kappa$ remains similar for different core masses (Section \ref{sect:tempevoSigma10})

\subsubsection{Impact of the monomer size}\label{sect:mp2008amonomer}
\begin{figure}
\begin{center}
\includegraphics[width=\columnwidth]{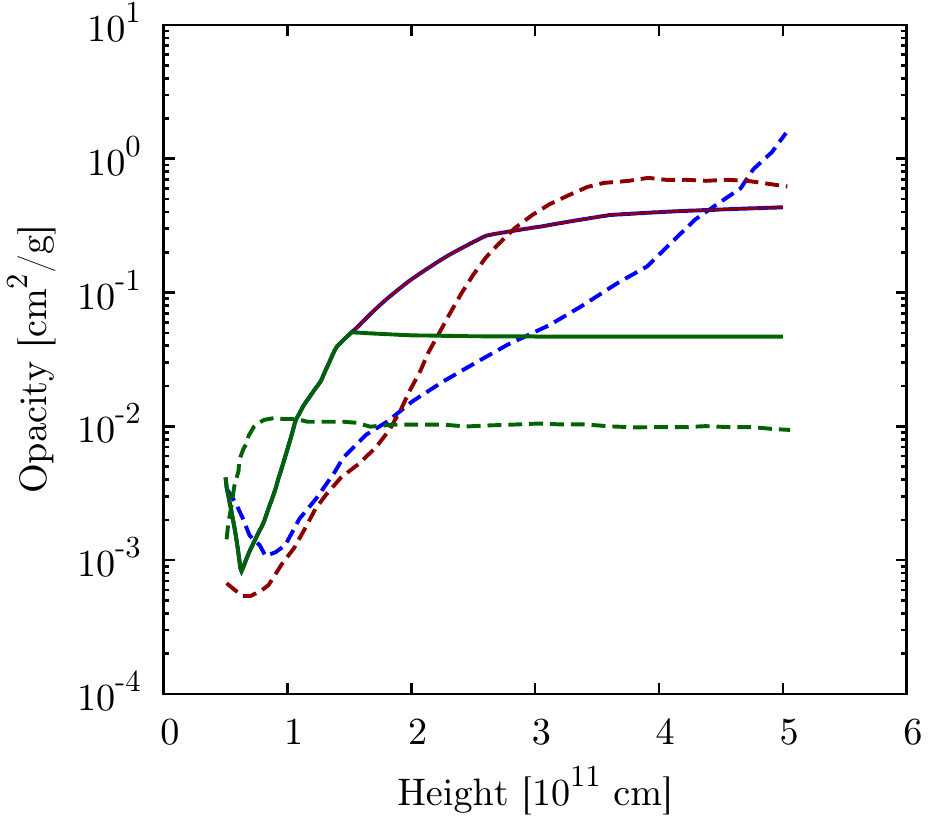}
\caption{\citetalias{movshovitzpodolak2008} comparison case. Impact of the monomer size on the opacity. Blue: 1 $\mu$m.  Brown: 10 $\mu$m. Green: 100 $\mu$m. Dashed lines are  from \citetalias{movshovitzpodolak2008}, solid ones from the analytical model where the 1 and 10 $\mu$m  case lead to an identical opacity so that the lines lay on top of each other.}\label{fig:mp08mono}
\end{center} 
\end{figure}

As \citetalias{movshovitzpodolak2008}, we have finally also investigated how the grain opacity changes if the assumed size of the monomers is varied. As mentioned (Sect. \ref{sect:calctypsizeopa}),  {in atmospheres where advection is not important as it is typically the case,} the size of the monomers  enters the analytical model only if the size of the grains calculated from the timescales is smaller than $a_{\rm mono}$. Figure \ref{fig:mp08mono} shows the opacity as a function of height for monomer sizes of 1, 10 and 100 $\mu$m. The results of both the analytical model and of \citetalias{movshovitzpodolak2008} are shown. 

For the analytical model, the opacity is identical for a monomer size of 1 and 10 $\mu$m. This is simply because the size of the grains found from the timescale arguments is everywhere larger than 10 $\mu$m (Fig. \ref{fig:mp2008fpgknre}). Only for $a_{\rm mono}=100$ $\mu$m, there is an impact, and over a large part of the outer atmosphere, the opacity becomes radially nearly constant at a low value of about 0.05 cm$^{2}$/g. This is quite similar to the result in the numerical model, even if the numerical value of $\kappa_{\rm gr}$ differs by roughly a factor five. In the numerical model, there is also a difference between the 1 and 10 $\mu$m case. The 10 $\mu$m case has a higher opacity and is clearly closer to the result of the analytical model than the $a_{\rm mono}=1$ $\mu$m case. This is not surprising; it is again a manifestation of the result that the analytical model overestimates the efficiency of grain growth in the outermost layers. Therefore, in the analytical model, the grains have also in the outer regions a size of $\gtrsim30$ $\mu$m.  {This means that the purely local and monodisperse description of the grain dynamics by the comparison of timescale alone is not sufficient to capture all effects seen in the numerical model, where  in particular the relative contributions from different sized grains matter}. As explained by \citetalias{movshovitzpodolak2008}, grains with a size of order 10 $\mu$m are  more efficient in causing a high opacity than both smaller and larger grains. If   $a_{\rm mono}$ is set to 10 $\mu$m instead of 1 $\mu$m in the numerical model, the opacity therefore becomes higher, and the radial shape is now more similar to the analytical model. 

The results illustrate that it is important to know the typical size of the grains when they are accreted into the protoplanet. If the grains in the protoplanetary disk are typically  already larger than $\sim$100 $\mu$m, this could  reduce the opacity in the protoplanetary atmosphere. A coupling of the grain evolution in the disk and in the protoplanetary atmospheres should therefore be included in future planet formation models.

\section{Simulations of Jupiter's formation coupled with the analytical grain growth model}\label{sect:combsims}
We now present the results of coupling the analytical grain evolution model with (giant) planet formation simulations. This is analogous to the work of \citetalias{movshovitzbodenheimer2010} where the numerical grain evolution model of \citet{podolak2003} and \citetalias{movshovitzpodolak2008} was coupled with the giant planet formation model of \citet{pollackhubickyj1996}. Our primary interest is to compare the analytical and numerical results for the giant planet formation timescale and the radial structure of the envelope. 

\subsection{Core accretion model and initial conditions}
The core accretion model used here was first introduced in Alibert, Mordasini \& Benz (\citeyear{alibertmordasini2004}), while a detailed description can be found in \citet{alibertmordasini2005} and \citet{mordasinialibert2012b}. We here only give a short summary. In \citet{mordasinialibert2012b}, simulations can be found  that resemble much the ones presented here with the difference that in this earlier work, the grain opacity was found by scaling the ISM opacity. Our code is based on the core accretion paradigm (see, e.g., \citealt{mordasiniklahr2010b} for an overview) and contains three computational modules that are very similar as in the \citet{pollackhubickyj1996} model.  Whenever possible, we have chosen parameters and settings that are identical to the ones in \citetalias{movshovitzbodenheimer2010}.
 
The first computational module calculates the core accretion rate $\dot{M}_{Z}$. It uses the same prescriptions for the planetesimal random velocities as \citet{pollackhubickyj1996}. The gravitational capture radius that includes the effect of the star is taken from \citet{greenzweiglissauer1992}. The size of the planetesimals is 100 km. The solid core is assumed to have a fixed density of 3.2 g/cm$^{3}$, in contrast to \citet{mordasinialibert2012c} where the core density is variable. All these setting are very similar or identical to those in \citetalias{movshovitzbodenheimer2010}.

The second computational module calculates the trajectories of the impacting planetesimals as they fly  through the protoplanetary envelope \citep{mordasinialibert2006}. This model is equivalent to the one of \citet{podolakpollack1988}. It yields the drag enhanced capture radius of the protoplanet which can be significantly larger than the core radius \citep{inabaikoma2003}. The second output from the impact model, namely the radial profiles of  mass and energy deposition, are currently not used. Instead, all mass is directly added to the core using the ``sinking'' approximation for the calculation of the core luminosity. The enrichment of the gaseous envelope is thus not considered, while it could be very important for a reduction of the critical core mass \citep{horiikoma2011}. This is because high enrichments of $Z\approx0.8$ are expected  for planets with masses between 1 and 10 $\mearth$ \citep{fortneymordasini2013}.

In our simulations, the only source of grains in the $\dot{M}_{\rm gr}$ term is the accretion of grains with the gas in the top layer of the atmosphere, i.e., we assume a ``deep'' deposition of the planetesimals (Eq. \ref{eq:grainaccrationrate}) for the reasons shown in Sect. \ref{sect:grainaccrrateplanetesimal}. In  \citetalias{movshovitzbodenheimer2010}, dust grains are in contrast injected both due to gas accretion and also planetesimal ablation, but  \citetalias{movshovitzbodenheimer2010} note that the planetesimal deposit their mass typically fairly deep in the gaseous envelope. In any case, based on the analytical results (Eq. \ref{eq:kappaDiffEpstein2}  and \ref{eq:kappaDiffStokes2}), a weak (or no) dependency of $\kappa_{\rm gr}$ on $\dot{M}_{\rm gr}$ is expected if the dominant grain growth mechanism is differential settling. This is indeed found to be the case, as will be discussed below. The grains are added assuming a dust-to-gas ratio in the disk $f_{\rm D/G,disk}=0.01$  {for all planetesimal surface densities as in \citetalias{movshovitzbodenheimer2010}.}  {Note that we thus ignore potential correlations of  $f_{\rm D/G,disk}$ and the planetesimal surface density.}  We have also run non-nominal simulations that use a $f_{\rm D/G,disk}$ of 0.001 and 0.1. We find that this has only a $\approx8$\% effect on the formation timescale (see Sect. \ref{sect:mbpl2010fdgamono}). The monomer size is 1 $\mu$m, but this only matters in the advection regime, which is itself not important (see Fig. \ref{fig:menvemcoreM1064}).  \citetalias{movshovitzbodenheimer2010} use a $a_{\rm mono}=1.26$ $\mu$m.

The third module solves the internal structure equations for the gaseous envelope in the quasi-static 1D approximation (see \citealt{bodenheimerpollack1986}) using the  SCvH EOS \citep{saumonchabrier1995}. We  simplify the equations by assuming that the luminosity $L$ is constant with radius $r$ inside the envelope and solve the temporal evolution based on an energy conservation approach \citep{mordasinialibert2012b}. This does not significantly influence the results, since  $\partial L/\partial r$ is very small in the radiative zones (see \citealt{mordasinialibert2012b} for details). The outer boundary conditions (radius, pressure, temperature) are the same as in \citetalias{movshovitzbodenheimer2010}. The opacity is calculated with the analytical model described above, it thus contains the contributions from the grains, plus the molecular and atomic opacities from \citet{freedmanmarley2008} and \citet{belllin1994}. The grain opacity can be calculated dynamically during a radial integration of the envelope structure based on the local properties in a layer and on $\dot{M}_{\rm gr}$ during one timestep. No special iterations or recursion on the opacity in the last timestep are necessary.  This means that the computational time is only insignificantly increased compared to the case where tabulated or fitted (ISM) opacities are used.
 
As in \citetalias{movshovitzbodenheimer2010}, the planet forms in situ at 5.2 AU in a disk with a gas surface density of 700 g/cm$^{2}$.  Disk evolution, orbital migration \citep{alibertmordasini2005} and the interaction with other embryos \citep{alibertcarron2013} are all neglected. The planetesimal accretion rate  is as mentioned calculated as in \citet{pollackhubickyj1996}, therefore it is likely too high for 100 km planetesimals \citep[e.g.,][]{fortieralibert2013}. This means that the simulations are not intended to show a necessarily realistic formation scenario of Jupiter. In any case, our interest here is to compare the analytical results with the numerical model of \citetalias{movshovitzbodenheimer2010}, and therefore to run simulations that are as similar as possible.

Finally, we consider three initial surface densities of planetesimals at the planet's position: 10, 6, and 4 g/cm$^{2}$, again as in \citetalias{movshovitzbodenheimer2010}, calling the three  simulations $\Sigma$10,  $\Sigma$6, and $\Sigma$4, respectively. 

\subsection{Time evolution of the core and envelope mass}
\begin{figure*}
\begin{minipage}{0.33\textwidth}
	      \centering
       \includegraphics[width=1\textwidth]{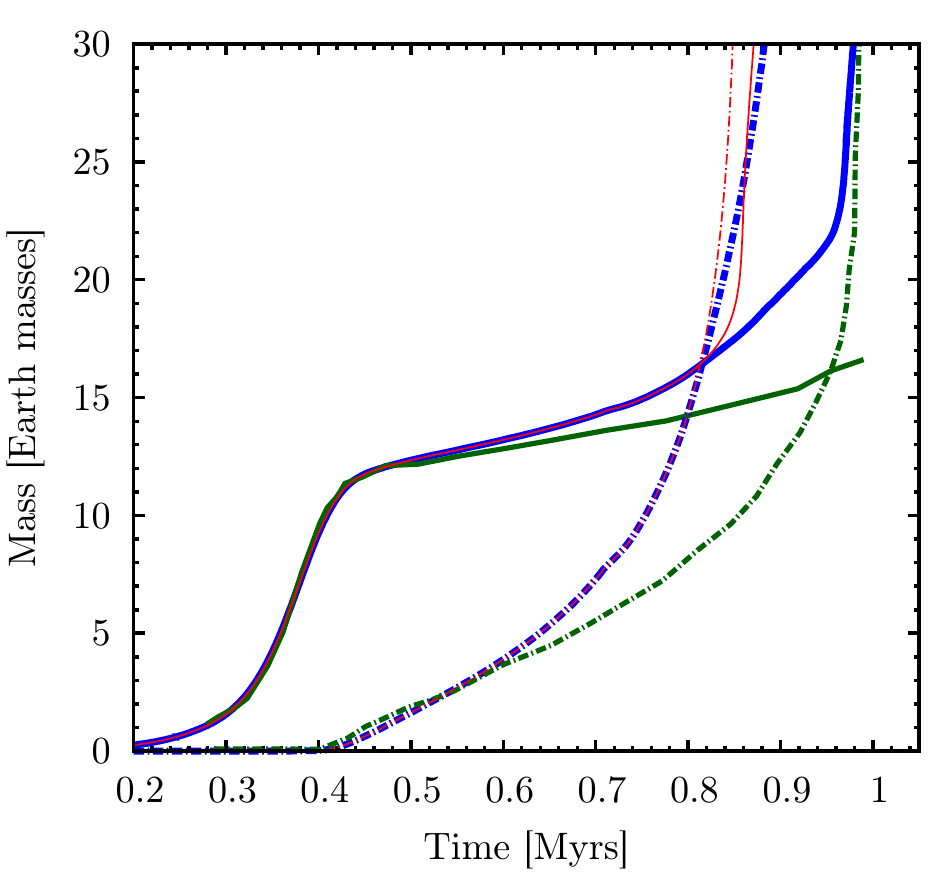}
     \end{minipage}
          \hfill
     \begin{minipage}{0.33\textwidth}
      \centering
                    \includegraphics[width=1\textwidth]{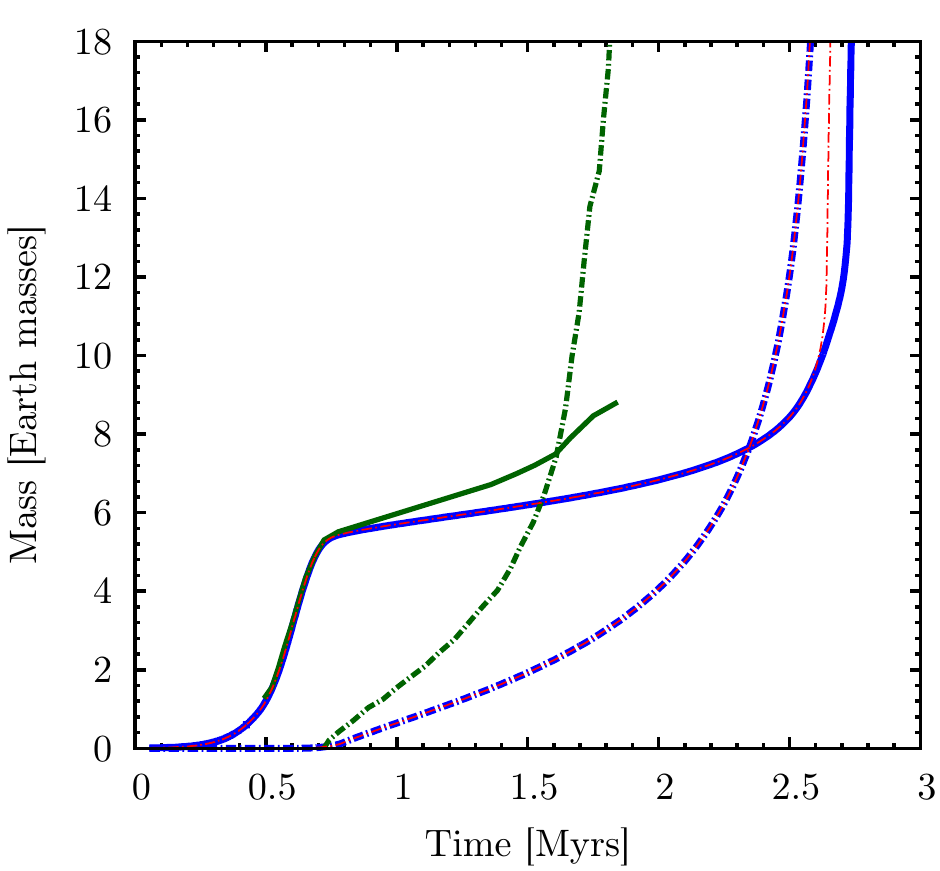}
     \end{minipage}
      \hfill
          \begin{minipage}{0.33\textwidth}
      \centering
                    \includegraphics[width=1\textwidth]{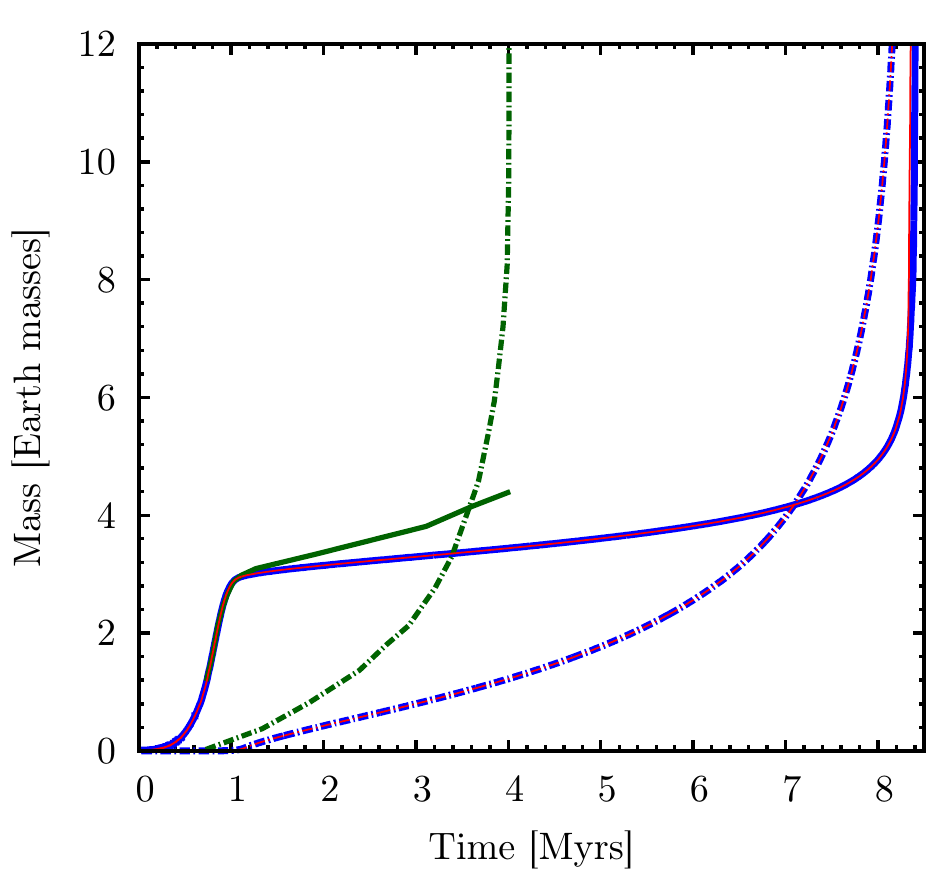}
     \end{minipage}
                  \caption{Core (solid line) and envelope mass (dashed-dotted line) as a function of time for Jupiter in situ formation. The  panels show the result for an initial planetesimal surface density of 10 (left), 6 (middle), and 4  g/cm$^{2}$ (right). The thick blue lines are calculated with the analytical model for the grain opacity, while the thinner green lines are  from the numerical model of \citetalias{movshovitzbodenheimer2010}.  The thin red lines are also from the analytical model, but without taking the advection regime into account. }
             \label{fig:menvemcoreM1064} 
\end{figure*}

The basic evolution of the core and envelope mass as a function of time for in situ formation has been described in many previous works (e.g., \citealt{pollackhubickyj1996}). The main result is that there are three different phases which we here only summarize very briefly. During phase I, the solid core forms rapidly (due to the assumption of low planetesimal random velocities). The luminosity of the planet is high and caused by the accretion of the planetesimals. The envelope mass is low. This first phase ends once the isolation mass \citep{lissauer1993} is reached, so that phase II starts.  

The phase II is a plateau phase that is characterized by a relatively slow increase in envelope and core mass, with the envelope growing somewhat faster than the core. The duration of the phase II depends on the grain opacity \citepalias[see, e.g.,][]{mordasiniklahr2014}. This is because in order to expand its feeding zone of planetesimals, the envelope mass needs to grow, and this is only possible if the gas already contained within the planet's Hill sphere can cool and contract. Phase II is the longest phase, hence its duration determines also the duration of the overall formation process. Phase II ends at the so-called crossover point when the core and envelope mass are equal and phase III begins. 

In this third phase, the gas accretion rate increases rapidly because the planet's cooling timescale decreases with increasing mass, leading to runaway gas accretion. Shortly after the crossover point (typically a few 10$^{4}$ yrs for a 10 $\mearth$ core), the gas accretion rate found by solving the internal structure equations becomes larger than the limiting gas accretion rate at which the disk can supply gas to the protoplanet (typically of order $10^{-2}\mearth$/yr). At this point, the planet detaches from the disk and contracts rapidly \citep{lissauerhubickyj2009,mordasinialibert2012b,ayliffebate2012}.

This fundamental behavior can be seen in Fig. \ref{fig:menvemcoreM1064}. It shows the core and envelope mass as a function of time for the three different surface densities. The results obtained with the model presented here and the results of \citetalias{movshovitzbodenheimer2010} are  included. In our model, the protoplanet starts with a tiny initial mass of 0.01 $\mearth$. We have set the embryo's starting time in a way that the evolution of the core mass during phase I overlaps in both models. This first phase is very similar in both simulations, as exactly the same equations are integrated, and only the evolution of the core (but not of the envelope and thus the opacity) matter. 

In the ideal case, the predicted temporal evolution found with the analytical and numerical model would be identical. A key figure that can be used to quantify the degree of agreement is the time of crossover. This is because the time to reach the crossover point directly depends on the opacity. It is also a key quantity determining the fundamental outcome of the planet formation process: if the time to reach crossover is shorter then the disk lifetime, then a gas dominated Jovian planet formation is likely to form. This is because the gas accretion after the crossover point occurs on a  short timescale of $10^{4}-10^{5}$ years which is one to two orders of magnitude shorter than a typical disk lifetime. In the other case, only a solid dominated Neptunian planet will form. It is therefore obvious that the crossover time as predicted with the analytical grain growth model is a key quantity if the model is used in planetary population syntheses. We therefore list in Tab. \ref{tab:compcross} the time of crossover, the crossover mass, and the luminosity at this point for the different simulations. Since phase I is nearly identical in both models, the variation in the time of crossover is in fact mainly given by a variation in the duration of phase II. This is the quantity that was already studied in \citetalias{mordasiniklahr2014}. 

\begin{table}
\caption{Characteristics of the Jupiter in situ formation simulations at the crossover point.}\label{tab:compcross}
\begin{center}
\begin{tabular}{lccc}
\hline\hline
Simulation & t [Myr]  & $M$ [$\mearth$] & $\log(L/L_{\odot})$  \\ \hline
10 g/cm$^{2}$, this work &  0.81 & 16.42 &  -5.20 \\
10 g/cm$^{2}$, \citetalias{movshovitzbodenheimer2010} &  0.97  & 16.09 & -5.54\\
6 g/cm$^{2}$, this work &  2.35  &  7.65 & -6.39 \\                                              
6 g/cm$^{2}$, \citetalias{movshovitzbodenheimer2010} &  1.63    &  7.50  & -6.27 \\ 
4 g/cm$^{2}$, this work &  7.09 & 4.18 & -7.43 \\
4 g/cm$^{2}$, \citetalias{movshovitzbodenheimer2010} &  3.62  &  4.09 & -7.13 \\  \hline
\end{tabular}
\end{center}
\end{table}

\subsubsection{The nominal $\Sigma10$, $\Sigma6$, and $\Sigma4$ simulations}\label{sect:comp1064sims}
The left panel of Figure \ref{fig:menvemcoreM1064} shows the $\Sigma10$ simulation. For a full ISM grain opacity, a crossover time of about 6 to 7.5 Myr has been found in previous studies \citep{pollackhubickyj1996,hubickyjbodenheimer2005,mordasinialibert2012b}. Taking into account the grain physics, the analytical model predicts a crossover time of only 0.81 Myr, while the numerical model of \citetalias{movshovitzbodenheimer2010} predicts  0.97 Myr. This corresponds to a difference of only about 16\%. Given the simplifications in the analytical approach, this represents a good agreement. It indicates that the analytical and numerical models predict grain opacities (or at least total atmospheric optical depths) that agree relatively well during the entire phase II. In the following section, we will compare the radial structure of the envelope including the opacity. Figure \ref{fig:menvemcoreM1064} and Table \ref{tab:compcross} show that the crossover mass  ($\approx 16$$\mearth$) is also very similar in the two models (difference of only 2\%). This is expected because the crossover mass is independent of the opacity for \citet{pollackhubickyj1996}-like calculations \citepalias{mordasiniklahr2014}. The luminosity is higher in the analytical model, which is expected given the shorter accretion timescale. 

The middle panel of Fig. \ref{fig:menvemcoreM1064} shows the $\Sigma6$ case. At  a full ISM grain opacity, a crossover time of more than 60 Myr is expected \citepalias[see][]{mordasiniklahr2014} which clearly exceeds protoplanetary disk lifetimes. With grain evolution, the crossover time is found to be 2.35 and 1.63 Myr in the analytical and numerical model, respectively. In contrast to the  $\Sigma10$ case, the analytical model thus here leads to a crossover time that is longer than found by \citetalias{movshovitzbodenheimer2010}, namely by about 0.7 Myr or about 40\%. While this is a larger difference, it is a consequence of the stronger sensitivity of the formation time on $\kappa$ at a lower core mass. This can be seen by the fact that in the $\Sigma10$ simulation, the reduction factor relative to the full ISM opacity is ``only'' a factor of about 6, whereas here the reduction factor is of order 30. This means that the details of the calculation of $\kappa$ are more important. The crossover masses of about 7.6 $\mearth$ agree again very well, while the luminosity reflects the difference  in the formation time.

The left panel finally shows the $\Sigma4$ simulation. The core mass is now very low, only about 3 $\mearth$  at isolation, so that we are studying gas accretion onto super-Earth-type planets. Such planets could be the progenitors of the recently observed low-mass, low-density planets \citep[e.g.,][]{rogersbodenheimer2011,lissauerjontof2013b}. At  full ISM opacity, the crossover point would only be reached after several 100 Myr \citepalias{mordasiniklahr2014}, which means that the formation of giant planets would be excluded for such disk conditions. Comparing the analytical and numerical model, a similar tendency as for the $\Sigma6$ case is seen, but the difference in the crossover time is now larger (7.1 instead if 3.6 Myr). This corresponds to a disagreement by a factor 2, and since is its of the same order as typical disk lifetimes, it can mean that in some cases, the analytical and numerical model would disagree on  whether giant planet formation is possible or not.  On the other hand,  except for the works of \citet{movshovitzbodenheimer2010} and \citet{rogersbodenheimer2011}, in basically all other planet formation calculations \citep[like][]{mizunonakazawa1978,stevenson1982,pollackhubickyj1996,papaloizounelson2005,hubickyjbodenheimer2005,tanigawaohtsuki2010,levisonthommes2010,horiikoma2011,ayliffebate2012,dangelobodenheimer2013} the grain opacity was essentially an unconstrained parameter which was varied over two or even three orders of magnitude. This would translate in a variation of the crossover time of the same magnitude in the linear regime \citepalias[see][]{mordasiniklahr2014}. In this sense, a difference by a factor 2 found from a first  simple analytical model is still a relatively small difference. But one should keep in mind that the analytical model has a tendency to overestimate the grain opacity in the atmosphere of low-mass protoplanets (see Fig. \ref{fig:crossM4}). The reason for this, and improved analytical models are currently being investigated (Ormel \& Mordasini, in prep.). One possible reason is that the analytical model overestimates the impact of the planet mass on the grain size. In the analytical model presented here, at crossover, the grains are in the $\Sigma10$ case a factor of about 3 to 6 (depending on height) bigger than in the $\Sigma4$ case. If the actual dependency is weaker, this could explain why the formation timescale is underestimated in the $\Sigma10$ case, but overestimated in the two other cases with a smaller core mass.

\subsubsection{Impact of the advection regime}
Figure \ref{fig:menvemcoreM1064} also shows simulations where the advection regime (Sect. \ref{sect:advect}) is not taken into account. One sees that this only appreciably affects the accretion  after the crossover point.  {Thus, the crossover time is nearly identical in runs with or without the advection regime (see also Tab. \ref{tab:compcrossfpg}). After crossover}, the gas accretion rate increases rapidly, so that the advection timescale becomes shorter (Sect. \ref{sect:advect}), to a point that in the outer atmospheric layers, it is less than the timescale of growth by differential settling. Therefore, the growth regime changes. This has the consequence that an outer layer of high opacity forms (see Fig. \ref{fig:crossM10} below). This lengthens the time between the crossover point and the moment when the limiting accretion rate is reached. For the $\Sigma6$ case, for example, the  time between crossover and the moment when the limiting gas accretion rate of $10^{-2}$ $\mearth$/yr is reached is about 0.38 and 0.29 Myr for the simulation with and without advection, respectively. In \citetalias{movshovitzbodenheimer2010}, the time between crossover and reaching a similar limiting gas accretion rate is 0.24 Myr. This indicates that the effect of advection is indeed overestimated in the analytical model as discussed in Sect. \ref{sect:advect}.  In any case, differences of order $10^{5}$ yrs are usually not important for the global growth history and final mass of a planet, because it is one order of magnitude less than typical disk lifetimes. 

\subsubsection{Impact of the disk dust-to-gas ratio and monomer size}\label{sect:mbpl2010fdgamono}
In Section \ref{sect:diffsetepstein} and \ref{sect:diffsetstokes} we have found that opacity in the dominating differential settling regime does not depend on the dust accretion rate or dust-to-gas ratio in the disk. To quantify the impact of $f_{\rm D/G,disk}$ on the overall formation history, we have recalculated the $\Sigma10$  {and $\Sigma4$} case also for  {other} $f_{\rm D/G,disk}$ besides the nominal 0.01. We find that the impact on the formation history is small.  {Table \ref{tab:compcrossfpg} shows the crossover time for different  $f_{\rm D/G,disk}$ for the $\Sigma10$ and $\Sigma4$ cases, and for simulations with/without the advection regime. For the $\Sigma10$ case with advection, the time of crossover is 0.88, 0.81, and 0.80 Myr for $f_{\rm D/G,disk}$=0.001, 0.01, and 0.1.} This means that a lower dust-to-gas ratio even leads to slightly larger formation timescale. Similar effects were already seen by \citetalias{movshovitzpodolak2008}. In any case, the difference corresponds to just  {9}\% of the formation timescale for a change of  $f_{\rm D/G,disk}$ by a factor 100. This means that the dependency is very weak. We remind that the analytical model tends to overestimate the efficiency of differential settling in the outermost layers (Sect. \ref{timescales}). If these layers are in reality dominated by Brownian motion, then at least some positive correlation of $f_{\rm D/G,disk}$ and the formation timescale is expected.  {The same is true for the impact of ``shallow deposition''. In the current analytical model, it remarkably only has a negligible effect in the $\Sigma10$ simulation, and even leads to a (mild) reduction of the crossover time in the $\Sigma4$ case as grains grow more massive. These large grains cause a $Q\approx2$ in the entire atmosphere, while otherwise $Q\approx3$ in the outer layers. The occurrence of Brownian motion (or potentially collisional grain fragmentation, see Sect. \ref{sect:perfectsticking}) could, however, significantly alter this finding.}

\begin{table}
\caption{ {Impact of $f_{\rm D/G,disk}$ on the crossover time in Myr. ``w/o'' stands for simulations neglecting advection. ``shallow'' stands for shallow deposition (Eq. \ref{eq:grainaccrationrate}). }}\label{tab:compcrossfpg}
\begin{center}
\begin{tabular}{lcccc}
\hline\hline
$f_{\rm D/G,disk}$ & $\Sigma10$  & $\Sigma10$ w/o & $\Sigma4$ & $\Sigma4$ w/o \\ \hline
$10^{-1}$&   0.80 & 0.79 & -  & 6.80 \\
$10^{-2}$ &  0.81  & 0.80 & 7.09 & 7.09\\
$10^{-3}$ &  0.88  &  0.83  &  - & 6.32 \\                                              
$10^{-4}$ &   -     &  0.79  & 5.44 & 5.49\\ 
shallow    &  -       &  0.79 & - &  6.60\\  \hline
\end{tabular}
\end{center}
\end{table}

 {The table also shows the consequences of $f_{\rm D/G,disk}$ for the $\Sigma4$ case. We see that the impact is clearly larger, with a reduction of the crossover time in simulations without advection from 7.09 in the nominal case to 5.49 Myr for $f_{\rm D/G,disk}=10^{-4}$ ($\approx$25\% difference). This is due to the following: in the $\Sigma4$ case, at low $f_{\rm D/G,disk}$, the grains in the outer layers stay so small ($\sim$1$\mu$m)  that their extinction efficiency $Q$ decreases to $\lesssim0.1$. The leads to a radially roughly constant low $\kappa\approx0.07$ cm$^{2}$/g in the outer parts of the atmosphere (the specific value might, however, be affected by our simplistic way of calculating $Q$). For higher $f_{\rm D/G,disk}$, $Q$ is in contrast always between 2 and 3. In the $\Sigma10$ case, the grains also become smaller at lower $f_{\rm D/G,disk}$ as seen in Sect. \ref{sect:effectfpg}. But in contrast to the $\Sigma4$ case, for this more massive planet they still remain large enough for $Q$ to be in the geometrical limit. As the grain growth mode is differential settling, this means that $f_{\rm D/G,disk}$ does nearly not affect $\kappa$  (Eqs. \ref{eq:kappaDiffEpstein1}, \ref{eq:kappaDiffStokes1}), so that the crossover time is almost independent of it.  Including advection slightly further reduces the crossover time to 5.44 Myrs in the $\Sigma$4 simulation. This is due to an modification of the radial opacity structure by  the same mechanism as observed for $f_{\rm D/G,disk}=10^{-4}$ in  the \citetalias{movshovitzpodolak2008} comparison case (Sect. \ref{sect:effectfpg}). The impact on the overall formation time is still not extremely large, as the reduction of $\kappa$ only occurs in the outer layers. The density there is low, so that the associated reduction of the total optical depth is not very high. }

We have also tested the  {implications} of   {different} monomer sizes $a_{\rm mono}$. For the \citetalias{movshovitzpodolak2008} comparison case, it was found that increasing the monomer size from 1 to 10 $\mu$m has no impact, while setting it to 100 $\mu$m leads to a reduction of the opacity in the outer layers (Fig. \ref{fig:mp08mono}). Simulating the $\Sigma10$ case with  $a_{\rm mono}=10$ and 100 $\mu$m instead of the nominal 1 $\mu$m has similar consequences: for  $a_{\rm mono}=10$ $\mu$m, the formation is virtually the same as in the nominal case. For $a_{\rm mono}=100$ $\mu$m, the time of crossover is shorter by about 10 \%. In particular, a higher gas mass for a given core mass is seen in phase I. This is because in phase I, the grain size found from the analytical estimates can be relatively small in the outer layers (less than 100 $\mu$m)

\subsection{Radial temperature, density, and opacity structure}
The ISM opacity scalings presented in \citetalias{mordasiniklahr2014}  lead with a calibrated grain opacity reduction factor of 0.003 to formation timescales that agree with those obtained with the numerical model of \citetalias{movshovitzbodenheimer2010}. However, a simple scaling of the ISM opacity suffers from the limitation that it does not lead to the same (or similar) radial structure of the opacity. Only the total optical depth is similar (see Fig. \ref{fig:mp08kappas}). This means that it is unclear if the reduction factors that where calibrated for certain initial conditions are applicable also for other situations.

In this respect, the analytical growth model has a significant advantage as it calculates the opacity dynamically based on microphysical considerations. It is therefore very interesting to compare how the radial structures of the envelopes look like, and how they compare with the results of \citetalias{movshovitzbodenheimer2010}.

\subsubsection{The $\Sigma10$ case: structure at crossover}\label{sect:Sigma10structcrossover}

\begin{figure}	      
\centering
       \includegraphics[width=1\columnwidth]{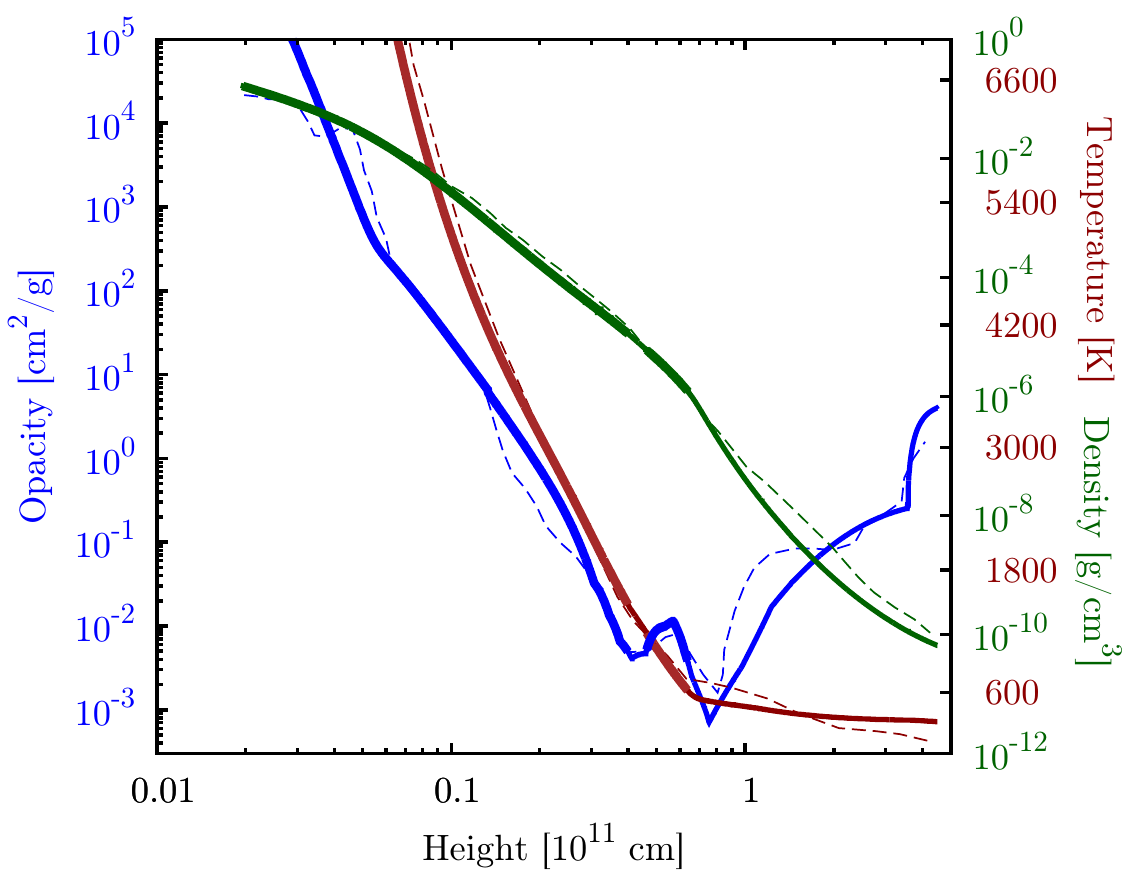}
                    \caption{Radial structure of the envelope in the $\Sigma$10 case shortly before crossover. The plot shows the opacity (blue), temperature (brown), and gas density (green)  as a function of height. Solid lines are the results form this work with convective parts  indicated by a thicker line. Dashed lines show the corresponding results from \citetalias{movshovitzbodenheimer2010} .}
             \label{fig:crossM10} 
\end{figure}

Figure \ref{fig:crossM10} shows the structure of the  gaseous envelope in the $\Sigma$10 case at a moment slightly before the crossover point. The core mass is 15.98 $\mearth$ while the envelope mass is 14.47 $\mearth$. The temperature, density, and opacity are shown. Note that in contrast to the situation in Sect. \ref{sect:compMPforgivenrhoT} where the density and temperature were externally given and only the opacity was then calculated, here the entire structure is the result of solving the coupled internal structure equations. This means that in contrast to the results in Sect. \ref{sect:compMPforgivenrhoT} , $T$, $\rho$, and $\kappa$ are self-consistent. 

The plot also shows the corresponding result of \citetalias{movshovitzbodenheimer2010} (their Fig. 4). In their simulation, the planet is at crossover and has a core mass of 16.09 $\mearth$.  One sees that the two models predict an internal structure that agrees well.  In \citetalias{movshovitzbodenheimer2010} the temperature at the core-envelope boundary is $18\,600$ K, while we find a temperature of 18\,040 K, corresponding to a difference of 3\%. The temperature structure consists of an outer zone that is radiative and roughly speaking isothermal (temperature increase by a factor $\sim$2), and a large inner convective zone where the temperature increases strongly (there is a tiny inner radiative zone between about 4 to $4.6\times10^{10}$ cm, but it remains without visible consequences for $\rho$ and $T$). In the  density structure, the radiative and convective parts can also be distinguished. 

For the opacity, the following layers can be seen, starting at the top: in the outermost parts of the atmosphere ($R\gtrsim3.6\times10^{11}$ cm), the growth occurs in the advection regime. This is a regime that only occurs at/after crossover (Fig. \ref{fig:menvemcoreM1064}). The grains remain relatively small, causing an opacity that is rather high, at least in the top layer where  $\kappa\approx3.6$ cm$^{2}$/g. This is comparable to typical ISM opacities.  The opacity structure of  \citetalias{movshovitzbodenheimer2010} shows a feature that might be related, but this is not entirely clear  {(see also Fig. 11 in \citealt{dangeloweidenschilling2014})}. For radii between $7.5  \times10^{10}$ and $3.6\times10^{11}$ cm, the growth mode is differential settling. This is the typical regime in the atmospheres. A slight change in the slope of $\kappa_{\rm gr}$ can be seen at $R\approx1.2\times10^{11}$ cm. This is where the drag regime changes from Epstein to Stokes drag. The grain opacity falls strongly with decreasing height (but is still larger than $\kappa_{\rm gas}$) which is very important for the ability of the planet to cool as noted by \citetalias{movshovitzbodenheimer2010}. This is clearly a consequence of the growth of the grains. The opacity reaches a global minimum of only about $7\times10^{-4}$ cm$^{2}$/g at $R=7.5\times10^{10}$ cm. Inside of this radius, the opacity is given by the molecular opacities rather than the grains, because now $\kappa_{\rm gr}<\kappa_{\rm gas}$. This is the same behavior as seen in Fig. \ref{fig:mp08kappas}, and was already found by \citetalias{movshovitzbodenheimer2010}.  The structure of the opacity below $R=7.5\times10^{10}$ cm is thus given by the \citet{freedmanmarley2008} data. This leads in particular to a second local minimum in $\kappa$ at about $4\times10^{10}$ cm. But in general, the opacity now increases with depth. The results of \citetalias{movshovitzbodenheimer2010} are very similar in this part of the envelope. This is not surprising as they use the same molecular opacity tables. 

The radius where the molecular opacities become larger than the grain opacity is further out than the radius where the grains evaporate. This grain evaporation radius is found to be at about  $4\times10^{10}$ cm as estimated from the evaporation timescale (Sect. \ref{sect:evap}). This means that grain evaporation is not important for the resulting opacity, which is an interesting result.  The radius where the molecular opacities become dominant over the grain opacity is also larger than the radius where convection first sets in (at about  $6\times10^{10}$ cm). This means that no convective regions exist where the opacity is given by the grains. This is important, since such region would make a special treatment necessary \citepalias{movshovitzbodenheimer2010}. This result is probably related to the finding of \citetalias{movshovitzbodenheimer2010} that including or neglecting the effects of convection on the grain growth process leads to negligible differences for the planet's evolution. 

\begin{figure*}
\begin{minipage}{0.5\textwidth}
	      \centering
       \includegraphics[width=0.97\textwidth]{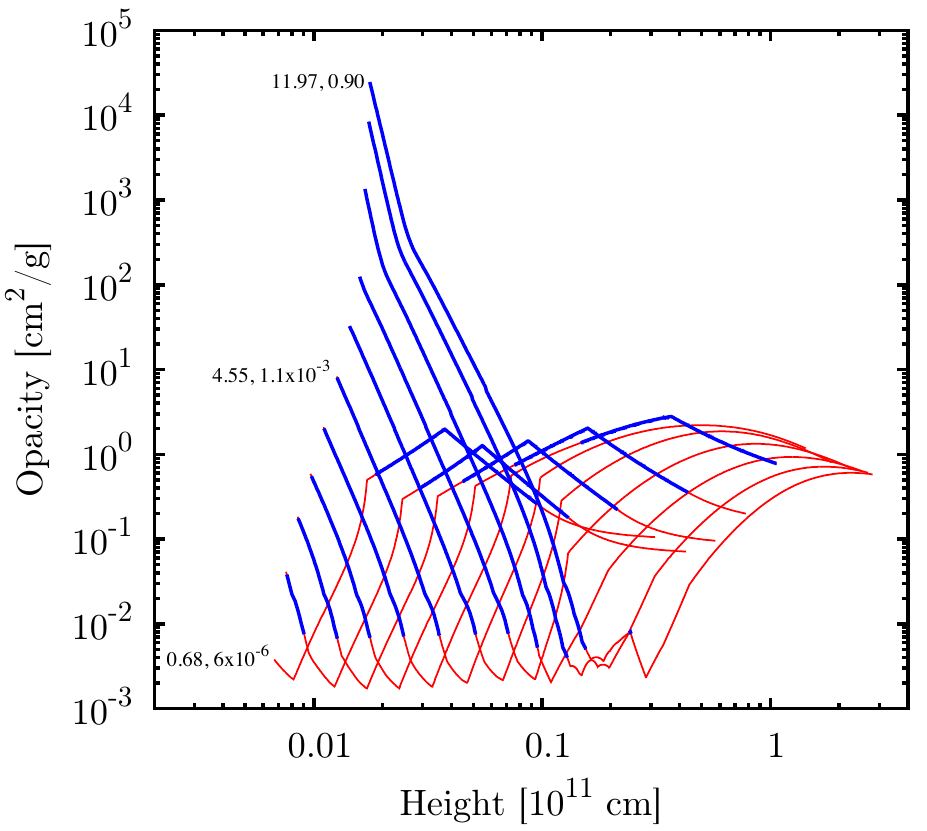}
     \end{minipage}
          \hfill
     \begin{minipage}{0.5\textwidth}
      \centering
                    \includegraphics[width=0.97\textwidth]{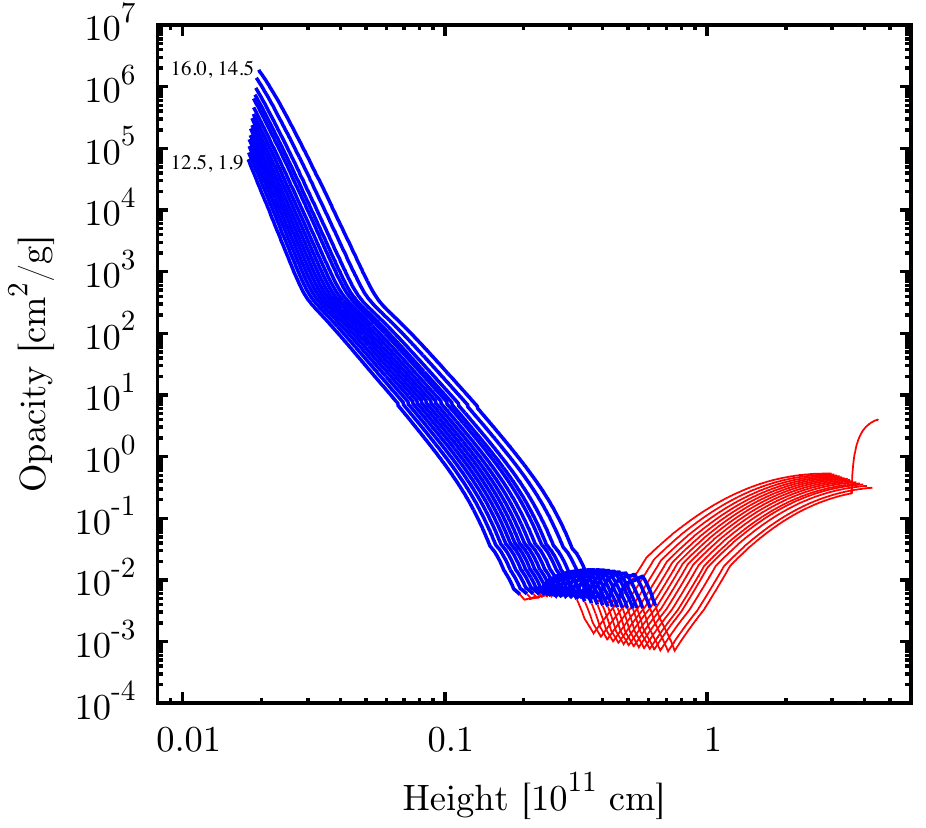}
     \end{minipage} 
                  \caption{Temporal evolution of the radial profile of $\kappa$ as a function of height (distance from the planet's center) in the $\Sigma10$ simulation. The left end of the lines corresponds to the core-envelope boundary while the right end is the planet's outer radius. Radiative layers are red, convective parts blue. The two numbers at the left give the corresponding core and envelope mass. The left panel shows the evolution during phase I. A sequence of 12 structures is show, covering a time interval of 0.27 Myr. During this time the core mass increases from 0.6 to  11.97 $\mearth$ while the envelope mass grows from $6\times10^{-6}$ to 0.9 $\mearth$. The left panels shows the evolution during phase II. A sequence of 15 structures is  shown, covering a time interval of 0.32 Myr.  During this time the core mass increases from 12.5 to  16 $\mearth$ while the envelope mass grows from 1.9 to 14.5 $\mearth$. This last structure is the same as shown in Fig. \ref{fig:crossM10}  and thus occurs just before crossover. It is the only structure where the advection regime occurs in the top layers, leading to an increased opacity.}
             \label{fig:evoM10} 
\end{figure*}

\subsubsection{The $\Sigma10$ case: temporal evolution of $\kappa(R)$}\label{sect:tempevoSigma10}
Figure \ref{fig:crossM10} shows the radial structure of $\kappa$ only at one specific moment in time. In order to understand how general this structure is, we  plot  in Figure \ref{fig:evoM10} the opacity as a function of height (covering the entire gaseous envelope) at 27 different moments in time. The left panel shows the temporal evolution during phase I, while the right panels shows phase II. A large range of core and envelope masses (and of luminosities) is covered, as indicated in the figure.

We see that nevertheless, the general structure of $\kappa$ is roughly speaking similar in all states.  It is of order 1 cm$^{2}$/g at the top, and then decreases to $\sim10^{-3}$ cm$^{2}$/g with increasing depth.  Inside of this minimum, the molecular opacities start to dominate, so that $\kappa$ now increases as we move further in. Somewhat inside of the $\kappa$ minimum, the energy transport mechanism changes from radiative diffusion to convection. There are some variations around this general picture. Very low-mass planets, for example, have an outer convective zone. It is therefore possible that the grain opacity calculated in these parts is affected by the fact that we do not take convection into account in the analytical model.  {These low-mass planets are also found to have a relatively large outer part where the opacity is of order 1 cm$^{2}$/g, followed by a rapid decrease of $\kappa$.}   

The last radial structure shown in the right panel of Fig. \ref{fig:evoM10} is the same as in Fig. \ref{fig:crossM10}, i.e., it is at crossover. As mentioned, the outer part is radiative down to $R\approx 6\times10^{10}$ cm. A second tiny inner radiative zone occurs  between 4.0 and $4.6\times10^{10}$ cm. It is clearly related to the local minimum in the molecular opacities in this region. Inside of $4.0\times10^{10}$ cm, the envelope is again convective. This sequence of convective and radiative zones is very similar to Fig. 3 of \citetalias{movshovitzbodenheimer2010} (which is, however, for a smaller envelope and core mass). At crossover, \citetalias{movshovitzbodenheimer2010} do not find the detached convective zone, and the radiative-convective boundary is at about $3.7\times10^{10}$ cm,  corresponding to our inner radiative-convective boundary at $4.0\times10^{10}$ cm. A comparison of the radiative and convective gradients in this region shows, however, that they are very similar, meaning that a small difference in $\kappa$ can easily change the classification of a layer as being radiative or convective. But the temperature gradient on the other hand will remain nearly the same, independent of the energy transport mechanism. 

 \begin{figure*}
\begin{minipage}{0.5\textwidth}
	      \centering
       \includegraphics[width=0.98\textwidth]{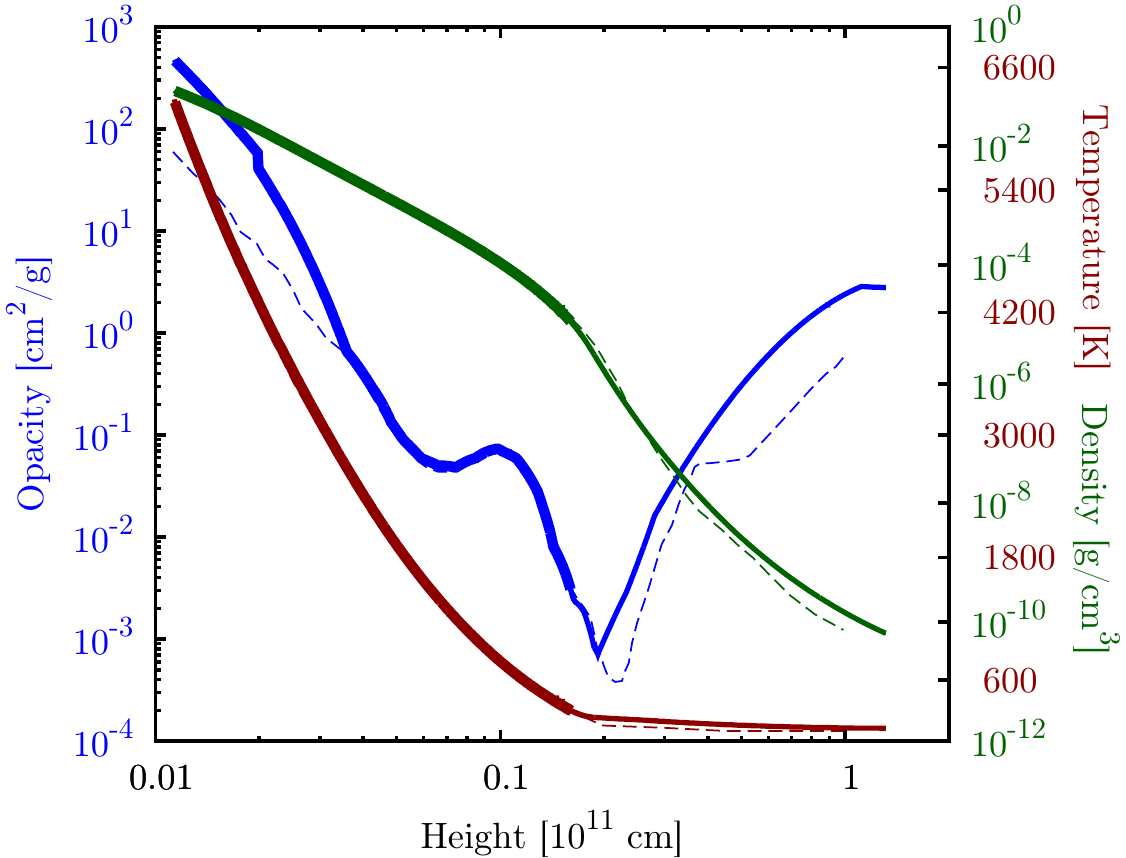}
     \end{minipage}
          \hfill
     \begin{minipage}{0.5\textwidth}
      \centering
                    \includegraphics[width=0.98\textwidth]{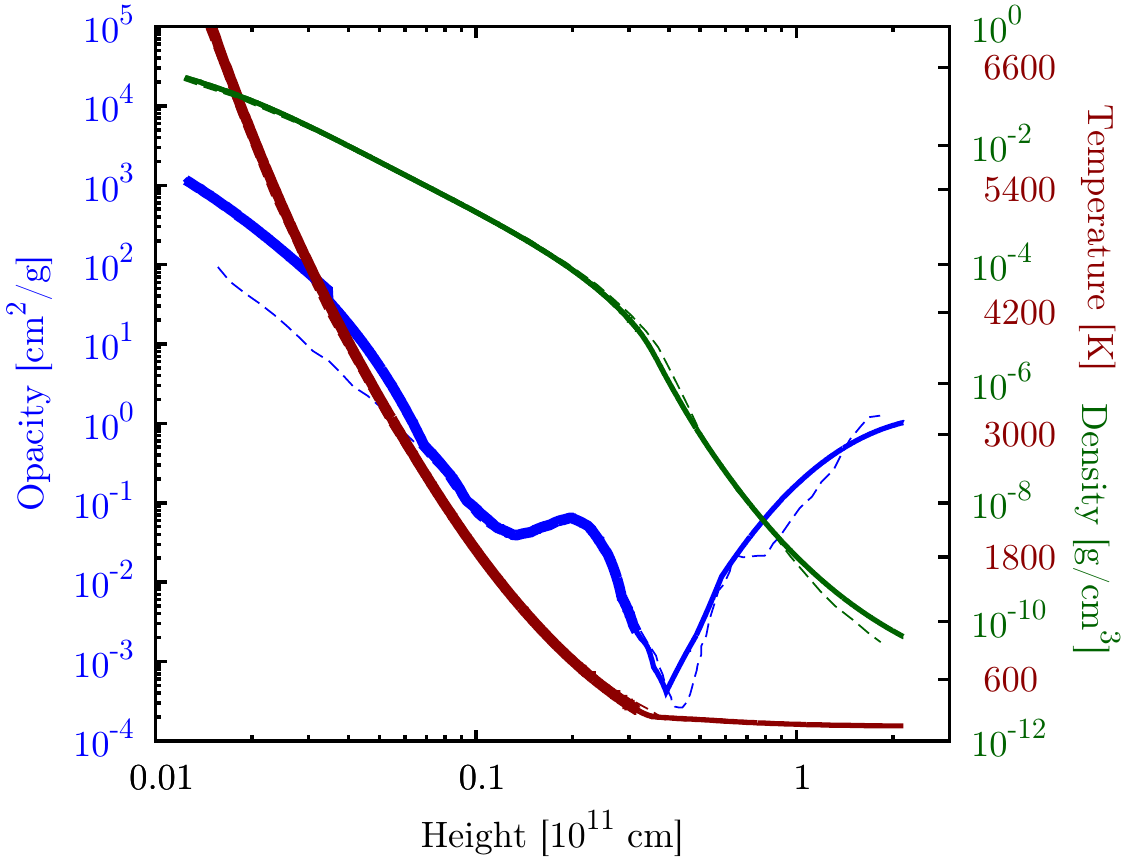}
     \end{minipage} 
                  \caption{Radial profile of the temperature, density, and opacity in the $\Sigma4$ simulation. Solid lines are from the analytical model with convective parts  indicated with by thicker lines. Dashed lines show the corresponding results of \citetalias{movshovitzbodenheimer2010}. \textit{Left panel}: when the core and envelope mass is 3.27 and 0.74 $\mearth$, respectively (3.21 and 0.74 $\mearth$ in \citetalias{movshovitzbodenheimer2010}). \textit{Right panel:} At a moment just before crossover. The core and envelope mass are  4.18 and 4.17 $\mearth$, respectively  (4.09 $\mearth$ in \citetalias{movshovitzbodenheimer2010}).}
             \label{fig:crossM4} 
\end{figure*}

In their $\Sigma10$ case (but not in the $\Sigma4$ simulation), \citetalias{movshovitzbodenheimer2010} find a convective zone in the outermost layers. In the layers where the advection regime applies, we also see a steepening of the radiative gradient due to the increased opacity. But it remains nevertheless a factor $\sim3$ shallower than the convective gradient so that we do not have such a top convective zone. On the other hand, further down where the differential settling determines the (low) opacity, the radiative gradient is shallower than the convective one by a factor 30, so that there is no doubt that these deeper layers are radiative. This means that also in the analytical model, the formation of a top convective layer could in principle occur for relatively small changes of $\kappa$ (or the EOS). 

Furthermore, the following   general properties are found: First, during phase II, the (outer) radiative zone is always roughly speaking isothermal (cf. \citealt{rafikov2006}). Second, the grain growth mechanism is always differential settling and not Brownian motion. This also holds for the $\Sigma6$ and $\Sigma4$ case. But we caution that the analytical model tends to overestimate the efficiency of differential settling in the outer layers, as mentioned in Sect. \ref{timescales}. Third, the grain size generally increases as the planet grows. During the early (later) stages of phase I, the grains have sizes of order 1 (10) $\mu$m in the outermost layers, and of 10 (100) $\mu$m in the inner atmospheric layers. During phase II, the grain size if of order 100 $\mu$m at the top, and 1000 $\mu$m (0.1 cm) at the place where the grains evaporate. 
 
\subsubsection{The $\Sigma4$ case}\label{sect:sigma4case}
This simulation is particularly interesting because first, it corresponds to a surface density that is only a factor 1.6 higher than in the MMSN of \citet{hayashi1981}. Second, the low core mass of only 3 to 4 $\mearth$  means that this case is interesting  also in the context of the formation of low-mass low-density planets which were recently discovered in large numbers (e.g., in the Kepler-11 system, \citealt{lissauerjontof2013b}; cf. also \citealt{rogersbodenheimer2011}).

Figure \ref{fig:crossM4}  shows the radial structure of the envelope at two moments in time as found by the analytical model and \citetalias{movshovitzbodenheimer2010}. The global structure of the opacity is similar as in the $\Sigma10$ case. The opacity is of order unity in the top layers and then decreases strongly to a minimum  where $\kappa$ is between a few times $10^{-4}$ and $10^{-3}$ cm$^{2}$/g. Inside of this minimum, the gas opacities dominate over the grains, and the opacity now increases rapidly towards the interior. Somewhat inside of the minimum, the energy transport changes from radiative diffusion to convection. The small inner radiative zone that was seen in the $\Sigma10$ case (and also the top convective layer in \citetalias{movshovitzbodenheimer2010}) is not found here. The envelope therefore simply consists of a deep convective part where almost all envelope mass resides ($\approx 98$\% at crossover) below a radiative atmosphere where the temperature only changes by a factor $\sim$2. This is in agreement with \citetalias{movshovitzbodenheimer2010}.

The figure in the left panel shows the envelope when the core and envelope mass is  3.27 and 0.74 $\mearth$, respectively. This is the same envelope mass as in \citetalias{movshovitzbodenheimer2010}, while the core mass is slightly lower (3.21 $\mearth$). \citetalias{movshovitzbodenheimer2010} find a temperature at the core-envelope boundary of 6230 K, while we find 6290 K which is very similar. However, the plot also shows that the opacity  in the atmosphere calculated by the analytical model is everywhere larger than in the numerical model (by a varying factor of $\sim2-5$). This is very likely the reason for the longer formation timescale found in Sect. \ref{sect:comp1064sims}. In the structure at crossover shown in the left panel, the agreement is better, but there is still a certain tendency of the analytical model to overestimate $\kappa$. For this structure, \citetalias{movshovitzbodenheimer2010} find a core/envelope interface temperature of 7680 K, while we have 7816 K corresponding to a difference of  about 2\%. The reason for the higher opacity is difficult to pinpoint without further information on the  state of the grains in the numerical model (like the grain size or dust-to-gas ratio as a function of height), but will be further investigated in upcoming work (Ormel \& Mordasini, in prep.).

\subsubsection{ {Perfect sticking vs. bouncing and fragmentation}}\label{sect:perfectsticking}
\begin{figure}	      
\centering
       \includegraphics[width=1\columnwidth]{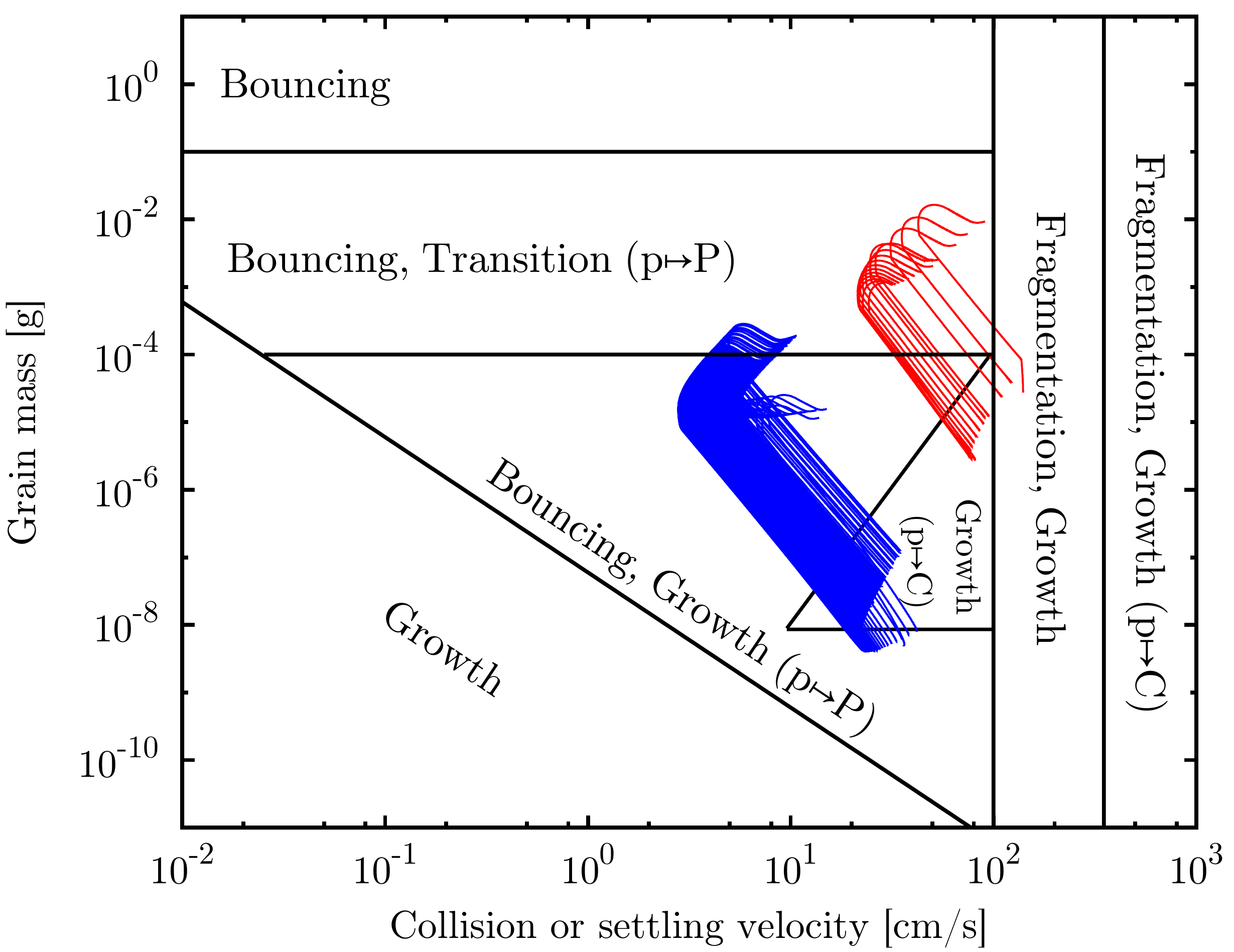}
                    \caption{ {Overlay of the settling velocity-mass relation and the outcome of laboratory experiments on dust collisions. The red and blue lines show the settling velocity versus the grain mass in the radiative part of the atmosphere during phase II for the $\Sigma$10 and $\Sigma$4 case, respectively. For the $\Sigma$10 case, these are the same 12 structures as shown in the right panel of Fig. \ref{fig:evoM10}. The black lines show in a simplified way the outcome of collision experiments \citep{guettlerblum2010} in the collision velocity-grain mass plane. The actual outcome depends also on the grain porosity and the mass ratio of the projectile and target. } }
             \label{fig:fragment} 
\end{figure}

 {An important simplification in the model is the one of perfect sticking. For grain growth in protoplanetary disks it is well known that collisions can in reality also lead to bouncing or even fragmentation (see the discussion in Sect. \ref{sect:mp08vsetdensqbounce}). It is therefore important to understand to what extent the assumption of perfect sticking is justified in protoplanetary atmospheres. Figure \ref{fig:fragment} shows an overlay of the relation of settling velocity versus grain mass predicted by the analytical model and the outcome of laboratory experiments on dust collisions  (a simplified version of Fig. 11 in \citealt{guettlerblum2010}). Most of these experiments were conducted with dust aggregates made of spherical monodisperse SiO$_{2}$ monomers of  about 1 $\mu$m in size.  The phase II of the  $\Sigma$10 and $\Sigma$4  simulation are shown. The actual collision velocity will be a fraction of the settling velocity. This overlay allows to study the fundamental outcome of the collisions of the grains (growth, bouncing, fragmentation). }

 {The collision outcome depends, however, not only on the grain mass and collision velocity. The mass ratio of the projectile and  target as well as the porosity of the two collision partners is equally important. Due to that, there are in principle eight different panels necessary to show the collisional outcomes for one fixed material type (see \citealt{guettlerblum2010} for details). Figure  \ref{fig:fragment}  therefore only gives a rough overview of the general behavior. A general result is first that low velocity collisions of small bodies lead to growth (triangle in the lower left corner labelled ``Growth''). This regime is important for grain growth in phase I (not shown in the figure). For  grain growth to fall into this regime also in phase II, the typical collisional velocities need to of order 1-10\% of the settling velocity for the $\Sigma$4 case, and $\sim$0.1 \% for the $\Sigma$10 case.  Second, if the collisional velocity is increased, bouncing becomes important (inverted triangle in the middle labelled ``Bouncing, Growth (p$\mapsto$P)''. However, growth can still occur if the collisions occur mainly between bodes of a different size that are both porous (represented as ``p$\mapsto$P'') and/or if small porous objects hit large compact ones (represented as ``p$\mapsto$C'').  If the collision velocities are of the same order as the settling velocities, this regime is dominant for the $\Sigma4$ case. The larger grains in the $\Sigma$10 simulation partially also fall in this regime, but also in the one labelled ``Bouncing, Transition (p$\mapsto$P)''. In this regime, the collisional outcome transitions from growth to bouncing also in the p$\mapsto$P case. This is an indication that bouncing is more important in more massive protoplanets. Third, if the impact velocity is further increased (beyond $\sim$100 cm/s in most regimes, $\sim$350 cm/s in the p$\mapsto$P case), fragmentation occurs, except for the p$\mapsto$C regime. In the $\Sigma$10 case this regime is marginally reached at crossover in the outer layers as shown by the rightmost structure which is the one at crossover. After crossover, the settling velocities continue to increase, reaching $\sim$600 cm/s at the onset of the disk limited gas accretion. For the $\Sigma4$ case such high velocities are only reached very shortly before the disk limited gas accretion. Together with grain advection, this could lead to an increased opacity in phase III, lengthening the duration of this phase.  }

 {In summary we conclude that the collision velocities in protoplanetary atmospheres are likely too low for grain fragmentation to be important, at least before crossover. The suppression of further grain growth due to bouncing could in contrast be  important especially for more massive cores. The actual outcome will depend on the typical mass ratio of the colliding grains, their porosity, and composition. Further studies describing these quantities \citep{podolak2003,ormelspaans2007,zsomormel2010} are thus needed to clarify to what extent the assumption of perfect sticking is applicable. } 

\section{Summary and conclusions}\label{sect:summaryconclusions}
In this work, we have derived a first simple but complete analytical model for the opacity due to grains suspended in the atmosphere (outer radiative zone) of forming protoplanets. The grain opacity $\kappa_{\rm gr}$ is a poorly known but central quantity for planet formation as it controls the protoplanet's gas accretion rate \citep[e.g.,][]{ikomanakazawa2000}. This determines the final bulk composition of  low-mass planets (their H/He mass fraction) as well as the time until runaway gas accretion occurs and a giant gaseous planet forms. Therefore, the magnitude of the grain opacity leads to potentially observable imprints in the planetary mass-radius relationship \citepalias{mordasiniklahr2014}.

Our general approach is similar to the one of \citet{rossow1978} and consists in the comparison of the  timescales of the microphysical processes that govern the grain dynamics. We consider  the following  processes (Sect. \ref{sect:relevanttimescales}): dust settling in the Epstein and Stokes drag regime, growth by Brownian motion coagulation and differential settling, grain evaporation, and the advection of grains due to the contraction of the gaseous envelope. With these timescales it is possible to determine the typical size of the grains in each atmospheric layer as well as the dust concentration. This allows to calculate the opacity. We derive analytical expressions for the grain opacity separately in five different regimes (Sect. \ref{sect:calctypsizeopa}).

We find that the dominating growth regime is differential settling. In this regime, the grain opacity takes a simple form and is given as $27 Q / 8 H\rho$ in the Epstein drag regime and $2 Q/ H\rho$ in the Stokes regime where $H$ is the atmosphere scale height and $\rho$ the gas density (Sects. \ref{sect:diffsetepstein} and \ref{sect:diffsetstokes}). Typically, the grains grow much larger than the relevant wavelengths of radiation, therefore the extinction coefficient $Q$ is simply equal to $\approx$2 under most circumstances (Sect. \ref{sect:extinctioncoeff}). These two equations are the most important direct result of this study.  These expression are in particular independent of the protoplanet's grain accretion rate $\dot{M}_{\rm gr}$.  

This first analytical model is simplified in many aspects (Sect. \ref{sect:limitations}):  important simplifications are a radially constant mass flux of grains through the atmosphere, the replacement of the grain size distribution by only one typical grain size per layer, the assumption of perfect sticking, and the approximation of the Rosseland mean opacity by the opacity evaluated at the wavelength associated with the maximum of the Planck function.
 
Despite this, we find  {in general} fair, and sometimes even surprisingly good  agreement with the numerical results of \citet{movshovitzpodolak2008} and \citet{movshovitzbodenheimer2010}, also on a quantitative level. Building on the model of \citet{podolak2003}, these works determine the grain opacity with a complex but computationally expensive numerical model. Numerous comparisons of the analytical and numerical results can be found in Section \ref{sect:compMPforgivenrhoT}.  {They also allow to understand the limitations of the analytical model, like an overestimation of the efficiency of growth in the top layers at least when advection is not important (Sects. \ref{timescales}, \ref{sect:mp2008amonomer}), or the absence of a high opacity zone around a layer of high planetesimal ablation (\ref{sect:effectfpg}).} 

 {But in general}, we confirm their key finding that grain growth is an efficient process in protoplanetary atmospheres. Starting with $\mu$m sizes in the outermost layers,  the grains can grow up to  $\sim$0.1 cm in the deeper layers (Sect. \ref{sect:mp2008amonomer}). Grain growth therefore leads to a characteristic radial opacity structure (Sects. \ref{sect:mp08structopa}, \ref{sect:Sigma10structcrossover}): in the outermost layers, $\kappa_{\rm gr}$ takes high ISM-like values  ($\sim$1 cm$^{2}$/g), but,  as we move deeper into the atmosphere,  it decreases strongly to very low values ($\sim$$10^{-3}$ cm$^{2}$/g). At this point, the grain opacity becomes so low that the grain-free molecular opacity of the gas becomes dominant.  This general structure is found for many different core and envelope masses (Sect. \ref{sect:tempevoSigma10}). This radial structure differs from the one obtained from scaling the ISM opacity by one uniform (and a priori arbitrary) reduction factor (Sect. \ref{sect:compism}). This was, nevertheless, the approach taken in many previous works (e.g., \citealt{mizunonakazawa1978,stevenson1982,pollackhubickyj1996,papaloizounelson2005,hubickyjbodenheimer2005,tanigawaohtsuki2010,levisonthommes2010,horiikoma2011,ayliffebate2012,dangelobodenheimer2013} and \citetalias{mordasiniklahr2014}).

A consequence of the grain growth is a total optical depth of the atmosphere that is much smaller than for a full ISM opacity, allowing rapid gas accretion. For example, in the \citet{movshovitzpodolak2008} comparison case ($M_{\rm core}$=12.6$\mearth$, $M_{\rm env}$=2.7$\mearth$), the optical depth is found to be about 90 and 50 in the numerical and analytical model, respectively (Sect. \ref{sect:compism}). For a full ISM grain opacity, it is in contrast about 4$\times$10$^{4}$. For comparison, in a grain-free atmosphere, an optical depth of about 25 is found. This indicates that grain-free opacities provide an optical depth that is closer to the grain calculations than the full ISM value. In retrospect, this suggests that classical studies like \cite{pollackhubickyj1996} should have considered the grain-free case  {as equally  meaningful as} the full ISM opacity case. Potentially, this would have meant that the classical timescale problem of the core accretion theory would not have been seen as an important drawback of this paradigm for several decennia.  

Analogous to \citet{movshovitzbodenheimer2010}, we have then coupled the analytical grain opacity model to our giant planet formation code. We have studied the in-situ formation of Jupiter for different planetesimal surface densities (Sect. \ref{sect:comp1064sims}) and compared the predicted formation timescales with the results of \citet{movshovitzbodenheimer2010}. For core masses of 16, 8, and 4 Earth masses, the analytical (numerical) model predicts a crossover time of 0.81 (0.97), 2.35 (1.63), and 7.09 (3.62) Myr. For comparison, at a full ISM grain opacity, the formation times would be about 6, 60, and $\gg$100 Myr, respectively. This means that the formation timescale predicted by the analytical model agrees very well with the numerical model for larger core masses. It is overestimated for low-mass cores, but, compared to the previous situation where not even the order of magnitude of $\kappa_{\rm gr}$ was analytically known, this is still an improvement.

We  have studied the impact of different dust-to-gas ratios in the accreted gas on the grain opacity (Sects. \ref{sect:effectfpg}, \ref{sect:mbpl2010fdgamono}). As indicated by the aforementioned equations for $\kappa_{\rm gr}$, we only found a very weak (or no) dependency, as already observed by \citet{movshovitzpodolak2008}. Our  analytical results yield the physical explanation which is that a higher dust input leads to a higher dust-to-gas mass ratio (which increases the opacity), but also larger grains (which decreases the opacity).  In contrast to Brownian coagulation, these effects cancel each other out in the dominating growth regime of differential settling. The dependency of $\kappa_{\rm gr}$ on different dust-to-gas ratios is therefore the litmus test to understand which growth process is dominant. 

This independence of $\kappa_{\rm gr}$ on the grain accretion rate is the result of this study with the most important wider implications for planet formation (Sect. \ref{sect:implicationsfehpebble}): First, it means that a higher stellar [Fe/H] (which presumably leads to more dust in the protoplanetary disk) should not strongly increase the grain opacity in protoplanetary atmospheres. This is an important insight for the core accretion theory. It means that a higher stellar [Fe/H] only favors giant planet formation as it presumably also leads to a higher surface density of planetesimals without being detrimental to it due to an increased opacity. This corroborates the result that the core accretion paradigm can explain the observed increase of the giant planet frequency with stellar [Fe/H].

Second, it is important for the recently suggested  formation of giant planet cores by the accretion of pebbles \citep{ormelklahr2010,lambrechtsjohansen2012,morbidellinesvorny2012}. Such small objects make high solid accretion rates possible. However, in contrast to classical 100 km planetesimals that deposit most of their mass in the convective zone or even directly on the core, such small objects strongly increase the grain input in the outer layers of the atmosphere due to ablation (Sects. \ref{sect:grainaccrrateplanetesimal},  {\ref{sect:impactgeometry}, \ref{sect:impactcompo}}). This could potentially increase the opacity, making giant planet formation harder.  Our study shows that this is --at least to first order-- not the case. An accretion of the core via small bodies should not strongly increase the opacity. Instead, the low random velocity, the large capture radius, the resulting high mean molecular weight of the envelope, and the low opacity should conspire to facilitate giant planet formation with small bodies.  {Note, however, that  higher grain opacities are expected if the protoplanet accretes bodies with a wide range of sizes: this leads to an input of small grains by ablation in all layers, which increases the opacity \citep{movshovitzpodolak2008,dangeloweidenschilling2014}}. 

The analytical grain opacity can be incorporated into numerical models calculating the gas accretion at a computational cost that is only insignificantly higher than tabulated opacities.   {Also the contemporaneous work of \citet{ormel2014} allows to estimate the grain opacity in an efficient fashion by adding an fifth equation to the normal planetary structure equations. This equation is numerically integrated in concert with the other four structure equations. In contrast to the analytical model here, it allows to take into account a radially varying grain input, and the effects of a bimodal grain size distribution. The basic conclusions obtained through this approach agree well with those found in the analytical model.}

Therefore, in upcoming work, we will repeat our population synthesis calculations presented in \citetalias{mordasiniklahr2014} but, instead of scaled ISM opacities, we will use  {physically motivated values of  $\kappa_{\rm gr}$}. This will make it possible to get a  better understanding of the role of grain opacity in controlling the planetary bulk composition, and the associated consequences for the planetary mass-radius relationship. This is very important in a time where it becomes possible to study for the first time the bulk composition of many exoplanets \citep[e.g.,][]{lopezfortney2013b}, and, in particular the transition from solid to gas-dominated planets \citep[e.g.,][]{marcyisaacson2014}.
  
\acknowledgements{I thank Th. Henning, G.-D. Marleau, P. Molli\`ere, K.-M. Dittkrist, H. Klahr,  Y. Alibert, W. Benz and C. W. Ormel for enlightening discussions and valuable input. I  thank the Max-Planck-Gesellschaft for the Reimar-L\"ust Fellowship.  {I also thank the referee Dr. Morris Podolak for an interesting and constructive report that helped to improve the manuscript.} }

\appendix

\section{Planetesimal mass deposition: impact geometry}\label{sect:impactgeometry}
The radial mass deposition profiles due to impacting planetesimals shown in Figure \ref{fig:massdepo} were calculated for an exact head-on impact geometry. However, it is clear that off-center impacts are much more likely than the (near) head-on case due to the larger collisional cross section. In this appendix we investigate the consequences of non-central impact geometries on the mass deposition profile. 

 \begin{figure*}
\begin{minipage}{0.6\textwidth}
	      \centering
       \includegraphics[width=0.98\textwidth]{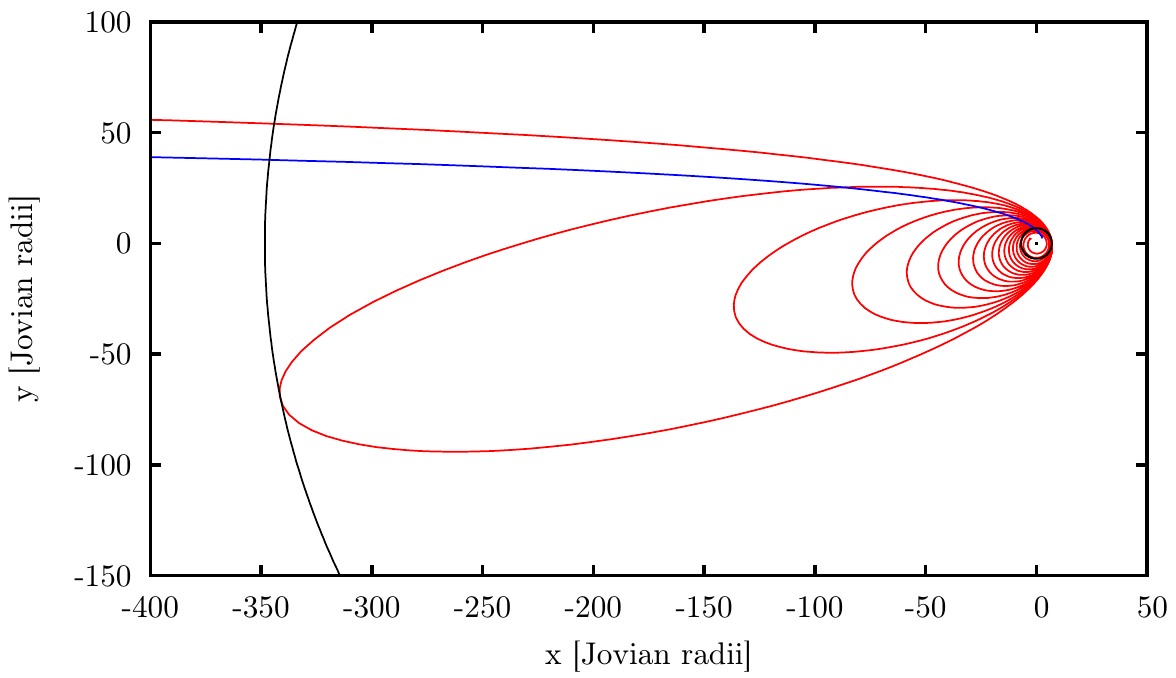}
     \end{minipage}
          \hfill
     \begin{minipage}{0.39\textwidth}
      \centering
                    \includegraphics[width=0.98\textwidth]{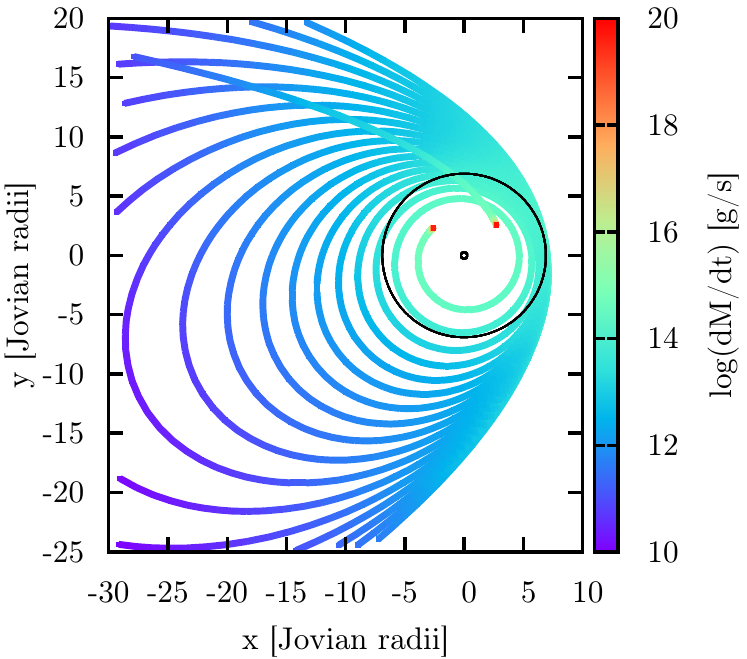}
     \end{minipage} 
                  \caption{\textit{Left panel:} trajectories of 100 km icy planetesimals in the envelope of the protoplanet in the $\Sigma$10 case at crossover for two different impact parameters. The critical case (red) and the mean case (blue) are shown. The three concentric black circles are the Hill sphere, capture, and core radius (from large to small). \textit{Right panel:} zoom-in onto the central regions of the two trajectories. The color code gives the instantaneous mass loss rate along the trajectory.  Almost all of the mass is deposited in the terminal explosion (red parts of the trajectory). The black circles are again the capture and core radius.}
             \label{fig:xy} 
\end{figure*}

The dissipative effect of gas drag increases the collisional cross section of a protoplanet that has a gaseous envelope beyond the cross section that is due to gravitational focussing only \citep{podolakpollack1988,inabaikoma2003}. The largest  impact parameter that leads to accretion is defined as the ``critical'' impact parameter $p_{\rm crit}$. It results in a planetesimal trajectory with an apocenter after the first encounter that is equal to the protoplanet's Hill sphere radius \citep{pollackhubickyj1996}. Such an orbit must inevitably lead to the accretion of the planetesimal (at least in the two-body approximation of protoplanet and planetesimal that is used here) while planetesimals with a larger impact parameter leave again the planet's Hill sphere and return under the dominant gravitational control of the star.  Assuming that planetesimals enter the protoplanet's Hill sphere isotropically then leads to a mean impact parameter of $p_{\rm mean}=p_{\rm crit}$/$\sqrt{2}$.  

Figure  \ref{fig:xy}, left panel, shows the trajectories of 100 km planetesimals in the envelope of the protoplanet in the $\Sigma$10 simulation at  the crossover point when the core and envelope mass is approximately 16 $\mearth$. The core radius (about 0.27 $\rj$), the capture radius (the pericenter of the critical orbit on the first encounter, about 6.9 $\rj$), and the Hill sphere radius (about 348 $\rj$) are also shown.

\begin{figure}
\begin{center}
\includegraphics[width=\columnwidth]{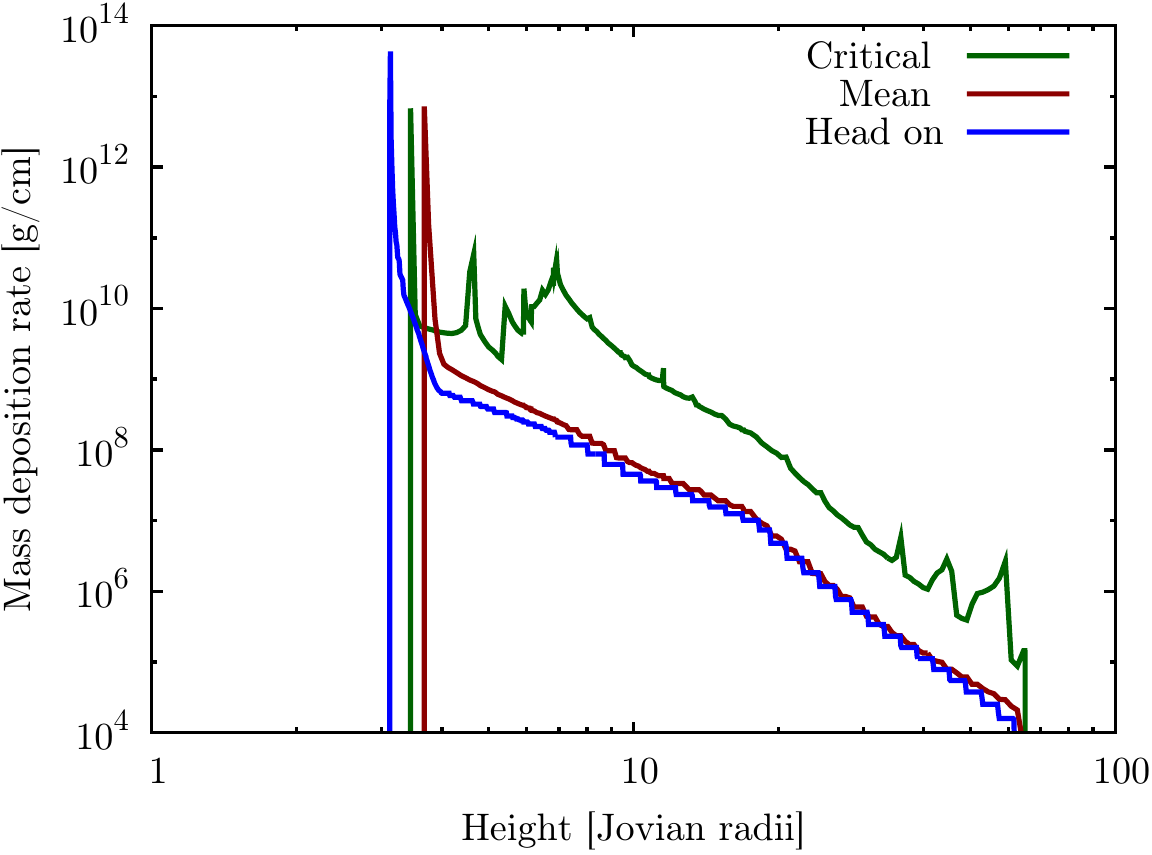}
\caption{Consequence of the impact geometry on the radial mass deposition profile of 100 km icy planetesimals. The mass deposition per length as a function of height above the planet's center is shown for a head-on collision (same line as in Fig. \ref{fig:massdepo}) and for the critical and mean impact parameter (corresponding to the trajectories of Fig. \ref{fig:xy}).}\label{fig:massdeprate100kmGeo}
\end{center}
\end{figure}

In the critical case, the planetesimal is destroyed after about 17 orbits around the protoplanet. In the mean case, the planetesimal is destroyed already on the first encounter. This indicates that the annulus of impact parameters that lead to spiraling trajectories is only relatively narrow.  The right panel of the figure shows a zoom-in onto the central regions of the two trajectories. The color of the lines shows the instantaneous mass loss rate. 

One sees that the mass loss rate is of order $10^{16}$ g/s during the first few pericenter passages. However, as soon as the aerodynamical disruption starts (at the very end of the trajectories), the mass loss rate increases very rapidly by approximately four orders of magnitudes. This is due to the rapid increase of the planet's cross section as the planetesimal is deformed by the aerodynamic load into a flat ``pancake'', leading to a terminal explosion \citep{zahnle1992,chybathomas1993}. This final destruction phase where more than 90\% of the mass is deposited has a duration of only $\sim$100 seconds. It is a violent process where a energy deposition rate of $\sim10^{31}$ erg/s is reached.  

Figure \ref{fig:massdeprate100kmGeo} shows the corresponding radial mass deposition profiles. The curve for the head-on case is the same as shown in Fig. \ref{fig:massdepo}. All three curves share the property that the mass loss rate is relatively small in the outer layers where only thermal ablation occurs, followed by sharp upturn when the aerodynamic disruption starts. The details differ however, and can be understood from the shape of the trajectories. On first notes that the aerodynamical disruption sets in the head-on and mean case at approximately the same altitude (about 4 $\rj$). This is due to the fact that both planetesimals approach the planet's core at a velocity $v$ approximately equal to the escape velocity. This means that they are exposed to approximately the same stagnation pressure ($\approx$ $1/2\rho v^{2}$ where $\rho$ is the atmospheric gas density). However, the head-on case penetrates about 0.5 $\rj$ deeper, since its velocity vector only has a radial component towards the core, while the mean case has a substantial tangential (along orbit) component. The critical case differs in the sense that its velocity is rather given by the Keplerian velocity around the planet than the escape velocity, i.e., it is about a factor $\sqrt{2}$ smaller. This means that the dynamic pressure becomes equal to the planetesimal's tensile strength only in a deeper layer.  Due to this, the terminal explosion occurs at a somewhat lower altitude (about 3.5 $\rj$). But since the radial component of the velocity vector is only very low, the radial range over which the mass is deposited is very thin, as visible in Fig. \ref{fig:massdeprate100kmGeo}. In this mass deposition profile, one also clear sees the consequences of the orbits around the planet before the final destruction. It leads to an increase of the mass deposition by about one, and, at special radii, two orders of magnitude. There is for example a local maximum at about 7 $\rj$. Clearly, this corresponds to the  region around the capture radius, where the planetesimal makes many orbital passes. 

It is interesting to compare the  cumulative mass deposition profile with the altitude $R_{\rm evap}$ where the ambient temperature of the atmosphere is sufficiently high to evaporate silicate grains. This altitude occurs at about 5 $\rj$ in the ``deep deposition'' case, see the ``evap'' line in Fig. \ref{fig:massdepo}. We neglect here that this distance is of course itself dependent on the opacity and thus mass deposition profile. Planetesimal material that is deposited above this height can contribute to grain formation.  We find that about 0.2\% and 8\% of the planetesimal's initial mass is deposited above this distance in the mean and critical impact geometry, respectively. At crossover, the gas accretion rate is about a factor 5 higher than the solid accretion rate \citep{pollackhubickyj1996}. Assuming an $f_{\rm D/G, disk}=0.01$, this means that the ratio of the effective grain accretion rate due to planetesimals relative to the grain accretion rate together with the gas is about  0.04 and 1.6, again for the mean and critical geometry. We conclude that in the typical impact geometry, grain accretion together with the gas is clearly dominant so that  the ``deep deposition'' assumption is indeed justified, while for the critical geometry both sources have approximately the same importance in the lower atmospheric layers. It is clear that we have here only analyzed an individual case.  A systematic  study is certainly warranted for future work. 

\section{Planetesimal mass deposition: planetesimal composition}\label{sect:impactcompo}

In this section we investigate the consequences of the planetesimal composition (ices versus silicates) for the radial mass deposition profile. One expects  that  icy planetesimals deposit much more mass in the upper layers compared to rocky impactors \citep[][]{podolakpollack1988}. In this section we confirm this general result, but we also find that the extent and cause of the difference can vary substantially depending on the impactor size.

\begin{figure*}
\begin{minipage}{0.33\textwidth}
	      \centering
       \includegraphics[width=1\textwidth]{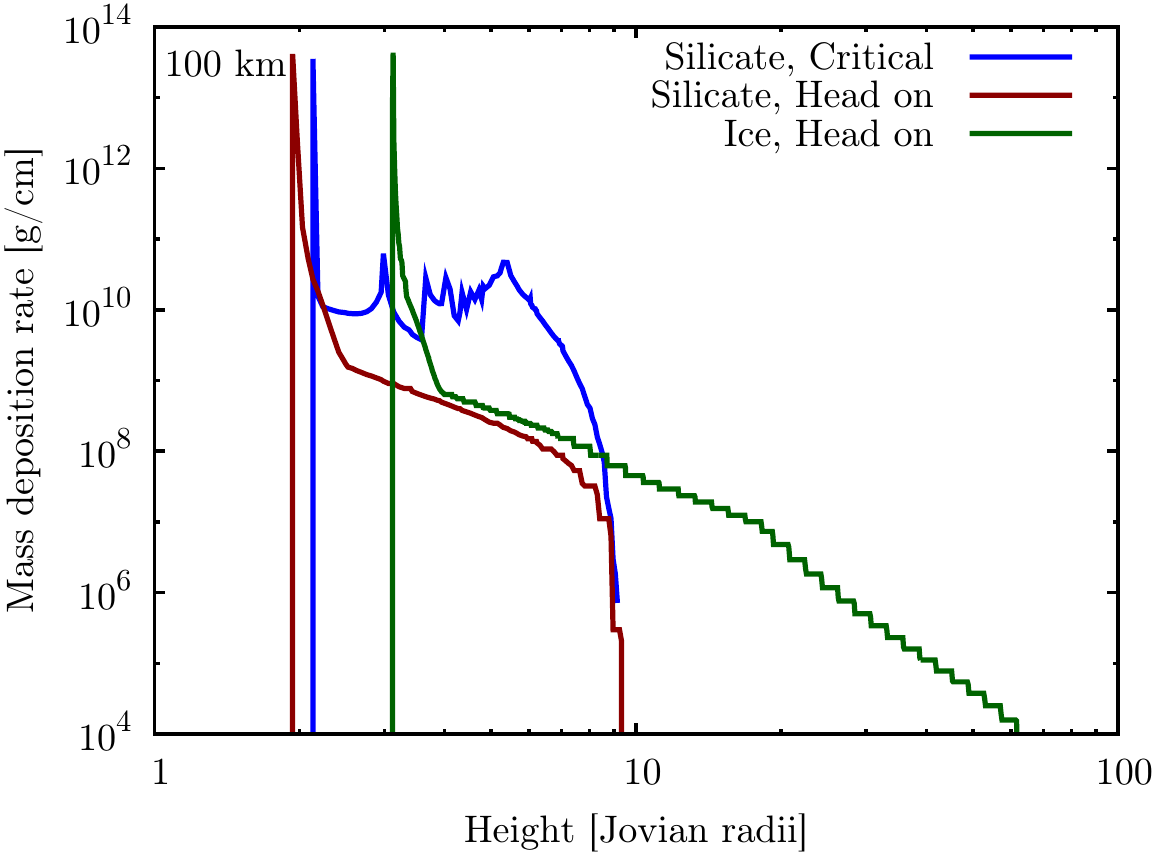}
     \end{minipage}
          \hfill
     \begin{minipage}{0.33\textwidth}
      \centering
                    \includegraphics[width=1\textwidth]{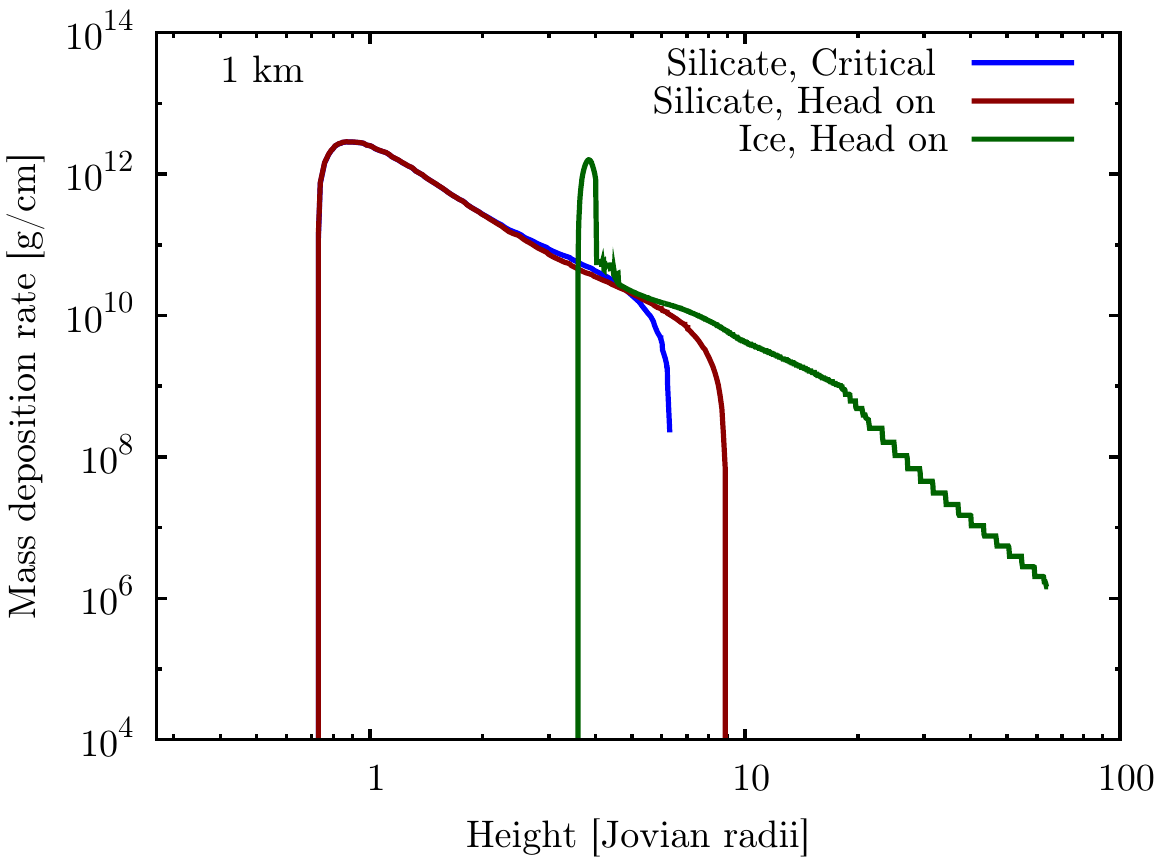}
     \end{minipage}
      \hfill
          \begin{minipage}{0.33\textwidth}
      \centering
                    \includegraphics[width=1\textwidth]{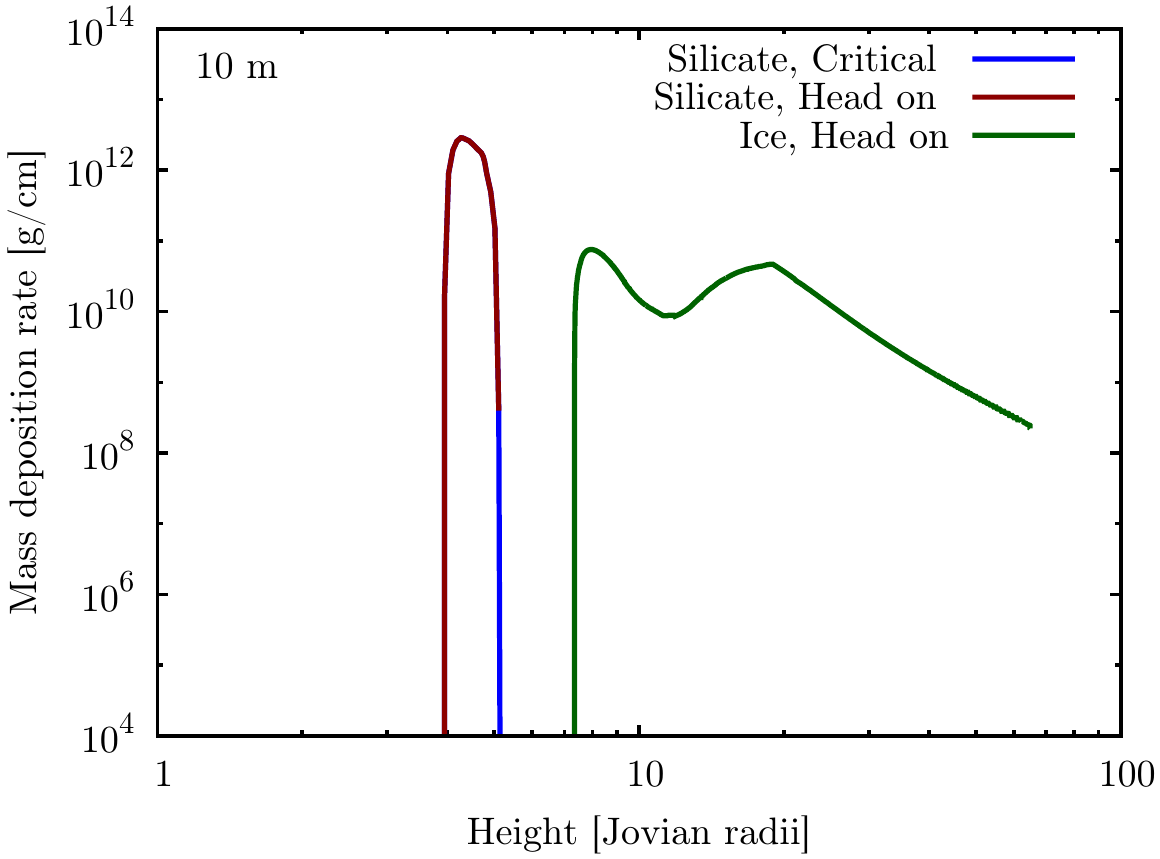}
     \end{minipage}
                  \caption{Radial mass deposition profiles of  100 km (left), 1 km (center) and 10 m planetesimals (right panel). The 1 km and 10 m curves are scaled so that the total mass deposition is the same as in the 100 km case. The planetesimal material and impact geometry is indicated in the plot. The head-on impact of  icy  planetesimals is the same as shown in Fig. \ref{fig:massdepo}.  The curves for 1 km and 10 m rocky planetesimals lay partially on top of each other for the two impact geometries .   }
             \label{fig:massdeposilicate} 
\end{figure*}

As in the previous section, we study the radial mass deposition profile in the $\Sigma$10 case at the crossover point. Note that the atmospheric structures at this moment are not identical for different planetesimal sizes and compositions  as the planet's accretion history  depends on these quantities \citep[e.g.,][]{pollackhubickyj1996}.  Figure \ref{fig:massdeposilicate} shows the radial profiles for planetesimals made of silicates for a head-on and critical impact geometry and for initial planetesimal sizes of 100 km, 1 km and 10 m. The curves for the latter two were scaled such that the total mass deposited is the same as in the 100 km case.  The profiles of head-on icy impactors, the same as in Fig. \ref{fig:massdepo}, are also shown for comparison. 

Three material properties are mainly responsible for the differences: the density (set to 1 and 3.2 g/cm$^{2}$ for ice and silicate, respectively), the tensile strength (4$\times$10$^{6}$  and 3.5$\times$10$^{8}$ dyn/cm$^{2}$, \citealt{svetsovnemtchinov1995}), and the surface temperature where thermal ablation becomes important.  This temperature depends on the vapor pressure law and is between 1300-2000 K for silicates \citep{opik1958} and 180-220 K for ices \citep{iselikuppers2002}. 
 
The left panel shows the mass deposition of 100 km impactors. For both compositions, the impact ends in a terminal explosion. We see that  it occurs for rocky planetesimals at about 2 $\rj$ instead of about 3 $\rj$ as for the icy impactors. This is a direct consequence of the different tensile strengths. The difference between the head-on and critical geometry is quantitatively equivalent as found in Fig. \ref{fig:massdeprate100kmGeo} for icy planetesimals. One also notes that in contrast to icy bodies, no mass at all is lost outside of about 10 $\rj$. This is a direct consequence of the higher temperature necessary for thermal ablation. For such large bodies, the main energy input driving ablation is shock wave radiation since before the terminal explosion, such big bodes always travel at a hypersonic velocity.  The terminal explosion happens much below the altitude where the background atmosphere is hot enough to vaporize silicate grain ($R_{\rm evap}\approx 5 \rj$), and pure thermal ablation is inefficient for such large bodies. This is due to the low heat transfer coefficient and the low surface-to-mass ratio. Therefore, only about 0.02\% and  2\% of the initial mass of the planetesimal could potentially contribute to grain formation. Grain accretion together with the gas is therefore clearly the dominant source in this scenario and the ``deep deposition'' scenario applies. 

The case of 1 km planetesimals (middle panel) is different. Here, the change of the material properties (namely the higher tensile strength) leads to a fundamentally different character of the impact. For icy composition, the 1 km planetesimal gets aerodynamically disrupted as it is the case for the 100 km impactors, so that it does not penetrate deeper than 3 to 4 $ \rj$. The rocky impactor initially also travels at a hypersonic velocity. Therefore the thermal mass loss starts at same height as in the 100 km case ($\approx 10 \rj$). But due to the lower surface-to-mass ratio, the 1 km planetesimal gets slowed down stronger at higher altitudes. This means that it never flies at a very high velocity in the dense lower layers. Therefore, the dynamic pressure never overcomes the tensile strength of the silicate impactor (but it does so for ice). The silicate impactor thus does not undergo aerodynamic disruption. Instead, it is slowed down to terminal velocity where the drag and gravitational force are in equilibrium. It then sinks down at subsonic velocity, undergoing  thermal ablation only due to the ambient radiation. Pure thermal ablation is still relatively inefficient for a 1 km sized body. Therefore, this impactor penetrates surprisingly deep into the envelope, namely to about 0.7 $\rj$, much deeper than larger (and smaller) bodies. Such 1 km silicate planetesimals thus belong to the class of intermediate sized objects that are too small to undergo violent aerodynamical fragmentation, but too large to be efficiently thermally ablated. They therefore have a surprisingly large penetration depth and form a  ``core-hit finger'' \citep[see][]{mordasinialibert2006} . 

The critical impact  makes about 9 orbits around the protoplanet before destruction. This is however not visible in the mass deposition profile as the temperature is too low in this part of the orbit. Ablation only sets in once the planetesimal is slowed down to terminal velocity and then radially sinks towards the core. Therefore, the mass deposition profile is nearly identical as in the head-on case. 

The 10 m impactor is shown in the right panel. One sees that the mass deposition profile of silicate impactors significantly differs from the one of icy bodies. Mass loss only occurs in a small annulus between 5 and 4 $\rj$. Such small bodies are slowed down to terminal velocity relatively high  in the atmosphere (both for a head-on and critical impact geometry) and then radially sink down undergoing thermal ablation due to ambient radiation. This is the same fundamental type of impact without aerodynamic disruption as for 1 km silicate planetesimals. The difference is that the 1 km body is initially heated by shock wave radiation and thus also looses some mass further up in the atmosphere (but sill only inside of 10 $\rj$). The 10 m object in contrast starts to loose mass only at about 5 $\rj$ where the undisturbed atmosphere become hot enough. It is not a coincidence that this is the same altitude as where silicate dust grains start to evaporate ($R_{\rm evap}$).  This means that this small silicate body does not contribute at all material that can form grains (at least in our approximation that no mechanical mass loss occurs at aerodynamic loads smaller than the tensile strength). This is an interesting result. It indicates that for cores inside of the iceline that accrete pebbles, accretion of grains together with the gas is the only source of material for grain opacity. Thus, again the ``deep deposition'' approximation would apply here.  
   
The large diversity of planetesimal impact scenarios in terms of destruction mechanisms, penetration depths, or the importance of the impact geometry and the implications for the mass deposition profile will be studied in a systematic way in future work.

\section{Determination of the grain size including the relative velocity gas-grains}\label{sect:numsolve}
In the analytical model, we consider two limiting cases, namely that the gas envelope is completely static (settling regime), or that the grains move at the full velocity of the gas (advection regime). This has the advantage that the grain size and thus opacity can be determined completely analytically. In this paragraph we demonstrate how the relative velocity grains-gas can be included directly in the determination of the grain size. The disadvantage is that the grain size can then only be determined by a numerical root search in each layer. On the other hand, it is then also possible to smoothly combine the two drag regimes into one single equation.  

The comparison of the drag force in the Epstein and Stokes regime shows that the two can be combined into one expression that reduces to the two separate expressions in the limits of very high or low Knudsen numbers (cf. \citealt{nayakshin2010}). Writing this drag law in terms of the relative velocity $v-v_{\rm gas}$ and equating with the gravitational force then yields terminal grain velocity $v_{\rm trav}$. It is given as
\beq\label{eq:vtrav}
v_{\rm trav}=\frac{3 v_{\rm gas} v_{\rm th} \ell \rho + a g \rho_{\rm gr} (2 a + 3 \ell)}{3  v_{\rm th} \ell \rho}.
\eeq
From this we get the timescale on which grains traverse one scale height, $\tau_{\rm trav}=H/v_{\rm trav}$. This expression is valid both in the Epstein and Stokes regime, and for settling and advection that had to be treated separately above. One thus has
\beq\label{tautrav}
 \tau_{\rm trav}=\frac{3 H v_{\rm th} \ell \rho}{3 v_{\rm gas} v_{\rm th} \ell \rho + a g \rho_{\rm gr} (2 a + 3 \ell)}.
 \eeq
For $v_{\rm gas}=0$, this reduces to Eq. \ref{eq:vsetEtausetE} and  \ref{eq:vsettausetstokes} in the appropriate limits of the Knudsen number while for a sufficiently high gas velocity it approaches $\tau_{\rm adv}=H/v_{\rm gas}$.  
 
Combining the expression for the growth timescale in the Brownian motion regime (Eq. \ref{eq:taucoagbrown}) with the expression for  $v_{\rm trav}$ and assuming again a radially constant grain flux yields the growth timescale 
\beq\label{taucoagcomb}
\tau_{\rm coag,comb}=\frac{2 \pi^{2} a^{5/2} R^{2} \rho_{gr}^{3/2}\left[3 v_{\rm gas} v_{\rm th} \ell \rho+ (2 a + 3 \ell) a g \rho_{\rm gr}\right]}{9 \dot{M}_{\rm gr} v_{th} \ell \rho \sqrt{3 k_{\rm B T}}}
\eeq
For $v_{\rm gas}=0$, this equation reduces to $\tau_{\rm coag,S}$ (Eq. \ref{eq:taucoagE}) for small grains and $\tau_{\rm coag,E}$ (Eq. \ref{eq:taucoagS}) for large ones, while for high gas velocities, it approaches Eq. \ref{eq:taucoagA}. 

For growth by differential settling, we can combine the growth timescale in the Epstein and Stokes regime into the equation
\beq
\tau_{\rm coal,comb}=\frac{3 \rho v_{\rm th}}{\pi n_{\rm gr} \rho_{\rm gr} g a^{3}}\left(\frac{9 \ell}{9 a + 8 \ell}\right)
\eeq
with Eqs. \ref{eq:fpggeneral}, \ref{eq:ngr}, and \ref{eq:vtrav} this becomes
\beq\label{taucoalcomb}
\tau_{\rm coal,comb}=\frac{48 \pi R^{2}\left[3 v_{\rm gas} v_{\rm th} \ell \rho + a g \rho_{\rm gr} ( 2 a + 3 \ell) \right]}{\dot{M}_{\rm gr} g ( 9 a + 8 \ell )}
\eeq
It is straightforward to show that this expression reduces to Eq. \ref{eq:taucoalE}, \ref{eq:taucoalS}, \ref{eq:taucoalAE}, and \ref{eq:taucoalAS} in the appropriate limits. 

To determine the grain size, one again sets Eq. \ref{tautrav} equal to Eq. \ref{taucoagcomb} and \ref{taucoalcomb} and numerically determines the grain size $a$ in the two regimes. Once it is determined, the process with the shorter timescale is chosen and the other quantities like in particular the opacity can be calculated in an analogous way as in the analytical model.

 \begin{figure*}
\begin{minipage}{0.5\textwidth}
	      \centering
       \includegraphics[width=0.98\textwidth]{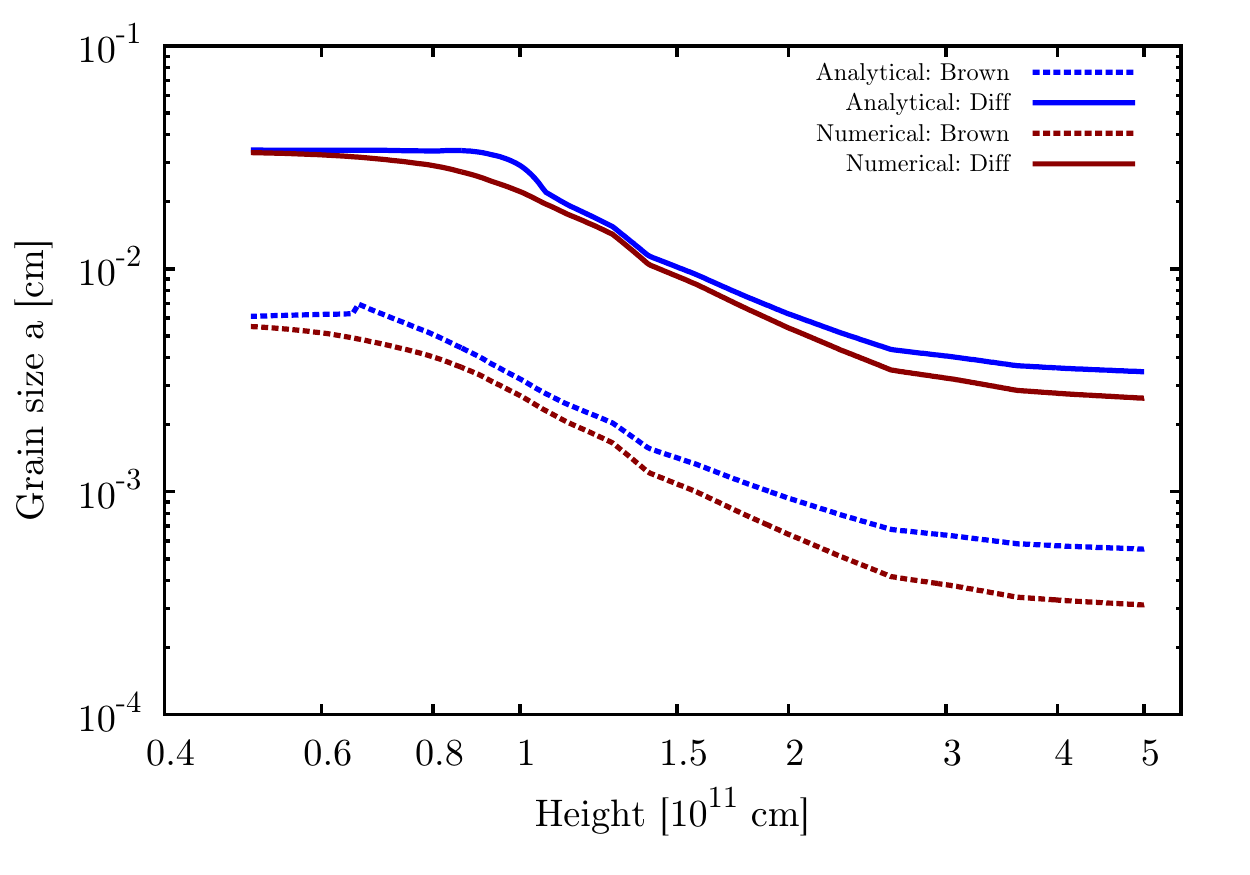}
     \end{minipage}
          \hfill
     \begin{minipage}{0.5\textwidth}
      \centering
                    \includegraphics[width=0.98\textwidth]{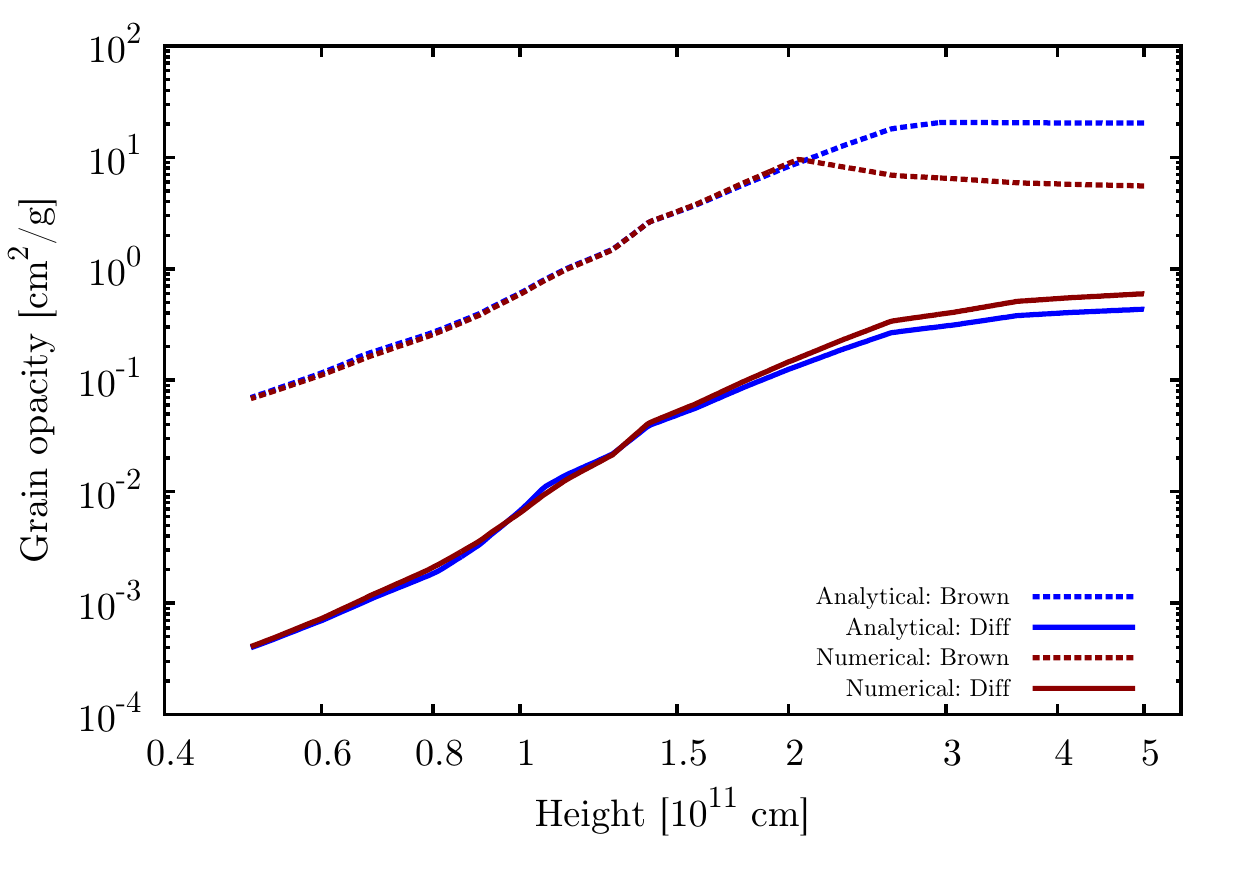}
     \end{minipage} 
                  \caption{\textit{Left panel:} Grain size as a function of height for  different growth mechanisms in the \citetalias{movshovitzpodolak2008} comparison case. The plot shows the size due to growth by Brownian motion coagulation (dotted) and differential settling (solid) as predicted by the analytical expressions and by solving numerally for the grain size. \textit{Right panel:} Corresponding grain opacity. }
             \label{fig:compNumAnMP2008}
\end{figure*}

The left panel of Figure \ref{fig:compNumAnMP2008} shows the grain size in the atmosphere of the \citetalias{movshovitzpodolak2008} comparison case (Fig. \ref{fig:2008rt}), comparing the size as found by the analytical expressions and the numerical solution for $a$. 

In the entire atmosphere, growth by differential settling is found to occur on a shorter timescale than by Brownian motion (Fig. \ref{fig:mp2008taucompcoagcoal}), leading to grains that are about one order of magnitude larger.  The grain size found analytically for differential settling is therefore the same as shown in Fig. \ref{fig:mp08fpg}. In this atmosphere, the advection regime as defined by a $\tau_{\rm adv}$ that is smaller than the growth timescale for $v_{\rm gas}=0$ does not occur. Therefore, advection is completely neglected in the analytical model. The figure shows that this leads to grain sizes that are slightly overestimated relative to the result when numerically solving for $a$. Here the small, but non-zero $v_{\rm gas}=0$ is properly taken into account with the consequence that grains cross one scale height on a somewhat shorter timescale. This reduces their grain size. We  also see the transition from the Epstein to the Stokes regime (at about $7\times10^{10}$ cm) occurs smoothly in the numerical solution, while at least for Brownian  coagulation, a small kink is visible in the analytical model. Both these differences  between the analytical and numerical model are expected. 

The right panel of the figure shows that the effect on the resulting opacity (the one given by growth due to differential settling) is however very small. Only a small increase occurs in the outer layers which do not contribute much to the total optical depth. This justifies that we the neglect advection in the analytical model for this atmosphere. The grain opacity in the upper layers as predicted from Brownian motion coagulation is lower in the numerical model. This comes from the decrease of $Q$ for  smaller grains. In the inner parts, the opacity is virtually identical. This is a consequence from the fact that the non-zero $v_{\rm gas}$ not only decreases the grain size, but also reduces the grain concentration.

To understand the global effects,  we have recalculated the $\Sigma10$ and $\Sigma4$ case, determining $a$ numerically. It is found that the difference is very small. For the $\Sigma10$ case, the crossover time is somewhat increased relative to the analytical model as expected, but only by about 10\%. For the $\Sigma4$ simulation, the difference  virtually vanishes (less than 1 \%). This shows again that advection is more important for more massive cores. It also means that it is not necessary to numerically solve for $a$, at least before crossover and for cores that are not much more massive than in the $\Sigma10$ case.

\section{Differential settling and Brownian motion in the advection regime}\label{sect:diffsetbrownadvection}
\begin{figure}
\begin{center}
\includegraphics[width=\columnwidth]{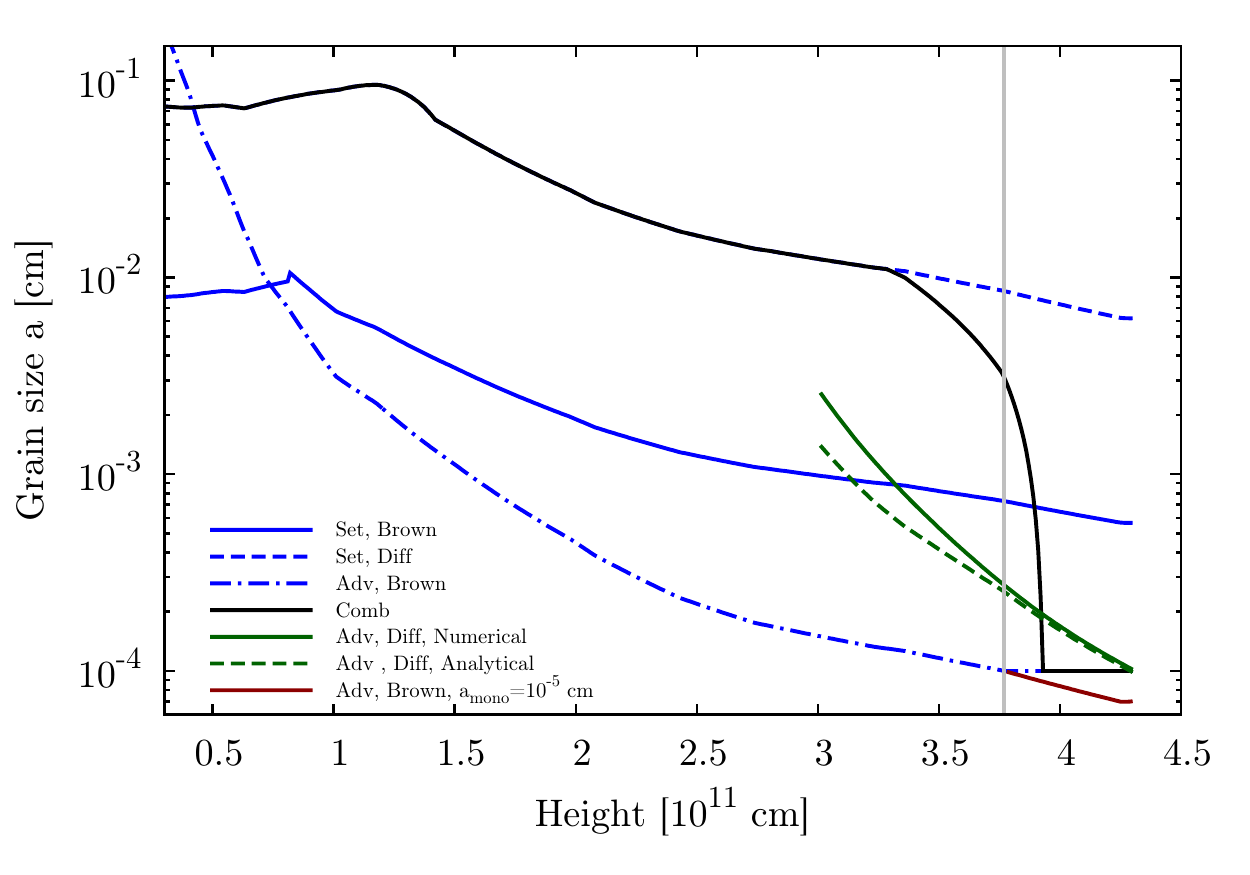}
\caption{Grains size as a function of height for different growth mechanism at crossover in the \citetalias{movshovitzbodenheimer2010} case (initial planetesimal surface density of 10 g/cm$^{2}$). }\label{fig:compNumAnM2010V3}
\end{center}
\end{figure}

Figure \ref{fig:compNumAnM2010V3} shows the grains size a function of height at crossover in the $\Sigma$10 case. The density and temperature of \citetalias{movshovitzbodenheimer2010} shown in Fig. \ref{fig:crossM10} is used to calculate the grain sizes together with a gas accretion rate of $\dot{M}_{\rm XY}=1.8\times10^{-4} \mearth$/yr.  In this atmosphere, the analytical model finds that the growth timescale by differential settling $\tau_{\rm coal,E}$ becomes longer than the advection timescale $\tau_{\rm adv}$ outside of about $3.8\times10^{11}$ cm. This is indicated by a vertical gray line.  Grain growth by differential settling neglecting the effect of envelope contraction (blue dashed line) leads to quite large grains of about 0.01 cm. In contrast, grain growth by Brownian motion coagulation in the advection regime (blue dashed-dotted line) is inefficient, so that grains remain at their smallest allowed size of $1\mu$m. The black line shows the combined interpolated grain size. The brown solid line shows the grain size if no lower limit of $1\mu$m is imposed. Then the grains would have  an even  slightly smaller size of $\gtrsim0.7$ $\mu$m.

The green solid line shows the grain size according to Eq. \ref{eq:acoagAEnum} for growth due to differential settling in the advection regime. It shows that at least according to the description in Sect. \ref{sect:advectbrown}, it seems not to be an extremely efficient process. While it does predict larger grain sizes, the difference to Brownian motion is only about a factor 2 or 3. In the normal regime (no advection), the grains due to differential settling (coalescence) are in contrast typically one order of magnitude larger than for Brownian motion. This means that therefore, even if coalescence is included in the advection regime, there is no large effect (reduction) of the opacity predicted by the analytical model in the outermost layers, at least with the description used here.  While the effect of advection before runaway is as mentioned small, we will nevertheless investigate this regime further (Ormel \& Mordasini in prep.).

Equation \ref{eq:acoagAEnum} gives an estimate of the grain size due to differential settling in the advection regime involving an integral.  To give for completeness an analytical expression also in this regime, we can use that the temperature is roughly speaking constant in the outermost layers (see Figs. \ref{fig:2008rt} and \ref{fig:crossM10}) while the density decreases rapidly towards the exterior roughly as $r^{-4}$. One can then calculate the integral (which gives then the column density) to find
\beq\label{eq:acoagAEanalytical}
a_{\rm coal A, E}\approx a_{0} \exp\left[\frac{4 \pi G \dot{M}_{\rm gr} M}{9 \dot{M}_{\rm XY}^{2}}\sqrt{\frac{\pi \mu m_{\rm H}}{k_{\rm B} T}}\frac{\rho R }{3} \left(1-\frac{R^{3}}{R_{0}^{3}}\right) \right].
\eeq
The grain size obtained with this equation is also shown in Fig. \ref{fig:compNumAnM2010V3} with the green dash line where it provides good approximation to the numerical result. It is however clear that this equation is not in general applicable, as it requires an a priori knowledge of the temperature and density structure.

\bibliographystyle{aa} 
\bibliography{biball2014v1}

\end{document}